\documentclass[superscriptaddress,amsmath,floatfix,aps,prc,twocolumn]{revtex4-2}
\DeclareUnicodeCharacter{03C9}{$\omega$}
\DeclareUnicodeCharacter{03C3}{$\sigma$}

\usepackage{amssymb}  
\usepackage{graphicx}
\usepackage{svg}
\usepackage{exscale}
\usepackage{textcomp}
\usepackage{enumerate}
\usepackage{amsmath}
\usepackage{amssymb}
\usepackage[compat=1.1.0]{tikz-feynhand} 
\usepackage{multirow}

\usepackage{hyperref}
\hypersetup{breaklinks=true, colorlinks=true, citecolor=blue}
\usepackage{color}
\usepackage[normalem]{ulem}  

\usepackage{graphics}
\usepackage{placeins}
\usepackage{xcolor}
\usepackage{natbib}

\usepackage{caption}
\usepackage{tabularx}
\usepackage{booktabs}
\usepackage{float}

\newcolumntype{L}{>{\raggedright\arraybackslash}X}

\usepackage{bm}

\newcommand{\mnras}{Monthly Notices of the RAS}
\newcommand{\apjl}{Astrophysical Journal Letters}
\newcommand{\aap}{Astronomy and Astrophysics Journal}

\newcommand{\araa}{Annual Review of Astronomy and Astrophysics}
\newcommand{\pasj}{Publications of the Astronomical Society of Japan}

\def\beq{\begin{equation}}
\def\eeq{\end{equation}}
\def\bea{\begin{eqnarray}}
\def\eea{\end{eqnarray}}

\def\phi{\mathrm{\varphi}}

\newcommand{\dd}{\mathrm{d}}
\newcommand{\Schw}{Schwarzschild }
\newcommand{\ce}{{\cal{E}}} 
\newcommand{\cl}{{\cal{L}}} 
\newcommand{\cb}{{\cal{B}}} 
 
\newcommand{\Ln}{{{\rm Ln}(r)}}
\newcommand{\LnD}{{{\rm Ln}^2(r)}} 
\newcommand{\coff}{{\rm Coff}} 
\newcommand{\risco}{{\rm rISCO}}
\newcommand{\tisco}{{\rm {\theta}ISCO}}

\newcommand{\SQRT}{{\triangle(r,b)}}

\begin{document}


\title[Relativistic dipole neutron stars and RR]{Radiative Back-Reaction on Charged Particle Motion in the Dipole Magnetosphere of Neutron Stars}
\author{Zden\v{e}k Stuchl\'{\i}k}
\author{Jaroslav Vrba}
\author{Martin  Kolo\v{s}}
\author{Arman Tursunov}

\email{zdenek.stuchlik@physics.slu.cz}
\email{jaroslav.vrba@physics.slu.cz}
\email{martin.kolos@physics.slu.cz}
\email{arman.tursunov@physics.slu.cz}

\affiliation{
 Research Centre for Theoretical Physics and Astrophysics,\\ Institute of Physics, Silesian University in Opava,\\ Bezru\v{c}ovo n\'am.~13, 746\,01 Opava, CZ }
 %

\begin{abstract}
The motion of charged particles under the Lorentz force in the magnetosphere of neutron stars, represented by a dipole field in the Schwarzschild spacetime, can be determined by an effective potential, whose local extrema govern circular orbits both in and off the equatorial plane, which coincides with the symmetry plane of the dipole field. In this work, we provide a detailed description of the properties of these "conservative" circular orbits and, using the approximation represented by the Landau-Lifshitz equation, examine the role of the radiative back-reaction force that influences the motion of charged particles following both the in and off equatorial circular orbits, as well as the chaotic orbits confined to belts centered around the circular orbits. To provide clear insight into these dynamics, we compare particle motion with and without the back-reaction force. We demonstrate that, in the case of an attractive Lorentz force, the back-reaction leads to the charged particles falling onto the neutron star's surface in all scenarios considered. For the repulsive Lorentz force, in combination with the back-reaction force, we observe a widening of stable equatorial circular orbits; the off-equatorial orbits shift toward the equatorial plane and subsequently widen if they are sufficiently close to the plane. Otherwise, the off-equatorial orbits evolve toward the neutron star surface. The critical latitude, which separates orbital widening from falling onto the surface, is determined numerically as a function of the electromagnetic interaction's intensity.
\end{abstract}


\maketitle

\section{Introduction}
Observations of the vicinity of black holes (BHs) and neutron stars (NSs) clearly indicate that such objects possess extremely strong magnetic fields. The magnetic field intensity near the surface of NSs ranges from $10^{8}~$G in the old binary systems, to $10^{15}~$G in young magnetars \cite{Popov:2023:IAUS:}. Around stellar-mass BHs in binary systems, the magnetic field intensity can approach $10^{8}~$G, while around supermassive BHs it can reach $10^{5}~$G \cite{Eat-etal:2013:Nat:,Gol-etal:2017:ApJ:,Dal:2019:ApJ:,Pio-etal:2020:MNRAS:}. Therefore, magnetic fields can be of enormous importance in physical processes around NSs, BHs or other compact objects \cite{Ruf-Tre:1973:APL:,Pra:1980:RNC:,Ruf-Wil:1975:PRD:,Dhu-Dad:1984:PRD:,Par-etal:1986:APJ:,Bla-Zna:1977:MNRAS:,Dad-etal:2018:MNRAS:,Stu-etal:2020:Uni:,Kol-Tur-Stu:2021:PRD:,Stu-Kol-Tur:2021:Uni:}. 

Of particular interest is the potential influence of the magnetic field on the epicyclic oscillatory motion of charged objects (considered in approximation of charged test particles) around stable circular orbits. These oscillations produce radiation that could explain the high-frequency quasi-periodic oscillations (HF QPOs) observed in X-ray emissions from binary systems containing NSs (quark stars) or BHs \cite{Stu-Kot-Tor:2013:ASTRA:,Kol-Stu-Tur:2015:CQG:,Kol-Tur-Stu:2017:EPJC:,Pan-Kol-Stu:2019:EPJC:,Stu-etal:2020:Uni:,Stu-Kol-Tur:2022:PASJ:}. \footnote{The tidal charge in the braneworld models of NSs \cite{Dad-etal:2000:PLB:} can be also well applied for fitting the data of HF QPOs in binary systems containing a NS \cite{Kot-Stu-Tor:2008:CQG:}}. 

The existence and interrelation of circular orbits both in and off the equatorial plane of a magnetized neutron star (NS), described by a simplified model using non-rotating Schwarzschild spacetime for its external gravitational field and a dipole magnetic field that defines the equatorial plane and symmetry axis, were studied in \cite{Vrb-Kol-Stu:2024:submitted:}. Their stability against radial and latitudinal perturbations, as well as the possibility of chaotic charged particle motion in belts concentrated around both types of circular orbits, were also examined. The frequencies of epicyclic (regular harmonic) oscillatory motion in both the radial and latitudinal directions around the circular orbits were determined, along with the orbital frequency of the circular motion. The possibility of fitting the twin-peak high-frequency quasi-periodic oscillation (HF QPO) X-ray data observed in NS sources was demonstrated for oscillations around both the in and off circular orbits using the geodesic model \cite{Tor-etal:2010:ApJ:,Stu-Kot-Tor:2013:ASTRA:} modified by the interaction of charged matter (hot spots treated as charged test particles) with the magnetic field \cite{Kol-Tur-Stu:2017:EPJC:,Stu-Kol-Tur:2022:PASJ:}. This holds for cases where the electromagnetic interaction is significantly weaker than the gravitational interaction. In binary systems with NSs exhibiting extremely strong magnetic fields, the assumed radiating hot spots (such as plasmoids or objects similar to ball lightning) can carry a very small specific charge, analogous to that of charged dust \cite{Vrb-Kol-Stu:2024:submitted:}. 

\footnote{The Lorentz force acting on electrons or protons (ions) in vicinity of the event horizon of magnetized rotating BHs, or the surface of magnetized rotating NSs can cause their enormous acceleration \cite{Stu-Kol:2016:EPJC:,Tur-Stu-etal:2020:ApJ:,Stu-etal:2020:Uni:,Stu-Kol-Tur:2021:Uni:} because of the electric component of the electromagnetic field generated by rotation of the compact object spacetime. For non-rotating compact objects such an extreme acceleration is not possible \cite{Kol-Stu-Tur:2015:CQG:}.}

In the present paper, we extend the previous study \cite{Vrb-Kol-Stu:2024:submitted:} by providing a detailed discussion of the properties and distribution of the off-equatorial circular orbits, as well as their energy in relation to the energy of the equatorial circular orbits with the same specific angular momentum. This relationship can be used to determine the characteristics of chaotic motion within the closed effective potential barriers concentrated around the stable circular orbits, as well as the potential for escape onto the neutron star (NS) surface due to the opening of these barriers. \footnote{Recall that in the charged Kerr-Newman-(de Sitter) spacetimes the motion of charged test particles is fully regular and can be given in terms of elliptic integrals \cite{Mis-Tho-Whe:1973:Gravitation:,Bal-Bic-Stu:1989:BAC:,Carter:1973:blho:,Stu:1983:BAC:,Kra:2005:CQG:,Kra:2021:EPJC:}.} 

However, the crucial focus of the present study is the inclusion of the influence of the back-reaction force acting on radiating charged test particles \cite{Tur-etal:2018:ApJ:,Stu-etal:2020:Uni:,Stu-Kol-Tur:2021:Uni:,Stu-Kol-Tur-Gal:2024:Uni:}. In the presence of strong magnetic fields around NSs, the fully general relativistic DeWitt-Brehme equations of motion under back-reaction forces \cite{DeW-Bre:1960:AnnPhys:} can be simplified to the much more manageable Landau-Lifshitz equation \cite{Lan-Lif:1975:Book:}, which is applied in this paper. To gain deeper insight into the nature of chaotic charged particle motion, we compare particle motion under the Lorentz force with that under the combined effect of the Lorentz force and the radiative back-reaction force. Our focus is on the evolution of charged particle motion, whether starting from stable in and off-equatorial circular orbits or from chaotic motion within belts, originating within a closed potential barrier and ultimately escaping through an open barrier onto the NS surface.

We track the influence of the back-reaction force and the possible conversion of chaotic motion, due to energy loss, into a direct fall of the test particle along magnetic field lines onto the NS surface. Conversely, we also examine the widening of circular orbits, as previously observed in the case of motion within an asymptotically uniform magnetic field around a BH \cite{Tur-Kol-Stu:2018:AN:,Stu-Kol-Tur:2021:Uni:,Stu-Kol-Tur-Gal:2024:Uni:}.

To test realistic astrophysical conditions, as in \cite{Vrb-Kol-Stu:2024:submitted:}, we consider the case of standard, non-extreme NSs, placing the NS surface at $R=3M$, which corresponds to the minimal non-extreme radius related to the photon sphere in the Schwarzschild spacetime.

Throughout this paper we use space-like signature \mbox{$(-,+,+,+)$}, and geometric system of units in which $G = c = 1$; we restore them when we need to compare our results with observational data. Greek indices run from $0-3$, Latin indices from $1-3$.

\section{Magnetized neutron stars}
In the present paper we apply the simplest fully general relativistic model of magnetized NS exterior by using the static and spherically symmetric Schwarzschild geometry \cite{Mis-Tho-Whe:1973:Gravitation:}, and the dipole magnetic field introduced in \cite{Pet:1974:PRD:,Was-Sha:1983:ApJ:,Bra-Rom:2001:ApJ:}. 

\subsection{Spacetime geometry}

In the standard Schwarzschild coordinates ($t,r,\theta,\varphi$) the line element of the Schwarzschild geometry takes the form 
\begin{equation}
	\mathrm{d}s^2=-f(r)\dd t^2+f(r)^{-1}\dd r^2 +r^2\left(\dd\theta^2 + \sin^2\theta \dd\varphi^2\right),
	\label{e:stint}
\end{equation} 
where
\begin{equation}
	f(r)=1-\frac{2M}{r}, 
	\label{e:lapsf}
\end{equation}
and $M$ denotes the mass of the neutron star.

\subsection{Magnetic field}

We assume the dipole magnetic field of considered NSs is defined as in \cite{Was-Sha:1983:ApJ:,Bra-Rom:2001:ApJ:} \footnote{This dipole field is identical to those generated by an electric current loop orbiting in the equatorial plane of the Schwarzschild geometry \cite{Pet:1974:PRD:,Kov-Stu-Kar:2008:CQG:,Pre:2004:CQG:}.} The four-vector electromagnetic potential $A^\mu$ of the dipole field of spherically symmetric neutron star, with gravitational field given by the \Schw metric, takes the form $A_\mu = (0,\, 0,\, 0,\, A_\varphi)$, where the only non-zero component is given by 
\begin{eqnarray}
    A_\varphi&=& - \cb 	\, h_1(r,\theta) = - \cb \, h(r) \, g_{\phi\phi}(r,\theta) 
	\label{e:aphi}
\end{eqnarray}
where
\begin{eqnarray}
   &&h(r)= \ln\left( 1 - \frac{2M}{r} \right) + \frac{2M}{r} \left( 1+ \frac{M}{r} \right),    
	\label{e:aphi-h}
\end{eqnarray}
the parameter governing the dipole magnetic field takes the form 
\beq 
\cb = \frac{3\mu}{8 M^3} \label{e:param}
\eeq
and $\mu$ is the magnetic dipole moment of the NS. Notice that at all $r>2M$ the function $h(r)<0$, and it vanishes for $r \to \infty$. 

Using the \Schw metric allows us to approach its event horizon. However, since we are considering NSs, it is important to note that the dipole magnetic field and the motion of the charged particles are constrained by the NS surface. In our study, we assume the NSs surface to be at $R=3M$, a value used in figures illustrating various phenomena related to the magnetized NSs.

The Maxwell tensor, $F_{\mu\nu} = A_{\nu,\mu}-A_{\mu,\nu}$ has only two non-zero components:
\bea
    F_{r\phi} &=& \frac{\partial{A_{\phi}}}{\partial{r}} = B^{\theta} \nonumber \\
    &=&  2 \cb \sin ^2\theta  \left(\frac{2 M (r-M)}{2 M-r}-r \ln \frac{r-2 M}{r}\right)
\eea
and
\bea
    F_{\theta\phi} &=& \frac{\partial{A_{\phi}}}{\partial{\theta}} = B^{r} \nonumber \\
    &=&-\cb \sin (2 \theta ) \left[r^2 \ln \frac{r-2 M}{r}+2 M (M+r)\right]
\eea
In the equatorial plane ($\theta=\pi/2$) $B^{r}$ vanishes, while on the symmetry axis ($\theta=0$) both $B^{r}$ and $B^{\theta}$ vanish. As a result, along the symmetry axis electromagnetic forces vanish, leaving only gravitational attraction. For this reason, no equilibrium points can exist along the axis of the magnetosphere with the dipole magnetic field. 

\begin{figure}
	\includegraphics[width=1\hsize]{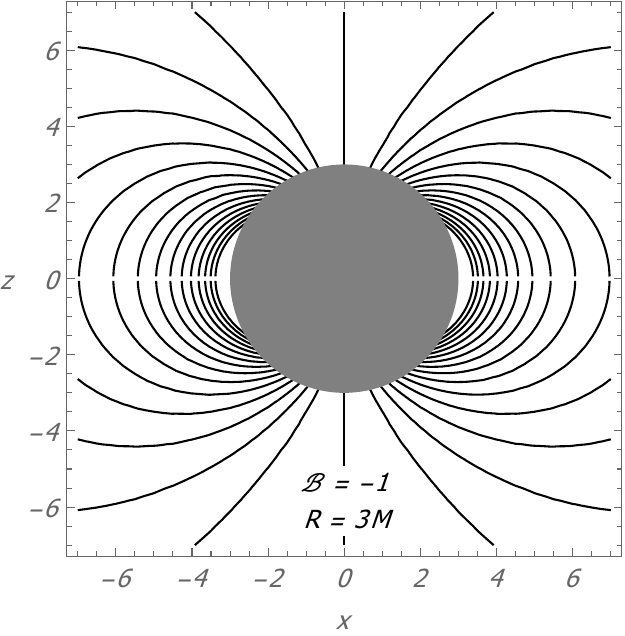}
	\caption{The equipotential surfaces of the 4-vector potential of the dipole magnetic field are constructed for the case of $\cb=-1$ and the surface of the neutron star at $R=3M$. \label{f:f0}}
\end{figure}

The electromagnetic potential (\ref{e:aphi}) of the dipole magnetic field is represented by the meridional sections of the equipotential surfaces in FIG.~\ref{f:f0}.

It is useful to express the magnetic dipole moment in terms of the magnetic field intensity $B=B^{\theta}$ measured in the equatorial plane at the NS surface $r=R$. Using the magnetic field intensity $B^{\hat{\theta}}$ measured by local static observers at the NS surface, the magnetic dipole moment can be expressed as \cite{Bak-etal:2010:CQG:}
\beq
    \mu = \frac{4M^3R^{3/2} \sqrt{R - 2M}}{6M(R - M) + 3R(R - 2M) \ln f(R)} B^{\hat{\theta}}.
\eeq

\subsection{Test model of magnetized neutron stars}

To illustrate the phenomena related to the charged particle motion in the NS magnetosphere, it is useful to choose a typical "test" model of the magnetic NSs.

In the Schwarzschild spacetime, particle motion can be considered and observed over the whole range $r>2M$. However, in this study, we limit our analysis to the region $r>R$, where $R$ represents the NS surface at radius $R>2M$. Realistic equations of state limit the surface radius to the range $3M<R<4M$ (see, e.g., \cite{Urb-Mil-Stu:2013:MNRAS:}), but the theoretical model of significantly simplified internal Schwarzschild spacetime with uniform distribution of energy density gives the innermost limit of $R=2.25M$. The more realistic Tolman VII model \cite{Tol:1939:PR:,Jia-Yag:2019:PRD:}, which uses a quadratic energy density distribution law, places the lower limit at $R=2.6M$ \cite{Stu-Vrb:2021:EPJP1:}, still considerably lower than the extremely compact NSs limiting value of $R=3M$ corresponding to the photon circular orbit of the external vacuum \Schw spacetime. In the extremely compact spacetimes extraordinary phenomena as centrifugal force inversion occur \cite{Abr-Mil-Stu:1993:PRD}. Polytropic equations of state also predict extremely compact objects corresponding to NSs having $R<3M$ \cite{Stu-Hle-Nov:2016:PRD:,Nov-Stu-Hla:2017:PRD:}. Standard realistic equations of state relating energy density and pressure typically yield the surface radii near $R=3.5M$ \cite{Gle:2000:BOOK:}, while the realistic model of the so-called Q-stars gives the limit $R=2.8M$ \cite{Bah-Bry-Sel:1989:NPB:}. Therefore, for the purposes of this study, we assume for our test NS applied in realistic astrophysical situations a surface radius of $R=3M$, which maximizes the standard region of $R$ predicted by the realistic models of NSs \cite{Web:2017:BOOK:,Urb-Mil-Stu:2013:MNRAS:}. 

Furthermore, we select for our test magnetized NS model the mass parameter $M=2M_{\odot}$, which is significantly above the "canonical" value of $M=1.4M_{\odot}$. This increase of the NS mass can be attributed to accretion from a companion star in observed old low-mass binary systems \cite{Klis:2000:araa:}. On the other hand, this choice remains well below the expected maximum mass $M_{\rm max}=2.5M_{\odot}$ \cite{Gle:2000:BOOK:,Web:2017:BOOK:,Urb-Mil-Stu:2013:MNRAS:}, as there are only exceptional observations indicating NS mass $M>2M_{\odot}$ \cite{Eek-Rez:2023:MNRAS:}. 

The magnetic field observed near the NS surface ranges from $10^8G$ to $10^{12})G$ in standard binary systems and can reach up to $10^{15}$~G or higher in magnetars \cite{Popov:2016:AAT:}. For our test magnetized NSs, we choose the value of $B_{\rm surf} = 10^{8}$~G, typical of many NSs in binary systems. 

Using the selected characteristic global parameters for the test magnetized NS, $R=3M$, $M_{\rm max}=2M_{\odot}$, $B_{\rm surf} = 10^{8}$~G, we obtain its dipole magnetic moment $\mu \doteq 0.74\times10^{-4} m^2$.

\section{Charged particle dynamics governed by the Lorentz equation of motion}
We first summarize and extend our previous study \cite{Vrb-Kol-Stu:2024:submitted:} on the conservative charged test particle motion under the Lorentz force. 

The Lorentz equation of the charged test particle motion is given by  
\begin{equation}
   \frac{\dd u^\mu}{\dd \tau} + \Gamma_{\alpha\beta}^{\mu} u^{\alpha} u^{\beta} = \frac{q}{m} F^\mu_{\alpha} u^{\alpha} ,
\end{equation}
where $u^\mu$ is the four-velocity of the particle with the mass $m$ and charge $q$, normalized by the condition
\mbox{$u_\alpha u^\alpha=-1$}. Here, $\tau$ is the proper time of the particle, and $F_{\mu\nu}$ is the antisymmetric Maxwell tensor of the electromagnetic field.

The Lorentz equation for the motion around magnetized NSs can be effectively treated within the Hamiltonian formalism \cite{Mis-Tho-Whe:1973:Gravitation:,Kol-Stu-Tur:2015:CQG:}. The Hamiltonian $H$ is expressed as:  
\begin{equation}
    H =  \frac{1}{2} g^{\alpha\beta} \left(\pi_\alpha - q A_\alpha\right)\left(\pi_\beta - q A_\beta\right) + \frac{m^2}{2},
    \label{e:ham}
\end{equation}
where canonical four-momentum $\pi_\mu$ is given by the relation $\pi_\mu= mu_\mu + q A_\mu$ and the Hamilton equations take the form 
\begin{equation}
    \frac{\dd x^\mu}{\dd \zeta} = \frac{\partial H}{\partial \pi_\mu}, \quad \frac{\dd \pi_\mu}{\dd \zeta} = - \frac{\partial H}{\partial x^\mu},
    \label{e:heq}
\end{equation}
with $\zeta=\tau/m$. The symmetries of the magnetized NS background (\Schw spacetime (\ref{e:stint}) and dipole magnetic field (\ref{e:aphi})) lead to the conservation of two components of the specific particle's canonical four-momentum: covariant specific energy and covariant specific axial angular momentum: 
\begin{equation}
    \ce = \frac{E}{m} = - \frac{\pi_t}{m}  = - g_{tt} u^t, \quad  \cl = \frac{L}{m} = \frac{\pi_\varphi}{m}  = g_{\varphi\varphi} u^\varphi + \bar q A_\varphi,
    \label{e:const}
\end{equation}
where $\bar q=q/m$ is the specific charge of the particle. Using conserved quantities (\ref{e:const}), the Hamiltonian (Eq.~(\ref{e:ham})) can be written in the form
\begin{eqnarray}
    \frac{H}{m^2} &=& \frac{1}{2}g^{rr} u_r^2 + \frac{1}{2}g^{\theta\theta} u_\theta^2 + \frac{1}{2}g^{tt} \ce^2 + \frac{1}{2} g^{\varphi\varphi} (\cl-\bar q A_{\varphi})^2 + \frac{1}{2}\nonumber\\
\end{eqnarray}
that can be separated into a dynamical part, $H_{\pm D}$, and a potential part, $H_{\rm P}$, given by the relations 
\begin{eqnarray}
&&H_{\mathrm{D}}= \frac{1}{2}\Big(g^{rr}u_r^2 +g^{\theta\theta}u_\theta^2\Big),\\
&&H_{\mathrm{P}}= \frac{1}{2}\Big[g^{tt}\mathcal{E}^2+g^{\varphi\varphi}\left( \cl - \bar q A_{\varphi} \right)^2+1\Big].\label{e:hpot}
\end{eqnarray}
The motion of charged particles is characterized by the turning points determined by the condition $H_P=0$, which governs the regions of spacetime available in the $r-\theta$ plane for particles with fixed motion constants $E,L$. 

\subsection{Effective potential of the motion, circular orbits and their stability}
For motion in the field of magnetized NSs governed by the Lorentz equation, we can introduce a 2D effective potential related to the particle's covariant energy. The effective potential acts as a barrier that determines the particle's motion based on the motion constants and the parameters of the background. It allows for a straightforward determination of the circular orbits and the epicyclic oscillations of harmonic character around the stable circular orbits. However, in general, the motion is chaotic \cite{Stu-etal:2020:Uni:}. 

In the following, unless otherwise stated, we simplify the calculations by putting $M=1$, expressing distances in units of the mass parameter $M$. 

The effective potential $V_{\rm eff}(r,\theta;\cl,b)$ is determined by the formula \cite{Kov-Stu-Kar:2008:CQG:,Kol-Stu-Tur:2015:CQG:}
\begin{eqnarray}
  &&  V_{\rm eff} (r,\theta) \equiv -g_{tt} \left[ g^{\varphi\varphi} \left( \cl - \bar q A_{\varphi} \right)^2 + 1 \right]\\  \nonumber
    &&= \left(1-\frac{2}{r} \right) \left[ \left( \frac{\cl}{r \sin(\theta)} - b \,h(r)\, r \sin(\theta) \right)^2 + 1 \right].
    \label{e:veff}
\end{eqnarray}

In a fixed background, the effective potential is thus determined by the motion constant $\cl$ and the "magnetic parameter" $b$, which defines the intensity of the particle interaction with the given magnetic field
\beq
    b = \bar{q} \cb = {\rm const} = \frac{q}{m} \frac{3\mu}{8M^3} .
\eeq 
For a particle with covariant specific energy $\mathcal{E}$, the turning points are determined by the condition $\mathcal{E}^2 = V_{\rm eff}(r,\theta;\cl,b)$. The effective potential is symmetric relative to simultaneous change of sign of the specific angular momentum $\cl$, and the magnetic parameter $b$. 

Circular orbits are governed by the local extrema of the effective potential, which is determined by vanishing of its derivatives with respect to the coordinates $r$ and $\theta$. Thus, the conditions for circular orbits are: $\frac{\dd V_{\mathrm{eff}}}{\dd r}=0$ and $\frac{\dd V_{\mathrm{eff}}}{\dd \theta}=0$. For circular orbits, both derivatives must vanish simultaneously. 

The orbital frequency of the circular motion, relative to the proper time of the orbiting particle, is determined by the relation 
\bea
\omega_{\phi}(r,\theta) = \frac{\dd \varphi}{\dd \tau} = \cl g^{\varphi\varphi}(r,\theta) - b  h_1(r,\theta).
    \label{e:freqgen}
\eea

Negative (positive) values of the magnetic parameter $b$ correspond to magnetic repulsion (attraction), where the radial component of the electromagnetic Lorentz force is directed outward (inward) for corotating particles with $\omega_{\phi}>0$. For counter-rotating charged particles with $\omega_{\phi}<0$, the force is directed oppositely. As shown in \cite{Kov-Stu-Kar:2008:CQG:,Vrb-Kol-Stu:2024:submitted:}, the background configuration of the magnetized NSs considered here allows for the existence of off-equatorial circular orbits in the case of the magnetic repulsion ($b<0$). 

The stable circular motion requires that the effective potential has a minimum, which is governed by the second derivatives of the effective potential with respect to the coordinates $r, \theta$.  For equatorial circular orbits, stability and frequencies of the associated epicyclic motion were discussed in \cite{Bak-etal:2010:CQG:} by using the method of perturbations of circular motion \cite{Ali-Gal:1981:GRG:}. The existence of off-equatorial circular orbits was demonstrated using the effective potential in \cite{Kov-Stu-Kar:2008:CQG:}, and their stability was addressed in \cite{Vrb-Kol-Stu:2024:submitted:}. In this work, we extend these studies using the method of the effective potential $V_{\rm eff}(r,\theta;\cl,b)$. For equatorial circular orbits, second derivatives with respect to the radial and latitudinal coordinates are sufficient. However, for off-equatorial orbits, a more comprehensive analysis is required \cite{Vrb-Kol-Stu:2024:submitted:} -- the local extrema of a function of two variables are governed by the matrix of second derivatives, known as the Hessian matrix:
\begin{equation}
H = \begin{bmatrix}
\frac{\partial^2 V_{\textrm{eff}}}{\partial r^2} & \frac{\partial^2 V_{\textrm{eff}}}{\partial r \partial \theta} \\
\frac{\partial^2 V_{\textrm{eff}}}{\partial \theta \partial r} & \frac{\partial^2 V_{\textrm{eff}}}{\partial \theta^2}
\end{bmatrix}.
\label{e:hess}
\end{equation}

The detailed form of the second derivatives of the effective potential can be found in \cite{Vrb-Kol-Stu:2024:submitted:}. Here, we give a brief summary of the results obtained for both equatorial and off-equatorial orbits, along with an extension illustrating the behavior of the effective potential relevant for describing the motion of charged particles in the belts surrounding the circular orbits. This is particularly important when considering the influence of the radiative back-reaction forces on particle motion. 

\subsection{Equatorial circular orbits}

Since \mbox{$\dd V_{\mathrm{eff}}/\dd \theta=0$} at $\theta=\pi/2$, independently on the radius $r$, the equatorial circular orbits are given by the condition \mbox{$\dd V_\mathrm{eff}/\dd r=0$} at fixed $\theta=\pi/2$.
Therefore, the relation for the axial component of the specific angular momentum of charged particles $\cl$ with a fixed magnetic parameter $b$ at circular orbits in the equatorial plane reads 
\begin{eqnarray}
  &&  \cl_{\rm c}= \frac{b}{r-3} \left[r^2 \, \Ln+2 r+6\right] \pm r\frac{ \SQRT}{r-3},
\end{eqnarray}
where 
\beq 
 \SQRT=\sqrt{b^2 \left[2 (r-1)+(r-2) r \, \Ln\right]^2+r-3} , 
\eeq
and
\beq
         \Ln = \ln (1 - \frac{2}{r}) . 
\eeq

The specific energy of the particles at these circular orbits, $\ce^2_\mathrm{c}=V_\mathrm{eff}(r,\cl=\cl_\mathrm{c})$, reads
\begin{eqnarray}
  &&  \ce_{\rm c}= r^{-1}\sqrt{1-\frac{2}{r}} \Bigg\{-b r^2 +\Bigg[r\frac{\pm \SQRT}{r-3} \\  &&+ \frac{b r^2 \, \Ln+2 b r+6 b}{r-3}-r^2 \, b\Ln-2b(r+1)\Bigg]^2\Bigg\}^{\frac{1}{2}}. \nonumber
\end{eqnarray}

Since we are considering the motion of charged test particles in the region $r>R=3$, we restrict our attention to the solutions for $\cl$ and $\ce$ given by the upper sign. Due to the symmetry of the effective potential with respect to the sign change of $\cl$ and $b$, we focus on the corotating orbits with $\omega_{\phi}>0$. These orbits generally have an axial component of the specific angular momentum $\cl>0$. However, there is a possible transition to negative values of $\cl$ for corotating orbits, which occurs at radii determined by the relation 
\bea
   && b_{\pm(\cl=0)}(r) =  \\ 
    && \pm \frac{r}{2}[\frac{1-r}{4}r^4\Ln^{2} - (r+1)(r^2+3) -(r^{4}+r^{2})\Ln]^{-1}.\nonumber 
\eea

For corotating orbits, this relation is relevant for the negatively charged particles, while for the counter-rotating orbits it applies to positively charged particles. 

The equatorial circular orbits can exist at $r>3$, though both the specific angular momentum and specific energy diverge as $r \to 3$; at $r=3$, the equatorial photon circular orbit is located. The reality condition 
\beq
    \left[2 b (r-1)+ b (r-2) r \ln \frac{r-2}{r}\right]^2+r-3 \geq 0
    \label{e:Lreal}
\eeq
governs the existence of circular orbits at the radii $2<r<3$, which is irrelevant for our study since we are focusing on the exterior of the magnetized NSs with a surface at $R=3$. 
\begin{figure*}
   \centering
	\includegraphics[width=0.67\hsize]{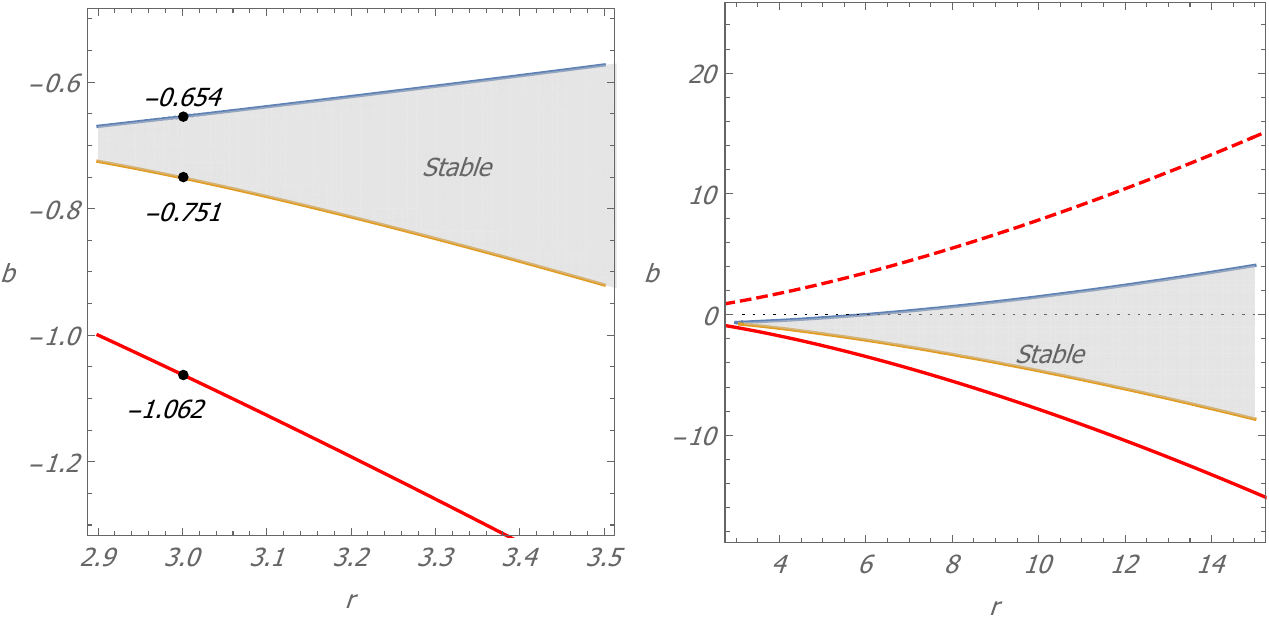}
	\caption{\label{f:f22}{Characteristic stability functions $b_\risco(r)$ (blue) and $b_\tisco(r)$ (yellow) determining the position of the equatorial circular orbits marginally stable against the radial and latitudinal perturbations. The stability region is shaded. The functions $b_{\pm\cl=0}(r)$ (red) determine vanishing of the specific angular momentum; the full line is relevant for the corotating orbits under consideration, while the dotted line is relevant for counter-rotating orbits. The left panel represents a zoomed area around the radius of the testing NS (R=3) and on the right panel is a global representative view.}}
\end{figure*}

\begin{figure*}
   \centering
	\includegraphics[width=0.7\hsize]{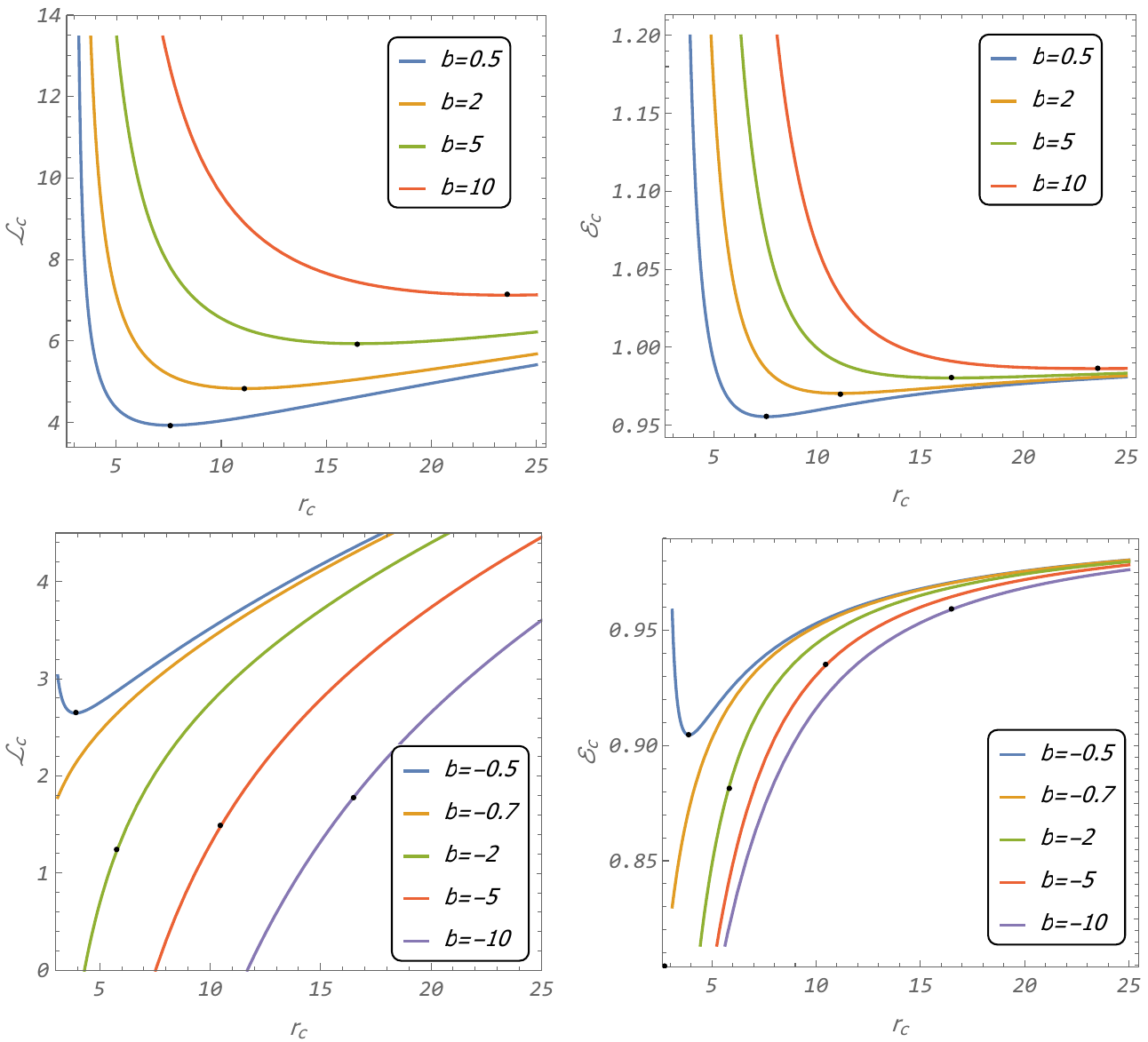}
	\caption{\label{f:f21}Radial profiles of the specific energy and the specific angular momentum at the equatorial circular orbits. Points depict the marginally stable orbits.}
\end{figure*}
The stability of the equatorial circular motion in the magnetized NS backgrounds, depending on the sign of the magnetic parameter $b$, is influenced not only by the radial perturbations in the equatorial plane (as is the case of uncharged particles) but also to off-equatorial, latitudinal perturbations. The $\theta$-instability significantly alters the situation and is directly linked to the existence of off-equatorial circular orbits \cite{Vrb-Kol-Stu:2024:submitted:}. 

The marginally stable equatorial circular orbits are determined by the conditions $\dd^2 V_\mathrm{eff}/\dd r^2 = 0$ and $\dd^2 V_\mathrm{eff}/\dd \theta^2 = 0$, which have to be evaluated at a given radius $r>R=3$, with $\cl = \cl_{c}(r)$ and $\theta=\pi/2$. Therefore, the marginal stability of the equatorial circular orbits is determined by two critical radial functions of the magnetic parameter $b$. 

The marginal stability with respect to radial perturbations, governed by $\dd^2 V_\mathrm{eff}/\dd r^2 = 0$, is characterized by the critical function $b_\risco(r)$, which can be determined from the condition \cite{Vrb-Kol-Stu:2024:submitted:}  
\begin{widetext}
\bea
&&0=(r-2) \left\{8 b^2 (r-1) \left[r (2 r-9)+3\right]+(r-6) (r-3)\right\} +2b\Bigg\{2 \left[(r-9) r+12\right] \times \\
&& \sqrt{4 b^2 (r-1)^2+b^2 (r-2) r \, \Ln \left[4 (r-1)+(r-2) r \, \Ln\right]+r-3}+\nonumber \\
&& (r-2)\, \Ln\Big[4 b r \left[r (2 r-13)+22\right]-36b+b (r-4) (r-2) r (2 r-3) \ln (r-2)+\nonumber\\
&& (r-6) \sqrt{4 b^2 (r-1)^2+b^2 (r-2) r \, \Ln \left[4 (r-1)+(r-2) r \, \Ln\right]+r-3}-b (r-4) (r-2) r (2 r-3) \ln r \Big]\Bigg\} \nonumber
\eea
\end{widetext}
Due to the complex nature of the condition presented above, the function $b_\risco(r)$ is determined using numerical methods. 

The marginal stability with respect to latitudinal perturbations, governed by $\dd^2 V_\mathrm{eff}/\dd \theta^2 = 0$, yields the critical function $b_\tisco(r)$ in a relatively simple analytic form 
\bea
b_\tisco(r) &=&-\frac{r}{2} \Big[-r^4 \LnD -4 r^3 \, \Ln \label{e:btheta}\\
&&-4 r^2\left(1+2 r^2 \, \Ln\right)-16 r-12\Big]^{-1/2}.
\nonumber 
\eea

The functions $b_\risco(r)$, $b_\tisco(r)$ and $b_{\pm(\cl=0)}(r)$ are represented in FIG.~\ref{f:f22}. In the region of interest, $r \geq 3$, the function $b_\risco(r)$ is increasing with radius starting from the initial point $b_\risco(r=3)=-0.654$, while the function $b_\tisco(r)$ is decreasing starting from the initial point $b_\tisco(r=3)=-0.751$. Note the distinction between the cases of Lorentz attractive forces (upper row) and Lorentz repulsive forces (lower row), as well as the special case $b=-0.5$, where the repulsive Lorentz force yields equatorial circular orbit properties similar to those in the Lorentz attraction case.

The function $b_{\pm(\cl=0)}(r)$ is increasing (decreasing) for particles with $b>0$ ($b<0$). At the limiting radius of the NS, the limiting values are $b_{-(\cl=0)}(r=3)=-1.062$ ($b_{+(\cl=0)}(r=3)=1.062$). For corotating orbits, the function $b_{-(\cl=0)}(r)$ is relevant in the case of magnetic repulsion $b<0$, while the function $b_{+(\cl=0)}(r)$, applicable in the region of magnetic attraction $b>0$, is relevant for counter-rotating particles. 

Illustrative radial profiles of the specific angular momentum $\cl(r;b)$ and specific energy $\ce(r,b)$ for corotating orbits and typical values of the magnetic parameter $b$ are presented in FIG.~\ref{f:f21}. For magnetic attraction ($b>0$), the radial profiles $\cl(r;b)$ and $\ce(r,b)$ resemble those of geodesic circular orbits in Schwarzschild spacetime, with minima corresponding to the rISCO. However, for magnetic repulsion with magnetic parameters $b<-0.751$ the radial profiles $\cl(r;b)$ and $\ce(r,b)$ differ significantly from the geodesic case -- the profiles increase over the entire range of radii $r>3$. Additionally, for corotating orbits with $\omega_{\phi}>0$, the function $\cl(r;b)$ can enter the region of negative angular momentum ($\cl<0$), in a region reaching $r=3$, while the point $\theta$ISCO is not an extremal point of the radial profiles for magnetic repulsion. 

The special behavior of $\cl(r;b)$ and $\ce(r,b)$ occurs in the region of small negative magnetic parameters ($0>b>-0.751$). In this range, the radial profiles exhibit behavior similar to that seen in the case of magnetic attraction for $0>b>-0.654$, while for $-0.654>b>-0.751$, only increasing radial profiles of $\cl(r;b)$ and $\ce(r,b)$ occur. 

\begin{figure*}

   \centering
	\includegraphics[width=\hsize]{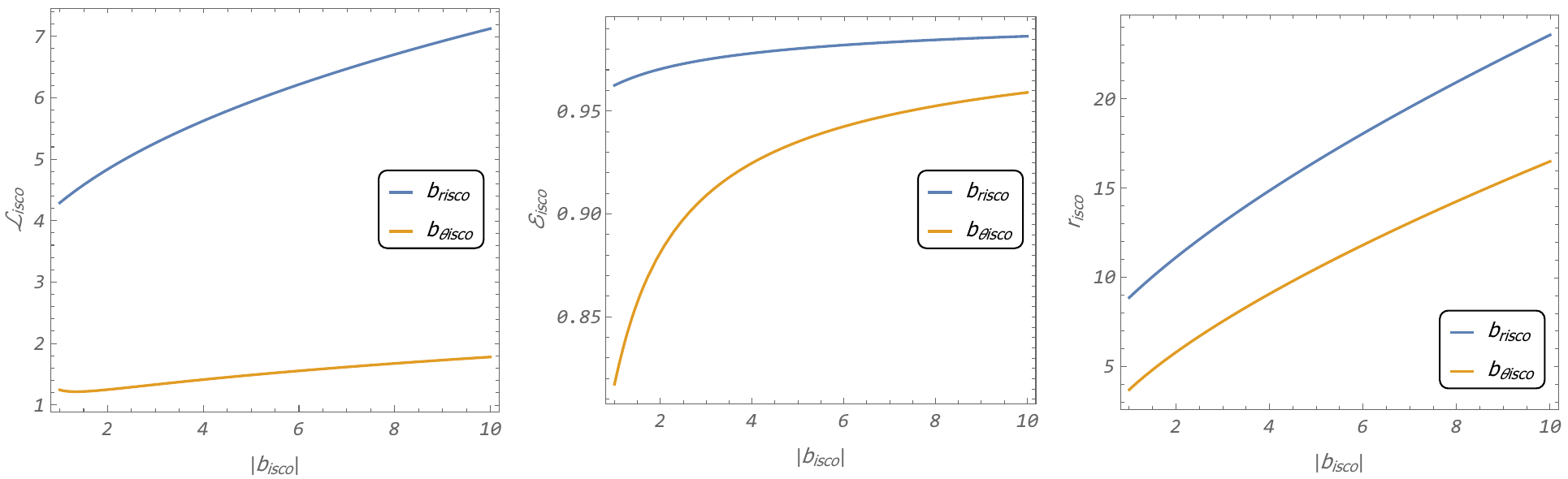}
	\caption{\label{f:f23}Specific angular momentum (left panel), specific energy (middle panel) and radial position (right panel) of the $r$ISCOs (blue curve) and the $\theta$ISCOs (yellow curves) are given as functions of magnitude of the magnetic parameter $b$. Blue curves depict the positive branch of $b$ (beginning of radial instability) and yellow curves correspond to the negative branch of $b$ (beginning of latitudinal instability).}
\end{figure*}

The equatorial circular orbits marginally stable relative to the radial (latitudinal) perturbations, represented by the radius $r_\risco(b)$ ($r_\tisco(b)$), as well as their corresponding specific energies, $\ce_\risco(b)$ ($\ce_\tisco(b)$), and specific angular momenta, $\cl_\risco(b)$ ($\cl_\risco(b)$), are illustrated in FIG.~\ref{f:f23}. 

The region of stability with respect to radial perturbations lies below the curve $b_\risco(r)$, while the region of stability with respect to latitudinal perturbations lies above the curve $b_\tisco(r)$. We can thus conclude that there are three regimes of stability of equatorial circular orbits at the region $r>3$. 

For magnetic parameter values $b>b_\risco(r=3)=-0.654$ and $b<b_\tisco(r=3)=-0.751$, two fundamentally distinct instability scenarios for equatorial circular orbits arise, corresponding to different types of perturbed particle motion. However, in the region $b_\risco(r=3)=-0.654>b>b_\tisco(r=3)=-0.751$, the equatorial circular orbits are stable against both radial and latitudinal perturbations for $r>3$. 

For $b>b_\risco(r=3)=-0.654$, stable circular orbits terminate due to radial instability at the standard rISCO in the equatorial plane, similarly to the case of circular geodesics around a Schwarzschild black hole; note that the function $b_\risco(r)$ has a zero point at $r=6$. For $r < r_\risco(b)$, no off-equatorial stable circular orbits are possible, and the particle falls radially onto the NS surface. As intuitively expected, the corotating equatorial circular orbits with marginal stability in the radial direction are shifted to greater (lesser) distance, as compared to the pure Schwarzschild case with $r_{ms}(b=0)=6$, depending on magnetic attraction ($b>0$) or magnetic repulsion ($b<0$). The angular velocity of these marginally stable orbits increases (decreases), in comparison with the Schwarzschild case, for increasing magnetic attraction (repulsion). 

For $b<b_\tisco(r=3M)=-0.751$, the situation differs due to a change in the behavior of the effective potential in the latitudinal direction. At $r_{\tisco}$, there is an inflection point of the effective potential $V_{\rm eff}(r,\theta;b,\cl)$ in the latitudinal direction. For $r<r_{\tisco}$ there is a maximum (minimum) of the effective potential in the latitudinal (radial) direction in the equatorial plane. Additionally, two symmetrically positioned minima of the effective potential appear outside the equatorial plane. The maximum corresponds to an unstable equatorial circular orbit, while the minima correspond to off-equatorial circular orbits whose stability is determined by the Hessian matrix (for details see \cite{Vrb-Kol-Stu:2024:submitted:}). The off-equatorial circular orbits are directly linked to the instability of the circular equatorial orbits in the vertical direction, starting at the radius $r_\tisco(b)$ and extending down to $r=3$. 

\subsection{Off-equatorial circular orbits and related belts of charged particles}
The corotating off-equatorial circular orbits are determined by the simultaneous solution of equations $\dd V_{\mathrm{eff}}/\dd r=0$ and $\dd V_{\mathrm{eff}}/\dd \theta=0$. The magnetic parameter $b$ and the specific angular momentum $\cl$ are related to the latitudinal coordinate $\theta$, which defines the circular orbit position along with the radial coordinate $r$ \cite{Stu-Kov-Kar:2008:CQG:,Vrb-Kol-Stu:2024:submitted:}. For an off-equatorial circular orbit at given coordinates $r,\theta$, the charged particle has magnetic parameter $b_\coff$ is fixed by the relation
\beq
b_{\coff}(r,\theta) = \pm\frac{-r \csc\theta}{2\sqrt{r^2 \, \Ln+2 r+2} \sqrt{r^2 \, \Ln+2 r+6}},\label{e:bc} 
\eeq
while the specific angular momentum $\cl_\coff$ and specific energy $\ce_\coff$ of the circular off-equatorial orbit are determined by the relations 
\begin{eqnarray}
	\cl_{\coff}(r,\theta) &=& \pm\frac{-r \sqrt{r^2 \, \Ln+2 r+2} \sin \theta}{2\sqrt{ r^2 \, \Ln+2 r+6}}.
	\label{e:lc} \\
	\ce_{\coff}(r,\theta) &=& 2 \sqrt{\frac{r-2}{r \left(r^2 \Ln+2 r+6\right)}}.
	\label{e:lc1}
\end{eqnarray}
Note that the specific energy of the off-equatorial circular orbits can be expressed as a function of the radial coordinate only. Of course, specifying the latitudinal coordinate of the orbit implies the magnetic parameter of the orbit (related to the specific charge of the particle) and the specific angular momentum.  

Restriction to corotating orbits with $\omega_{\phi}>0$, we find that off-equatorial circular orbits can exist only for charged particles with $b<0$, i.e., under magnetic repulsion.\footnote{Particles under magnetic attraction, with $b>0$, have to follow the counter-rotating off-equatorial orbits with $\omega_{\phi}<0$.} 

\begin{figure*}
	\centering
	\includegraphics[width=1\hsize]{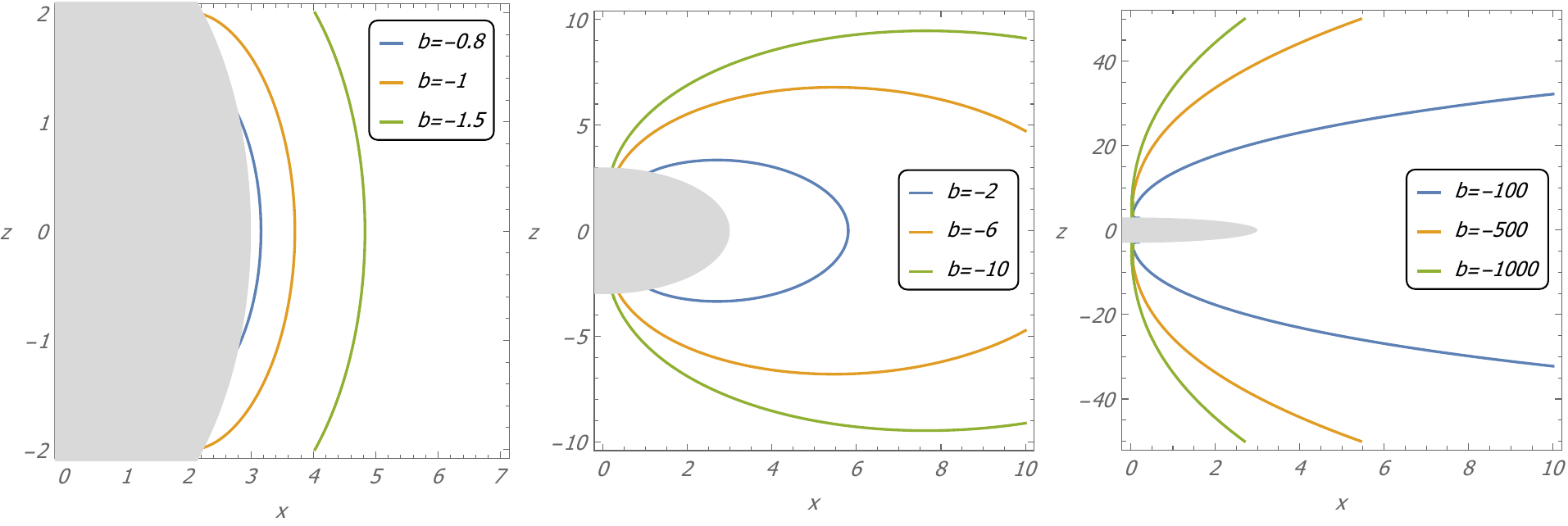}
	\caption{\label{f:f3}Position of the off-equatorial circular orbits for various series of the magnetic parameter $b$. Notice that for small negative values of the magnetic parameter, the off-equatorial orbits can be fully or partially inside the NS surface.}
\end{figure*}

The function $b_{\coff}(r,\theta)$, which is symmetric relative to the equatorial plane, directly determines the position of circular orbits located off the equatorial plane. The possible positions of the off-equatorial circular orbits, as given by Eq.~(\ref{e:bc}), are represented for selected values of the magnetic parameter $b<-0.751$ in FIG.~\ref{f:f3}, where Cartesian coordinates $x,z$ are used instead of the spherical coordinates $r,\theta$. For small values of $b$, the entire family of the off-equatorial circular orbits is concentrated in vicinity of the NS surface. However, for very large values of $b$, the off-equatorial circular orbits near the equatorial plane ($\theta \sim \pi/2$) are located very far from the NS surface, but they rapidly approach the surface as their location approaches the symmetry axis ($\theta \to 0$), or as Cartesian coordinates are approaching $x \to 0$ and $z \to 3$. In the region of intermediate values of $\theta$, the variation of $x$ is rapid, while variation of $z$ is slow. As a result, the surfaces defined by $b_{\coff}(r,\theta)=const$ resemble oblate ellipsoids deformed near the NS and constrained by its surface. 

For particles under magnetic attraction, corotating off-equatorial circular orbits do not exist -- there are no extrema of the effective potential, meaning no simultaneous solutions to the equations $\dd V_{\mathrm{eff}}/\dd r=0$ and $\dd V_{\mathrm{eff}}/\dd \theta=0$ exist for $\omega_{\phi}>0$ and $b>0$. 

\begin{figure*}
   \centering
	\includegraphics[width=0.66\hsize]{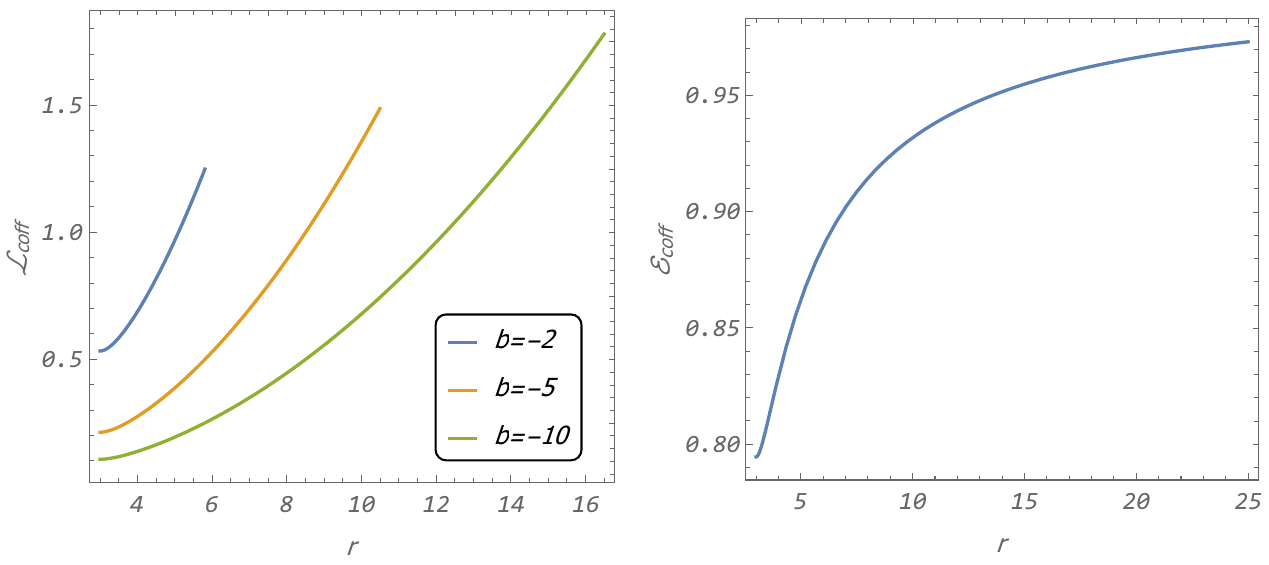}
	\caption{\label{f:f24}Radial profiles of specific energy and specific angular momentum of the off-equatorial circular orbits. Notice that the specific energy profile is independent of $b$, while the specific angular momentum depends on $b$.}
\end{figure*}

The specific energy profile must be calculated for the appropriate combination of $b$ and $\cl$ functions, and the positive root should be taken, as we are considering the positive root states \cite{Bic-Stu-Bal:1989:BAC:,Bal-Bic-Stu:1989:BAC:}. The functions $\cl_{\coff}(r,b)$ where the magnetic parameter $b$ is used instead of coordinate $\theta$, and $\ce_{\coff}(r)$ are shown in FIG.~\ref{f:f24}. 

\subsubsection{Properties of the off-equatorial circular orbits}

The stability of the off-equatorial circular orbits is determined by the determinant of the Hessian matrix, given by Eq.~(\ref{e:hess}). The determinant takes a simple form \cite{Vrb-Kol-Stu:2024:submitted:} 
\bea
\det H =\frac{32 (3-r) \left(r^2 \, \Ln+2 r+2\right) \cot ^2\theta}{r^4 \left(r^2 \, \Ln+2 r+6\right)^2}
\eea

The stability condition  
\beq
\det H > 0 
\eeq
is satisfied throughout the region of interest $r>3$. Therefore, all off-equatorial circular orbits located above the surface of our test NS are stable.  

The region of off-equatorial circular orbits is initiated by the equatorial circular orbits that are marginally stable against the vertical perturbations at $\theta$ISCO. For particles with a fixed magnetic parameter $b < -0.743$, the radius $r_{\tisco}(b)$, implicitly given by the curve $b_{\tisco}(r)$ in FIG.~\ref{f:f22}, determines the limiting radius of the off-equatorial circular orbits of particles with this fixed $b$. This radius decreases with decreasing $\theta$ until reaching the NS surface at $r=3$.

\begin{figure}
    \centering
	\includegraphics[width=0.66\hsize]{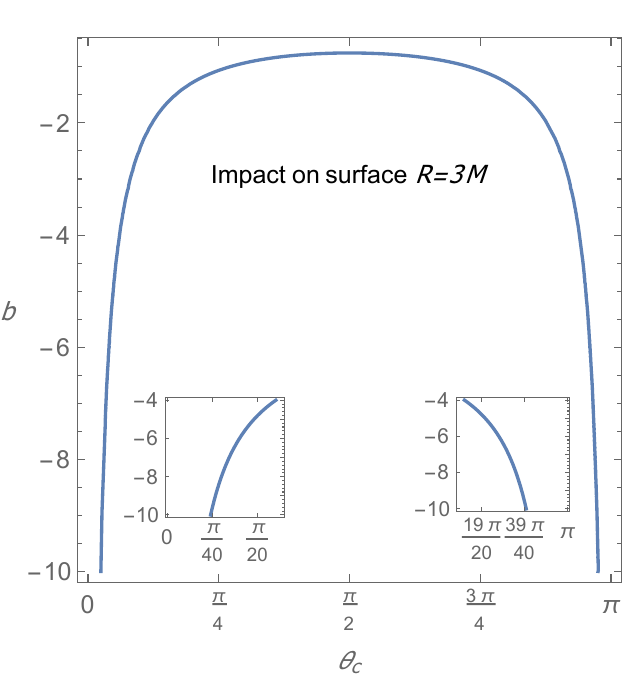}
	\caption{\label{f:f20}The latitude coordinate $\theta_{\rm cr}(b;r=3)$ of the off-equatorial circular orbit at the surface of the testing NS with the radius $r=R=3$, given as a function of the parameter $b$.}
\end{figure}

FIG.~\ref{f:f3} shows the distribution of off-equatorial circular orbits for various values of the dipole magnetic field parameter $b$ in the allowed region of $b<-0.751$. However, we must account for the limits imposed on the function $b_{\coff}(r,\theta)$ by the NS surface located at $r=3$. For $b=-0.751$, the radius $r_{\tisco}$ is located just at $r=3$, meaning no off-equatorial circular orbits exist above the NS surface in this case. For lower values of $b$, off-equatorial orbits appear above the NS surface, being restricted to a critical latitude $\theta_{cr}(b,r=3)$ corresponding to the NS surface. The function $\theta_{cr}(b;r=3)$ is illustrated in FIG.~\ref{f:f20}. 
This critical latitude $\theta_{cr}(b;r=3)$ changes rapidly near the "starting" value of the magnetic parameter $b=-0.751$, which marks the onset of the off-equatorial circular orbits. For large negative values of $b$, the critical angle $\theta_{cr}(b;r=3)$ approaches but never reaches $\theta=0$, but changes slowly. 

For particles with a fixed parameter $b<-0.751$, the radii of off-equatorial orbits are given by the function $r_{\coff}(\theta,b)$, implicitly determined by the function $b_\coff(r,\theta)$. For a fixed value of $b$, the region of the off-equatorial circular orbits is limited by the condition $r_{\coff(\theta,b)}=3$, which defines the limiting function $\theta_{cr}(b;r=3)$, and by the radius $r_{\tisco}(b)$ at $\theta=\pi/2$. 

The function $\cl_\coff(r,\theta)$, determined by Eq.~(\ref{e:lc}) and considered along the curve $b=b_\coff(r,\theta)$ as illustrated in FIG.~\ref{f:f24}, represents the profile of the specific angular momentum of the off-equatorial orbits in the $r-\theta$ (or $x-z$) plane. This profile ends at the surface radius $r=3$ and $\theta=\theta_{cr}(b;r=3)$, where the marginally stable off-equatorial circular orbits is located. It begins at $\theta=\pi/2$ at the radius $r=r_{\coff(\theta=\pi/2,b)}$, which is determined by the function $b_\tisco(r)$. We can express this function in a convenient form $\cl_{Coff}(r,b)$, to provide the radial profiles for fixed values of $b$, where the coordinate $\theta$ corresponding to a given radius $r$ is determined by $b=b_\coff(r,\theta)$. 

The function $\ce_\coff(r)$, which represents the profile of the specific energy of the off-equatorial circular orbits (determined by Eq.~(\ref{e:lc1}) and illustrated in the right panel of FIG.~\ref{f:f24}) is limited by the inner radius $r=3$, corresponding to the marginally stable off-equatorial orbit. It starts at the radius $r=r_{\coff(\theta=\pi/2,b)}$, corresponding to the equatorial circular orbit that is marginally stable with respect to latitudinal perturbations, and is given by the function $b_\tisco(r)$. 

Observations of toroidal structures around the Vela pulsar \cite{Gun-Ost:1969:Nat:} suggest a possible resemblance to the stable off-equatorial circular orbits of charged particles predicted in magnetospheres with dipole magnetic fields. Similar features observed in the remnant of SN1987A \cite{Che-Fra:1987:Nat:} hint at the influence of magnetic fields in shaping these structures.

\subsubsection{Belts}

The function $b_{\coff}(r,\theta)$ allows us to directly determine the position of the circular orbits located off the equatorial plane. The motion constants $\cl_{\coff}(r,\theta)$ and $\ce_\coff(r)$ are given by Eqs (\ref{e:lc}) and (\ref{e:lc1}). The off-equatorial motion of charged particles within belts is then governed by the effective potential, assuming that for a given particle with magnetic parameter $b=b_{\coff}(r,\theta)$ and specific angular momentum $\cl=\cl_\coff(r,\theta)$, the specific energy $\ce>\ce_\coff(r)$. 

To understand the behavior of particles bound near an off-equatorial stable circular orbit with parameters $b$, $\cl$ and located at radius $r_{Coff}(b,\cl)$, it is helpful to determine the specific energy $\ce_{c-eq}$ of the unstable equatorial circular orbit that has the same values of the magnetic parameter $b$ and specific angular momentum $\cl$, and is located at the radius $r_c$ satisfying the relation $r_{Coff}(b,\cl) < r_c < r_{\tisco}$. 

\begin{figure*}
    \centering
	\includegraphics[width=\hsize]{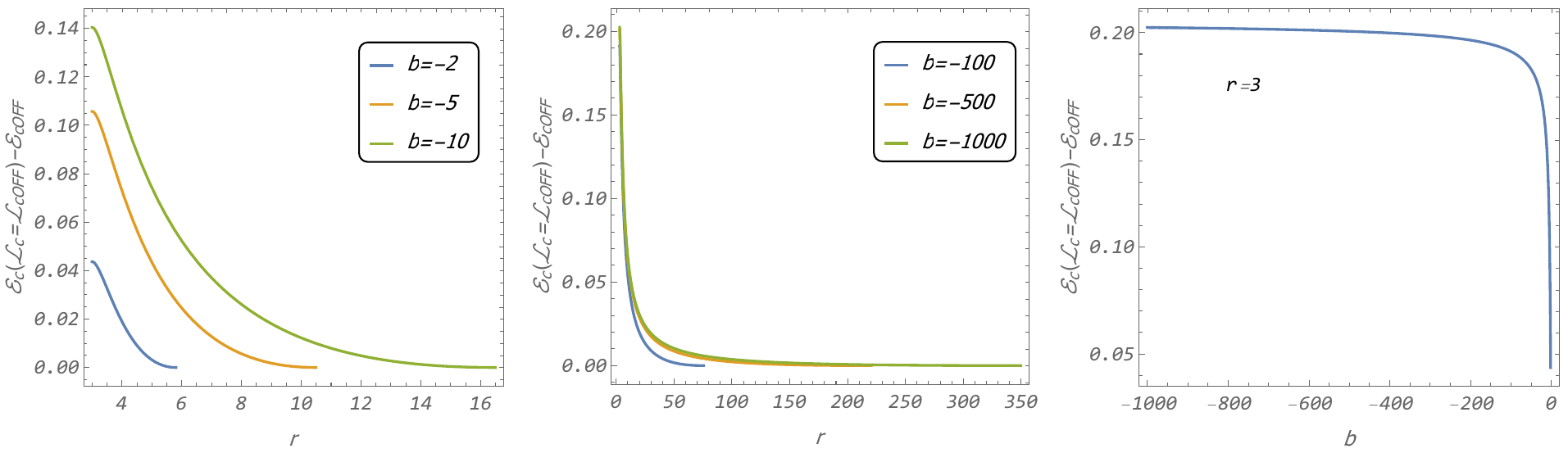}
	\caption{\label{f:f2}The difference of specific energies of the unstable equatorial and stable off-equatorial circular orbits for the same specific angular momentum are given for several fixed values of the magnetic parameter $b$ on the left and middle panel. The difference of these specific energies at the surface of the testing NS (R=3M) is given on the right panel.}
\end{figure*}

This situation is illustrated in FIG.~\ref{f:f2}. For selected values of the parameter $b$, the specific energy of the off-equatorial orbits $\ce_{Coff}(r=r_{Coff}(b,\cl))$ is compared with the specific energy of the corresponding unstable equatorial circular orbit $\ce_c(b,\cl)$, where both $b$ and $\cl$ are the same as those for the off-equatorial circular orbit under consideration. The calculation starts at $r=r_{\tisco}$ and ends at $r=3$. Additionally, the energy difference over this radius interval is shown -- indicating that as the magnitude of $b$ increases, the energy difference also increases.   

\begin{figure}
    \centering
	\includegraphics[width=0.7\hsize]{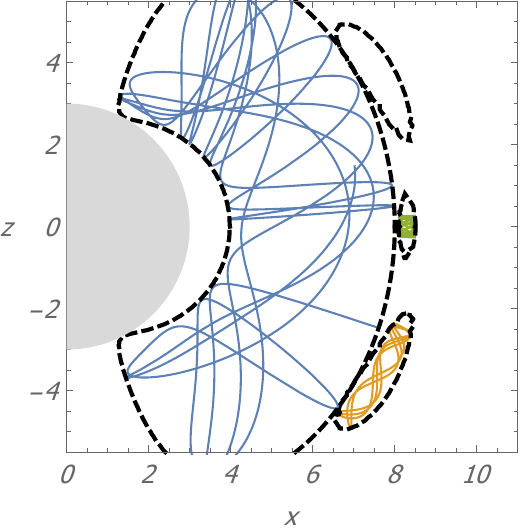}
	\caption{\label{f:f10}Different types of bound closed orbits represent the motion in belts. Blue, green and orange trajectories are given for $b=-2,-3,-4$ respectively. The specific energy ($\ce$), the specific angular momentum ($\cl$), and the specific energy of the related circular orbit ($\ce_\mathrm{c}$) are given for considered configurations in the following way\newline $b=-2$: $\ce = 0.883$, $\ce_\mathrm{c}= 0.927$ and $\cl= 0.802$,\newline $b=-3$: $\ce = 0.92104$, $\ce_\mathrm{c}=0.92103$ and $\cl=1.664$, \newline $b=-4$: $\ce = 0.918$, $\ce_\mathrm{c}=0.914$ and $\cl=1.114$. }
\end{figure}

For particles with fixed values of $b$ and $\cl$, corresponding to a stable off-equatorial circular orbit, and energy $\ce > \ce_{c-eq}(b,\cl)$, the off-equatorial motion can cross the equatorial plane into the opposite hemisphere of the spacetime. This can lead to the formation of large belts of charged particles, with motion centered around the off-equatorial circular orbits, similar to the Van Allen radiation belts observed around Earth and some of the extraterrestrial systems \cite{2023Sci...381.1120C}. These belts occur around the radius where the stable equatorial circular orbits of particles with specific charge $\bar{q}$ (corresponding to the magnetic parameter $b$) enter the region of latitudinal instability, as illustrated for several negative values of parameter $b$ in FIG.~\ref{f:f10}. 

The large belts that cross the equatorial plane arise for $\ce > \ce_{c-eq}$; however, there can also be belts restricted to regions either above or below the equatorial plane, where the specific energy is limited by the condition $\ce_{Coff} < \ce < \ce_{c-eq}(b,\cl)$. 

Realistic belts, composed of particles characterized by a fixed magnetic parameter $b$ include particles with specific angular momentum values $\cl$ within an interval corresponding to the family of the off-equatorial circular orbits, $\cl_{Coff}(b,r=3) < \cl < \cl_{\tisco}(b)$. As a special case, one can consider smaller regions that allow for oscillatory motion around the stable equatorial orbits. 

The illustrative cases of motion around the off-equatorial circular orbits in the close vicinity of the NS, presented in FIG.~\ref{f:f10}, correspond to "test particles" of very low specific charge, such as dust grains or even large objects as plasmoids, in the magnetic field around the test NS under consideration. These cases are also applicable for protons and ions in much weaker magnetic fields. 

The \textit{belts} illustrated in FIG.~\ref{f:f10} correspond to dust grains with mass $m \sim 10^{-13}g$, orbiting around the test magnetized NS with mass $M \sim 2M_{\odot}$ and surface magnetic field strength $B=10^{8}G$. The equatorial radial distance of these belts can be approximated by the radius $r_{\tisco}(b)$. As the latitudinal coordinate of the motion approaches the symmetry axis ($\theta \to 0$), the belts approach $r=3$, consistent with the nature of the off-equatorial circular orbits. 

Belts composed of protons, ions, or counter-rotating electrons, are located at much larger distances from the NS near the equatorial plane ($\theta \sim \pi/2$). These form a "nearly flat cloud" above the equatorial plane at intermediate latitudes, approaching the NS surface near $\theta\sim 0$ in structures nearly parallel to the symmetry axis. 

A detailed discussion of these dynamics, including the role of radiation back-reaction force is postponed to the following section.  
%

Considering electrons, protons, iron ions, and charged dust as constituents of the belts, the $b$ parameter values for the test NS are as follows: $|b_\mathrm{elec}|\doteq 6.4\times 10^{10}$, $|b_\mathrm{prot}|\doteq 3.5\times10^7$, $|b_\mathrm{iron}|\doteq 6.3\times10^5$ and $|b_\mathrm{dust}|\doteq 5.9\times10^{-4}$, respectively. For stronger magnetic fields on the NS surface, such as $10^{12}G$ ($10^{15}G$), the magnetic parameters $b$ increase by $4$ and $7$ orders, respectively, compared to the case of $10^{8}G$. 

The magnetic parameter $b$ determines the equatorial distances of these belts, expressed in units of the NS mass $M$. The radii, approximated by the orbits marginally stable against the vertical perturbations, read: $r_\mathrm{elec}\doteq 5.6\times10^7$, $r_\mathrm{prot}\doteq 3.7\times10^5$ and $r_\mathrm{iron}\doteq 1.2\times10^4$. For the dust grains with mass $m \sim 10^{-13}$ g, no belts can form at $r>3M$, because the small value of $b_\mathrm{dust}$ does not allow for off-equatorial circular orbits ($|b_\mathrm{dust}|<|b_\mathrm{min}|$). 

\begin{figure}
	\centering
	\includegraphics[width=0.7\hsize]{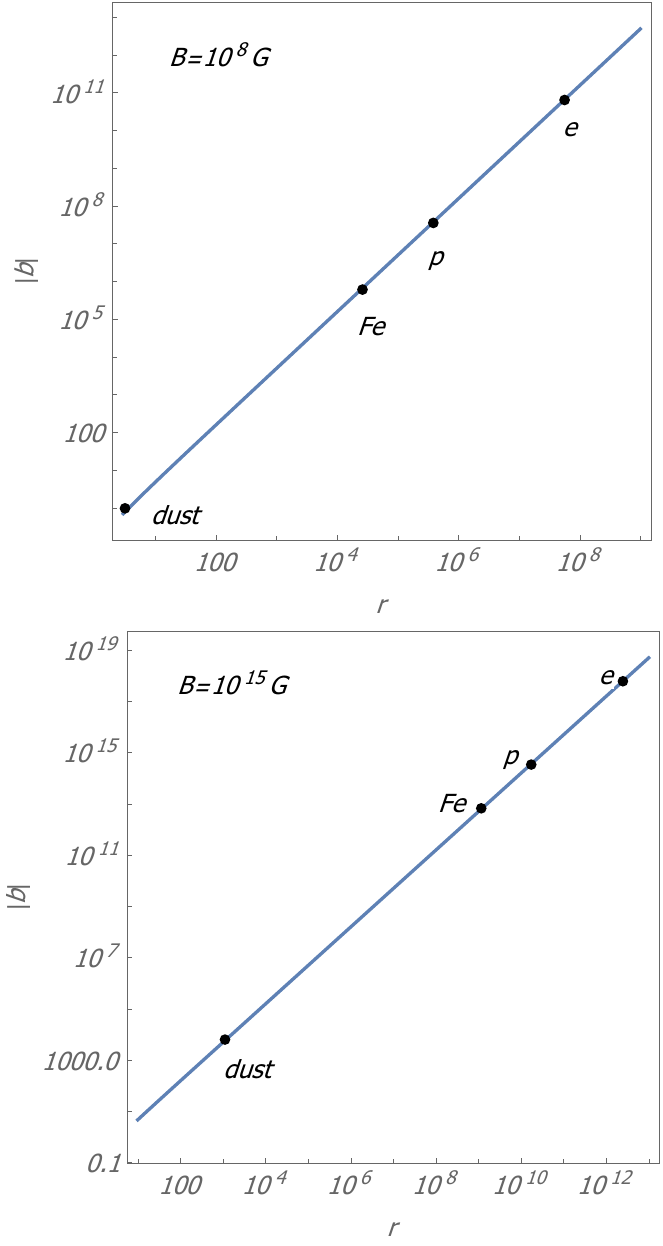}
	\caption{\label{f:f14}Radial \textit{equatorial} distance of belts estimated by $r\sim r_{\tisco}$ for large values of (negative) parameter $b$.}
\end{figure}

Dust belts can form above the NS surface, if the specific charge is significantly increased, or the magnetic field strength is enhanced up to $B\sim 10^{12}G$. For the magnetic field strength related to magnetars, $B\sim 10^{15}G$, the length scales increase significantly for all types of particles.  
Estimates of the distance of belts formed by various types of charged particles near the equatorial plane are shown in FIG.~\ref{f:f14}, both for the test NS and for an extreme case with magnetic field strength corresponding to magnetars, $B\sim 10^{15}G$.  

To estimate the equatorial distance of belts created by protons, electrons, or ions, which are located at large distances in the realistic magnetic fields of NSs, we can use an approximation for $b_{\tisco}(r)$ at large distances. This leads to the following approximate formula: 
\beq
b_\tisco = -\frac{1}{8} \sqrt{\frac{3}{2}} r^{3/2}+\frac{5 \sqrt{r}}{32 \sqrt{6}}-\frac{29}{1280 \sqrt{6} \sqrt{r}}.
\eeq
Using the leading order of this approximation, we can express the approximate belt radius as a function of the magnetic parameter $b$ in the form
\begin{equation}
    r_\tisco = 4 \sqrt[3]{\frac{2}{3}} (-b)^{2/3}. 
\end{equation}

Finally, we have to emphasize the difference of the motion of charged particles in a uniform magnetic field, where a particle falling on a central object is not driven towards the poles as it falls but can penetrate magnetic field lines near the equatorial plane \cite{Kol-Stu-Tur:2015:CQG:}, in stark contrast to the motion in a dipole magnetic field where the motion is oriented to the pole of the magnetic field.


\subsection{Three regimes of the charged particle motion}

The relative magnitude of the gravitational and electromagnetic forces acting on a charged test particle defines three regimes of their motion. As demonstrated by Frolov and Shoom \cite{Fro-Sho:2010:PRD:} for the case of Schwarzschild BHs immersed in asymptotically uniform magnetic fields, in the combined gravitational and magnetic fields in the vicinity of compact objects two characteristic frequencies of the orbital motion can be introduced: the Keplerian frequency related to the gravitational field and given by the relation 
\beq
         \Omega_K = \frac{r_{g}^{1/2}c}{r^{3/2}} , 
\eeq
where $r_{g}=MG/c^2$ is the gravitational radius of the compact object, and the cyclotron frequency related to the Lorentz force acting in the considered magnetic field and given by the relation 
\beq
         \Omega_c = \frac{qBc}{E}
\eeq
where the energy of the particle $E=\gamma m c^2$. The cyclotron frequency can be positive or negative (as related to the Keplerian frequency considered to be positive) in dependence on the orientation of the magnetic field and the particle electric charge sign. 

For particles in the vicinity of the gravitational radius, $r \sim r_g$, we find $\Omega_K \sim c^3/(GM)$; assumption of their Lorentz factor $\gamma \sim 1$ implies $\Omega_c = \frac{qB}{mc}$. The relation of magnitudes of these frequencies defines the three possible regimes of motion in the vicinity of the compact object. The gravitational regime where $\Omega_K >> \Omega_c$, magnetic regime where $\Omega_K << \Omega_c$, and chaotic regime where $\Omega_K \sim \Omega_c$. 

The ratio of these frequencies, $\Omega_c/\Omega_K$, gives the dimensionless parameter 
\beq
             b_u = \frac{qBMG}{mc^4}
\eeq 
characterizing the relative strength of the Lorentz and gravitational forces acting on the particles in the close vicinity of a compact object immersed in an asymptotically uniform magnetic field. Such a parameter coincides, after introduction of some properly chosen coefficient, with the parameter $b$ introduced in the present paper for the dipole magnetic field, if the particles are moving near the surface of the NSs. With increasing distance of the motion from the NS surface, the locally defined parameter $b_{loc-dip}(r)$ (connected with the locally measured magnetic field related to the local gravity strength) has to be modified due to the character of the dipole magnetic field; at large distances, we can expect that $b_{loc-dip}(r) \sim 1/r$, since the dipole magnetic field vanishes in the equatorial plane as $1/r^3$, while the gravitational field as $1/r^2$. Nevertheless, in calculations of the motion we apply the constant "global" magnetic parameter $b$ governed by the magnetic field strength at the NS surface. 

In terms of the magnetic parameter $b$, we can define three regimes of the charged particle motion in the following way: gravitational regime means $b<<1$, chaotic regime means $b \sim 1$, magnetic regime means $b>>1$. Clearly, these regimes reflect nature of the motion near the NS surface, because the parameter $b$ is a constant representing the magnetic moment of the dipole field, reflected by the magnetic field strength at the NS surface. 

However, these regimes have to be modified with increasing distance from the NS surface. For NSs with a fixed dipole magnetic field, we have to take into consideration also the decreasing of the "locally" defined parameter $b_{loc-dip}(r)$. This means that the magnetic regime of the motion near the NS surface can be shifted to the chaotic regime at sufficiently large distances. In other words: if in the vicinity of a NS surface a gravitational regime of the motion is relevant, it is kept to be relevant at all distances from the NS; if the regime is chaotic near the NS surface, it will be changed to the gravitational regime in sufficiently large distances; if the magnetic regime occurs near the NS surface, it is shifted to the chaotic regime at sufficiently large distances, and finally the gravitational regime enters the play at very large distances. 

We can categorize the motion of charged particles into two fundamental groups: free and bound. Particles following the free (bound) motion are escaping (cannot escape) to infinity. Particles following the bound motion are restricted by a potential barrier preventing escape to infinity; the barrier can be closed -- the motion is restricted to a closed region, or can be open -- the motion terminates at the NS surface. These barriers will be discussed later when the influence of the radiative back-reaction on the motion will be included. Note that in close vicinity of the minima of the effective potential corresponding to stable circular orbits the harmonic epicyclic motion is possible, with frequencies governed by the second derivatives of the effective potential against the radial and latitudinal coordinates. These frequencies could be relevant in explaining the twin HF QPOs observed in the binary systems containing a neutron star, as demonstrated in \cite{Vrb-Kol-Stu:2024:submitted:}. 

\subsection{Lorentz factor of the particle motion} 

The Lorentz factor of a moving particle can be in a given spacetime expressed relative to any observer. In our case, it is natural to consider the static observers of the Schwarzschild spacetime. 

Any local static observer endowed with an orthonormal reference frame ($e_{(a)}$, with $a = t, r, \theta, \phi$), can express a test particle $4-momentum$ $p^{\mu}$ in the standard special-relativistic form 
\beq
    p^{\mu} = mu^{\mu} = m\gamma (1, v),        \gamma = (1 - v.v)^{-1/2} , 
\eeq
where $m$ is the particle mass, $u^{\mu}$ denotes the $4-velocity$ and $v$ is the $3-velocity$ measured in the reference frame of the observer, and $\gamma$ is the Lorentz factor of the particle. 

In the Schwarzschild spacetime, the orthonormal reference frame consists of $4-vectors$ 
\bea
     e_{(t)} &=& ({\frac{1}{f(r)^{1/2}}, 0, 0, 0}) , \\
	e_{(r)} &=& ({0, f(r)^{1/2}, 0, 0}) , \\
		 e_{(\theta)} &=& ({0, 0, \frac{1}{r}, 0}) , \\
		 e_{(\phi)} &=& ({0, 0, \frac{1}{r sin\theta}, 0}) . 
\eea

The $3-velocity$ $v$ has components $v^{i}$ where $i = r, \theta, \phi$. In the dual reference frame ($e^{(a)}$) the three components are determined by relation 
\beq
            v^{(i)} = \frac{u^{\mu}e_{\mu}^{(i)}}{u^{\mu}e_{\mu}^{(t)}} . 
\eeq
The conserved covariant energy can then be expressed in the form 
\beq
        E = -p_{t} = -p^{(a)}e_{t(a)} = m \gamma f(r)^{1/2} . 
\eeq 
The Lorentz factor of the particle motion related to the static observers is thus given simply by the covariant specific energy $E/m$ of the particle reduced by the metric coefficient due to expression 
\beq
         \gamma =  \ce f(r)^{-1/2} . 
\eeq
For particles with $E/m >1$, reaching infinity is the Lorentz factor related to static observers at infinity given directly by the specific energy. Here we are giving the Lorentz factor for particles reaching the NS surface, using the relation 
\beq
         \gamma_{surf} = 3^{1/2} \ce . 
\eeq
We compare the Lorentz factor obtained for particles under the pure Lorentz force with those obtained for the Lorentz force combined with the RR force. As we shall see in the next section, Lorentz factor $\gamma$ of a charged particle motion influences its radiative back-reaction force through associated combinations of its $4-velocity$. 

\subsection{Possible role of quantum effects}

It is useful to shortly comment possible role of the quantum effects that are expected in strong magnetic fields of NSs, especially in the case of the magnetars. We mention here two interesting phenomena that should be detailed study in the future. 

The first one is related to the relativistic pair plasma that could influence the motion of charged particles because of possible polarization effects -- these phenomena should be treated at least in the framework of the kinetic theory in the limit of collision-less plasma that is closest to our treatment of the charged particle motion. Nevertheless, it should be stressed that the ultra-strong magnetic field strength exceeding the quantum critical threshold $B_{quant} = 4.4x10^{13}G$ \cite{Dun-Tho:1992:ApJ:,Usov:1992:Nature:} occurs only in the vicinity of magnetars surface, where $B \sim 10^{15}G$, but it is irrelevant for the NSs observed in standard atoll sources where the surface magnetic field strength is in the range of $10^{8}G-10^{12}G$, significantly lower than $B_{quant}$. 

The other special issue requiring a detailed study is the role of quantum mechanics in the energy spectrum of charged particles in strong magnetic fields. According to laws of quantum mechanics, the energies of charged particle cyclotron orbits are quantized creating so-called Landau levels \cite{Lan-Lif:Book:1977}. These are degenerate levels, with a number of electrons (protons, ions) directly proportional to the magnetic field intensity. The Landau level effect corresponds to a quantum harmonic oscillator with an energy spectrum given by the relation 
\beq
          E_n = (n + \frac{1}{2})h\omega_L + \frac{p^2}{2m}
\eeq
where $n \geq 0$ is the quantum number, $p$ is particle $3-momentum$ measured along the magnetic field lines, $h$ is the Planck constant, and $\omega_L = \frac{q}{m}B$ -- the energy quantum $\Delta E = h\omega_L$ is thus governed by the cyclotron (Larmor) frequency. The sets of wave functions having the same number $n$ are called Landau levels. The application of this quantum effect in astrophysics requires a detailed study. 

The Landau quantization of the motion in the strong magnetic fields is a crucial ingredient of both the integer and fractional quantum Hall effect playing a significant role in astrophysics. There are studies of the Hall effect related to a non-relativistic electron-ion plasmas including the electron-inertia effect \cite{Loeb:1956:Science:,Kim-Mor:2014:PP:} or the relativistic Hall effect neglecting the electron-inertia effect \cite{Com-Ase:2014:PRL:,Kawazura:2017:PRE:,Yos-Hir-Hat:2024:arx:}.

\section{Equations of motion including radiative back-reaction forces}

The accelerated motion of charged particles in an electromagnetic external field leads to electromagnetic radiation of the synchrotron or bremsstrahlung type \cite{Lan-Lif:1975:Book:,Jackson:1999:Book:}; in this work, related to the motion in dipole magnetic field combined with spherically symmetric gravitational field, the synchrotron radiation is crucial. The equations of motion including the radiative back-reaction then contain two types of forces of electromagnetic origin
\beq
         \frac{du^{\mu}}{d\tau} = f_{L}^{\mu} + f_{R}^{\mu}, 
\eeq 
where the Lorentz force is $f_{L}^{\mu}=(q/m)F^{\mu\nu}u_{\nu}$, and the radiative back-reaction force in flat spacetime and the non-relativistic limit takes the form 
\beq
    f_{R}^{\mu} = \frac{3q^2}{2m} \frac{d^{2}u^{\mu}}{d\tau^2}. 
\eeq

The motion of a relativistic particle with charge $q$ and mass $m$ in a curved spacetime is governed by the Lorentz-Dirac equation \cite{Poisson:2004:LRR:,Tur-etal:2018:ApJ:}
\beq \label{cureqmogen1}
   \frac{D u^\mu}{\dd \tau} \equiv \frac{\dd u^\mu}{\dd \tau} + \Gamma^\mu_{\alpha\nu} u^\alpha u^\nu = \frac{q}{m} F^{\mu}_{\,\,\,\nu} u^{\nu} + \frac{q}{m} {\cal F}^{\mu}_{\,\,\,\nu}  u^{\nu}, 
\eeq
where the gravitational interaction is described by the Christoffel symbols $\Gamma^\mu_{\alpha\nu}$, the Lorentz forces (LF) are governed by the external electromagnetic field tensor $F_{\mu\nu}$, and the radiative back-reaction force is governed by the particle's own electromagnetic field ${\cal F}_{\mu\nu}$ and its interactions with the spacetime background. The explicit form of Eq.~(\ref{cureqmogen1}) in curved spacetimes is given in the DeWitt-Brehme form \cite{DeW-Bre:1960:AnnPhys:,Poisson:2004:LRR:}
\bea 
&& \frac{D u^\mu}{\dd \tau} = \frac{q}{m} F^{\mu}_{\,\,\,\nu} u^{\nu} 
+ \frac{2 q^2}{3 m} \left( \frac{D^2 u^\mu}{d\tau^2} + u^\mu u_\nu \frac{D^2 u^\nu}{d\tau^2} \right) \nonumber \\ 
&& + \frac{q^2}{3 m} \left(R^{\mu}_{\,\,\,\lambda} u^{\lambda} + R^{\nu}_{\,\,\,\lambda} u_{\nu} u^{\lambda} u^{\mu} \right) + \frac{2 q^2}{m} ~f^{\mu \nu}_{\rm \, tail} \,\, u_\nu, 
\label{eqmoDWBH}  
\eea 
where the last term in Eq.~(\ref{eqmoDWBH}) is the so-called tail integral
\beq
f^{\mu \nu}_{\rm \, tail}  = \int_{-\infty}^{\tau-0^+}     
D^{[\mu} G^{\nu]}_{ + \lambda'} \bigl(z(\tau),z(\tau')\bigr)   
u^{\lambda'} \, d\tau'.
\eeq
The Ricci tensor vanishes in the vacuum metrics. The integral in the tail term is evaluated over the past history of the charged particle with primes indicating its prior positions. The existence of the "tail" integral in Eq.~(\ref{eqmoDWBH}) implies that radiation reaction in curved spacetime has a non-local nature, meaning the motion of the charged particle depends on its entire history, not just its current state. A detailed derivation of the equations of motion for radiating charged particles can be found in \cite{Poisson:2004:LRR:}. 

The radiation field ${\cal F}_{\mu\nu}$ in Eq.~(\ref{cureqmogen1}), emitted by the charged particle, interacts with the curvature of the background spacetime and comes back to the particle with a delay corresponding to the tail integral in Eq.~(\ref{eqmoDWBH}). In this sense, the radiated electromagnetic field of charged particle carries information about the history of the particle. Even in the absence of external forces, such as the Lorentz force, the free trajectory of a charged particle does not follow the spacetime geodesics -- this is one of the most important consequences of Eq.~(\ref{eqmoDWBH}). The significance of the tail term in realistic astrophysical situations is a complex issue because of the subtle relations of the forces involved in the motion of charged particles, and the strong dependence of these forces on the background parameters and the properties of the charged particles themselves \cite{Stu-Kol-Tur-Gal:2024:Uni:,San-Car-Nat:2023:PRD:,San-Car-Nat:2024:PRD:}. 

The radiative reaction (RR) force, due to the synchrotron radiation of charged particles moving around compact objects, strongly affects their motion. Its influence has been discussed in the case of charged particles moving around black holes immersed in an external magnetic field, specifically in cases involving asymptotically uniform magnetic field \cite{Tur-etal:2018:ApJ:,Tur-Kol-Stu:2018:AN:,Tur-etal:2020:ApJ:,Kol-Tur-Stu:2021:PRD:,Stu-etal:2020:Uni:,Stu-Kol-Tur:2021:Uni:}. 

In this study, we examine the role of the RR force for charged particle motion in a dipole magnetic field around a NS modelled by the Schwarzschild geometry. As a result, we can neglect terms containing the Ricci tensor. Under these assumptions, the DeWitt-Brehme equation of the charged particle dynamics simplifies to the Landau-Lifshitz form (for details see \cite{Tur-etal:2018:ApJ:})  
\bea \label{curradforce}
&& \frac{\dd u^\mu}{\dd \tau} + \Gamma^\mu_{\alpha\nu} u^\alpha u^\nu
  = \frac{q}{m} F^{\mu}_{\,\,\,\nu} u^{\nu} +  \frac{2 q^2}{m} ~f^{\mu \nu}_{\rm \, tail} \,\, u_\nu \\
&& + \,   \frac{q\,k}{m} \left(F^{\alpha}_{\,\,\,\beta ; \alpha} u^\beta u^\mu + \frac{q}{m} \left( F^{\alpha}_{\,\,\,\beta}
F^{\beta}_{\,\,\,\mu} +  F_{\mu\nu} F^{\nu}_{\,\,\,\sigma} u^\sigma u^\alpha \right) u^\mu \right), \nonumber
\eea

The Lorentz force is characterized by the magnetic parameter $b$, while the RR force is characterized by the radiative reaction (RR) parameter 
\bea \label{RRp}
k \equiv \frac{2}{3}\frac{q^2}{mGM} . 
\eea
For our test NS, we assume $M \sim 2M_{\odot}$. The RR parameter for electrons is $k_{\rm NS} \sim 10^{-18}$, and it is by factor $m_p/m_e \sim 1/1836$ lower for protons. For electrons moving around BH of stellar origin $k \sim 10^{-19}$, for supermassive BH with $M \sim 10^{9}M_{\odot}$ $k \sim 10^{-27}$, while for hypothetical primordial BH of mass $M \sim 10^{15}g$ we find $k \sim 1$.  


\section{Relevance of the radiation process in strong magnetic fields}

The relevance of the radiation process is a complex issue depending on many parameters and their interplay. The crucial role is played by the combination of the role of the RR parameter, influenced by the mass $M$ of the central object, and the magnetic parameter $b$ that governs the influence of the magnetic field strength. 

The role of the RR force, discussed in detail in \cite{Stu-Kol-Tur-Gal:2024:Uni:}, can be for the motion in vicinity of a NS surface illustrated by relating the three components of the RR force to the gravitational force fixed to be $F_{\rm G} \sim 1$ and the Lorentz force $F_{\rm L} \sim b$, as presented in Table \ref{t:tab0}.  

The three RR components have strongly different magnitudes in dependence on the parameters $k$ and $b$: $F_{\rm tail} \sim k$, $F_{\rm RR1} \sim kb$, $F_{\rm RR2} \sim kb^2$. Table \ref{t:tab0} provides estimates of the magnitudes of the RR forces ($F_{\rm tail}$, $F_{\rm RR1}$, $F_{\rm RR2}$) and the Lorentz force ($F_{\rm L}$), given for representative values of the magnetic field strength $B$ and the stellar mass objects with $M \sim M_{\odot}$. However, the RR force is also strongly influenced by the Lorentz factor $\gamma$ of the moving particle through its $4-velocity$ that enters in a complex way Eq.~(\ref{curradforce}) governing the RR force. Clearly, the role of $\gamma$ factor is large near the NS surface but decreases with increasing distance from the NS.  
%
\begin{table*}[ht]

\centering
\begin{tabularx}{\textwidth}{L L L L L L}
\toprule
 & \quad $\boldsymbol{F_{\rm L}}$ \quad & \quad $\boldsymbol{F_{\rm tail}}$ \quad & \quad $\boldsymbol{F_{\rm RR1}}$ \quad & \quad $\boldsymbol{F_{\rm RR2}}$  \quad \\	
 \textbf{B [Gs]}  & \quad $\boldsymbol{\cb}$ & \quad $\boldsymbol{k}$ & \quad $\boldsymbol{{}k\cb}$ & \quad ${}\boldsymbol{k\cb^2}$ \\	
\midrule
 $10^{15.8}$ & ${\sim}10^{18}$ & ${\sim}{}10^{-18}$ & ${\sim}1$ & ${\sim}10^{18}$ \\ 
 $10^{12}$ & ${\sim}10^{14}$ & ${\sim}{}10^{-18}$ &${\sim}10^{-4}$ & ${\sim}10^{10}$\\  
 $10^8$ & ${\sim}10^{10}$ & ${\sim}{}10^{-18}$ &${\sim}10^{-8}$ & ${\sim} 10^{2}$ \\
$10^{6.4}$ & ${\sim}10^{8}$ & ${\sim}{}10^{-18}$ &${\sim}10^{-10}$ & ${\sim}10^{-1}$ \\
 $10^4$ & ${\sim}10^{6}$ & ${\sim}{}10^{-18}$ &${\sim} 10^{-12}$ & ${\sim}10^{-6}$ \\ 
 $10^0$ & ${\sim}10^{2}$ & ${\sim}{}10^{-18}$ &${\sim}10^{-16}$ & ${\sim} 10^{-14}$ \\
 $10^{-2.9}$ & ${\sim}1$ & ${\sim}{}10^{-18}$ &${\sim} 10^{-18}$ & ${\sim}10^{-18}$ \\
 $10^{-5}$ & ${\sim}10^{-3}$ & ${\sim}{}10^{-18}$ & ${\sim} 10^{-21}$ & ${\sim}10^{-23}$\\ 
\bottomrule
\end{tabularx}
\caption{Magnitudes of different forces acting on a radiating charged particle in motion governed by the dimensionless form of Eq.~(\ref{curradforce}) are given for typical values of the magnetic field strength. The~estimates correspond to a relativistic electron in the vicinity of a NS with $M={M}_{\odot}$. Electron electromagnetic self-force (tail term) and gravitational interaction is the same for all cases, $F_{\rm tail}\sim{}10^{-18}$ and $F_{\rm G}=1$. 
\label{t:tab0}}
\end{table*}

Given that we are considering a NS with a very strong magnetic field and a surface at $R=3M$, we can neglect the tail term in Eq.~(\ref{eqmoDWBH}), as the reflective barrier of the Schwarzschild spacetime is hidden beneath the NS surface and any radiation entering the NS surface is captured. \footnote{The tail term enters the play for extremely compact stars having a radius at $R<3M$.} Furthermore, we also disregard the influence of NS radiation on charged particle motion. 

Eq.~(\ref{curradforce}) represents the final form of the equations of motion for modelling the influence of the radiative back-reaction force on charged particle motion around a NS with a dipole magnetic field. To better understand the role of the RR force, we compare the resulting trajectories with those obtained from the equations of motion that only account for the Lorentz force due to the external magnetic field. The motion under the Lorentz force was thoroughly studied and classified in our previous paper \cite{Vrb-Kol-Stu:2024:submitted:}, where we described equatorial and off-equatorial circular orbits, as well as chaotic off-equatorial chaotic motion in belts related to these circular orbits. 

In this study, we investigate how various trajectories, classified based on the Lorentz force motion, are modified by the presence of the RR force. This allows us to classify the influence of the radiative back-reaction force. The motion in the gravitational regime, where $b<<1$, was discussed in detail in the previous paper \cite{Vrb-Kol-Stu:2024:submitted:}. Therefore, in the present paper, we focus attention to the motion in the chaotic regime, where $b \sim 1$, and in the magnetic regime, where $b>>1$. 

We construct the trajectories of charged test particles influenced by the Lorentz and the RR forces, governed by the LL equations of motion for stable circular orbits located "in" and "off" the equatorial plane. These trajectories include oscillatory motion around these circular orbits, as well as chaotic motion of particles under magnetic repulsion (concentrated around off-equatorial circular orbits) or magnetic attraction (concentrated around equatorial circular orbits). 

We test particles starting from circular orbits, as well as particles starting near these orbits, which exhibit epicyclic (nearly harmonic) motion around the circular orbits when only the Lorentz force is considered. We always compare the trajectories with the RR forces included to those with the RR forces omitted. For chaotic motion off the equatorial plane, we construct particle trajectories with initial energy that allows motion to cross the equatorial plane, as well as for those confined to motion above or below the equatorial plane. 

The parameters $b$, $\cl$ govern the effective potential of the motion under the Lorentz force, while the specific energy $\ce$ determines the allowed regions for motion under the Lorentz force. The RR force is governed by the parameter $k$, which we adjust to illustrate the nature of the motion governed by the LL equation. In realistic scenarios, the radiative parameter $k$ is much smaller, and the influence of the RR force is more subtle but behaves similarly to the phenomena presented here. 

Table \ref{t:tab0} characterizes the interplay between the Lorentz and RR forces in realistic settings around strongly magnetized NSs. The most important part of the RR force is represented by the $F_{\rm RR2}$ component which can be comparable with the Lorentz force $F_{\rm L}$ force in the case of magnetars, and is very important for the standard NSs observed in the atoll sources with magnetic field strength approaching $B \sim 10^{12}G$. On the other hand, the role of the tail term $F_{\rm tail}$ is always negligible. To reflect also the role of the $F_{\rm RR1}$ term, we assume in our calculations the RR parameter with a value of $k=0.1$ that is irrelevant for protons orbiting the NSs (but can be relevant for protons moving around primordial BHs). To illustrate the role of the magnitude of the RR parameter, we also add trajectories constructed for parameter $k=0.01$, clearly shows that the RR force acts along longer timescales as $k$ decreases. 

\section{Radiation reaction influence on charged particles in a chaotic regime of motion}

The motion of charged particles at the exterior of NSs with radius at $R=3M$ is most interesting and reaches the chaotic regime of the motion, where the condition $|b| \sim 1$ is assumed. Numerical calculations show that the chaotic regime enters the gravitational regime at $|b_{grav}| = 0.1$, where the motion starts to be concentrated around geodesic trajectories, and the magnetic regime at $|b_{mag}| = 10$, where the motion starts to be concentrated around magnetic field lines. 

It is natural to separate the discussion of the motion according to the orientation of the Lorentz force in the equatorial plane. In the attractive case ($b>0$), only instability of equatorial circular orbits against radial perturbations is possible. In the repulsive case ($b<0$), the behavior is more complex: equatorial circular orbits can be radially unstable for sufficiently low negative values of $b$, but they become stable for an intermediate interval of negative values of $b$. For sufficiently high negative values of $b$ the equatorial circular orbits can be unstable against vertical perturbations. At the region of vertical instability, the existence of the off-equatorial circular orbits is allowed. This difference in stability of equatorial circular orbits results in a completely different character of the particle motion under the pure influence of the Lorentz force, and especially when the RR force is included; interesting unexpected phenomena arise under the influence of the RR force. 

\subsection{Attractive Lorentz force}

For positively valued magnetic parameter $b$, the Lorentz force in the equatorial plane is attractive and supports the gravitational force. In this case, unstable equatorial circular orbits with respect to radial perturbations are possible, if they are located under the $rISCO$ orbit. The off-equatorial circular orbits are not allowed in such a case. As a result, chaotic motion under the Lorentz force, which extends over a wide range of latitudinal coordinate $\theta$, always crosses the equatorial plane. 

We focus on the scenarios where the effective potential barrier is either closed around the circular orbits or open to the NS surface. The opening of the barrier can occur either radially or vertically. 

In both the attractive and repulsive Lorentz forces, there exists a situation where the effective potential barrier is open to infinity. 
This special case allows for particles to escape to infinity, but we do not explore it in detail here, as it typically results in escape under the influence either the Lorentz force or when the RR force is included. Some results concerning the influence of the RR force on open motion toward infinity in the context of magnetized black holes can be found in \cite{Stu-etal:2020:Uni:,Stu-Kol-Tur:2021:Uni:}. 

\subsubsection{Equatorial circular orbits}

\begin{figure*}
    \centering
	\includegraphics[width=0.9\hsize]{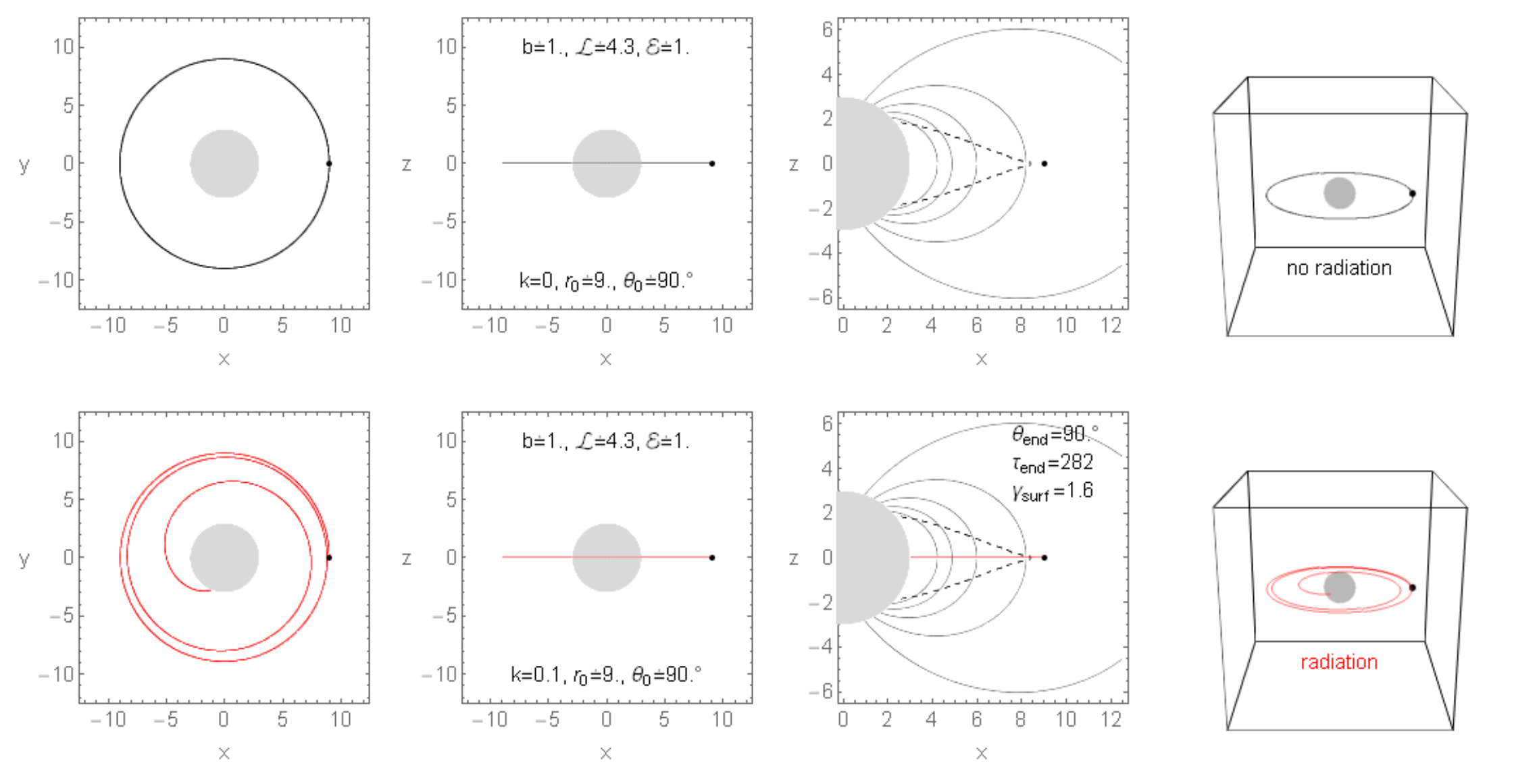}
	\caption{Equatorial circular orbits under attractive Lorentz force influence. The motion starts from a stable circular orbit. Black are trajectories under pure Lorentz force and red are trajectories with additional RR force influence. The parameters of the orbits are presented in the figure. The central shaded region corresponds to the testing NS. \label{f:f28a}}
\end{figure*}

The RR force acting on particles following unstable or near-marginal stable equatorial circular orbits under the magnetic attraction, where the effective potential is open in the radial direction, always leads to the particle's immediate onto the NS surface. The fall occurs along a trajectory confined to the equatorial plane -- see FIG.~\ref{f:f28a}.

\subsubsection{"Smoothing" of epicyclic near-circular motion}

\begin{figure*}
    \centering
	\includegraphics[width=0.9\hsize]{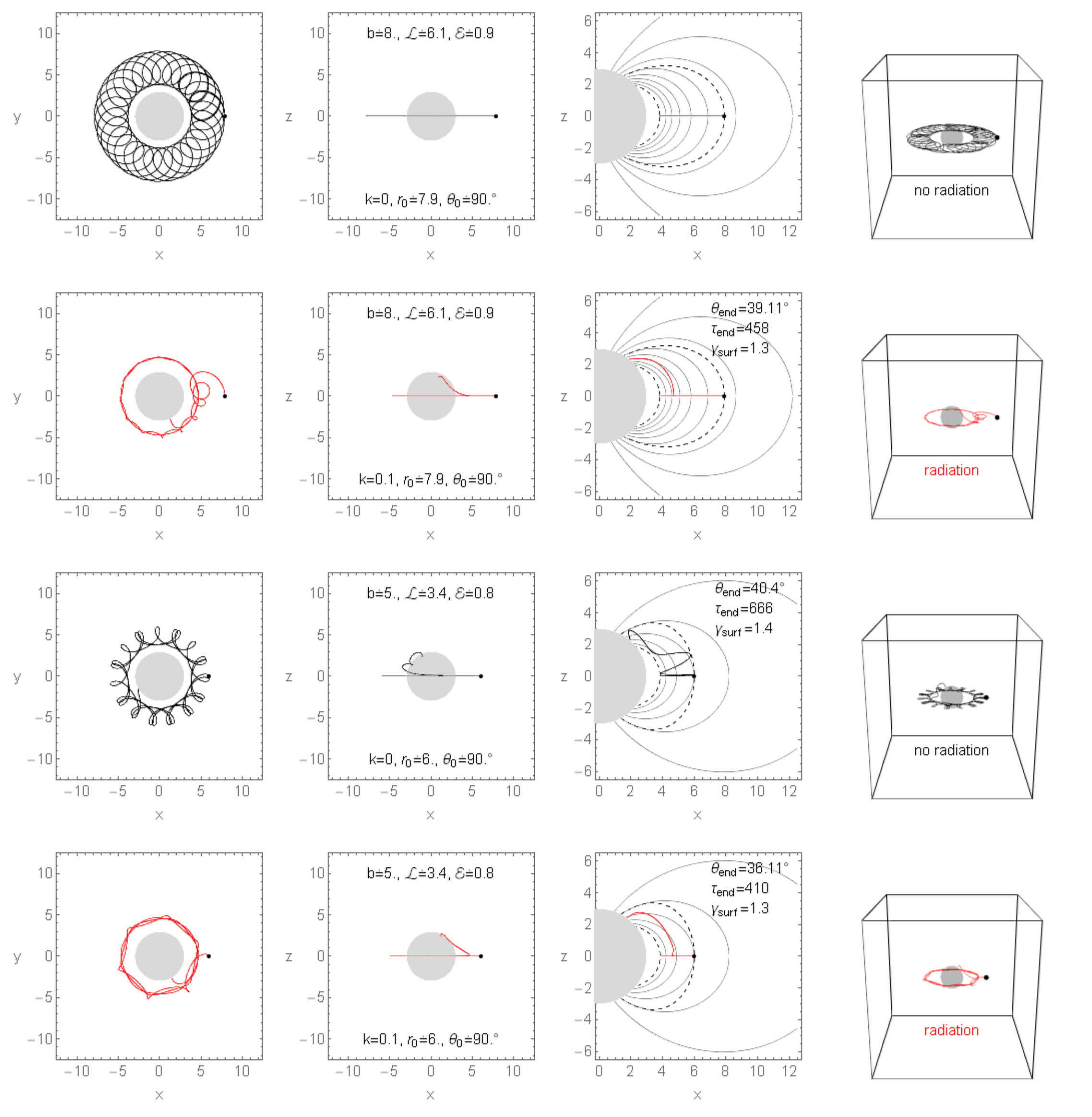}
	\caption{\label{f:f28b}RR influence on the particles moving under attractive Lorentz force in effective potential barrier open in the latitudinal direction. Black are trajectories without and red are trajectories with the RR force. Two types of the Larmour precession are illustrated.}
\end{figure*}

Particles slightly displaced from the position of a circular orbit in the equatorial plane experience different outcomes depending on the nature of the effective potential barrier corresponding to motion under the Lorentz force. When the barrier is restricted in the radial direction, but open vertically toward the NS surface, epicyclic oscillatory motion (such as Larmor precession) is gradually "smoothed" by the RR force. Initially, the oscillatory motion is reduced to near-circular motion, after which the particle falls onto the NS surface (FIG.~\ref{f:f28b}). This behavior is observed for both cases of the Larmor precession. The particle always falls in the vertical direction and the time of the fall is approximately the same whether the motion is governed solely by the Lorentz force or includes the RR force.

%

\subsubsection{Chaotic bound orbits}

The chaotic motion of particles off the equatorial plane, within large regions around the equatorial circular orbits, is explored in two cases: (i) when the effective barrier under the Lorentz force is open to the NS surface in the radial direction, and (ii) when there is an island barrier allowing for trapped motion under the Lorentz force alone. 

\begin{figure*}
    \centering
	\includegraphics[width=0.9\hsize]{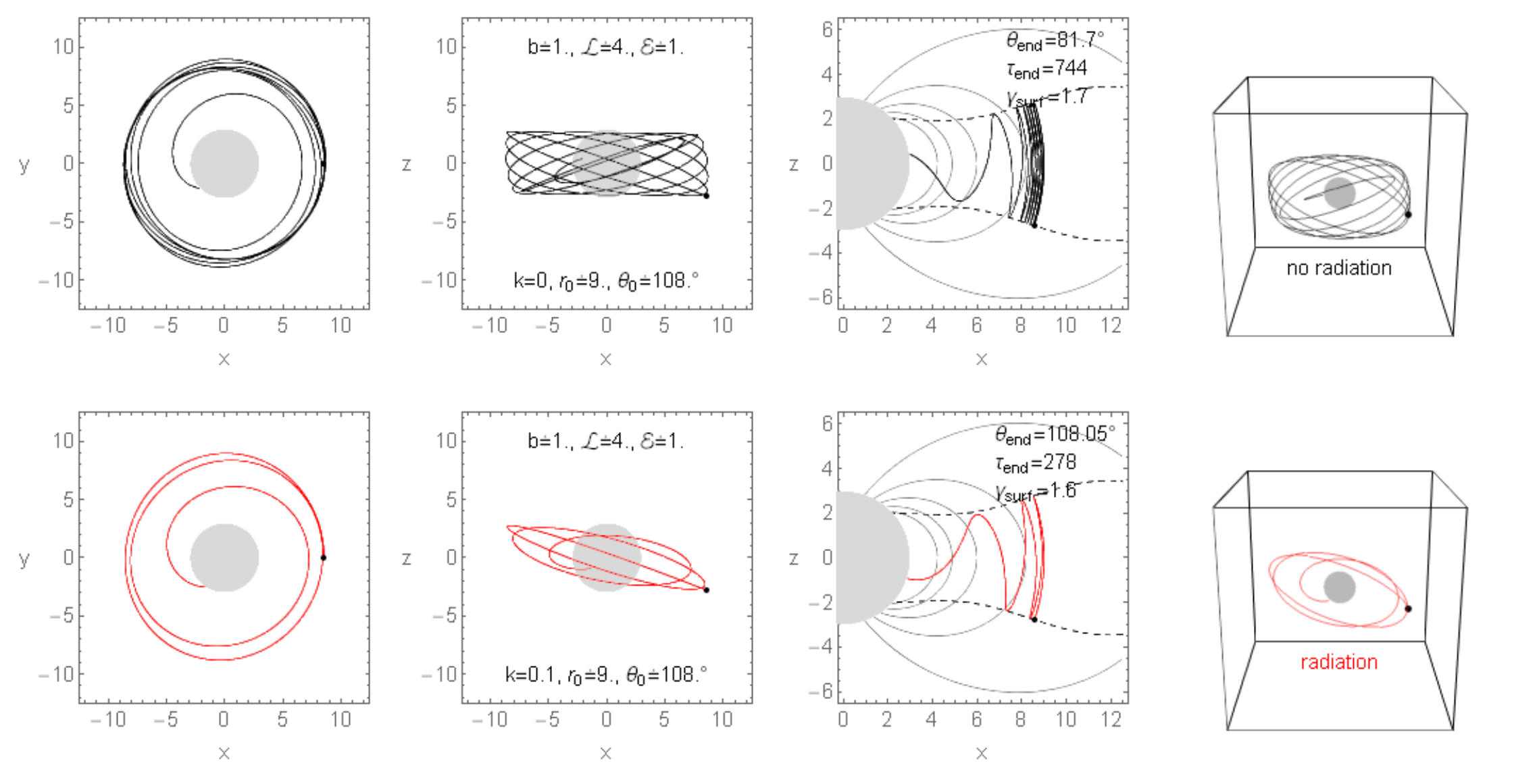}
	\caption{\label{f:f29}The RR influence for the motion under attractive Lorentz force with effective potential barrier open in the inward radial direction.}
\end{figure*}

\begin{figure*}
    \centering
	\includegraphics[width=0.9\hsize]{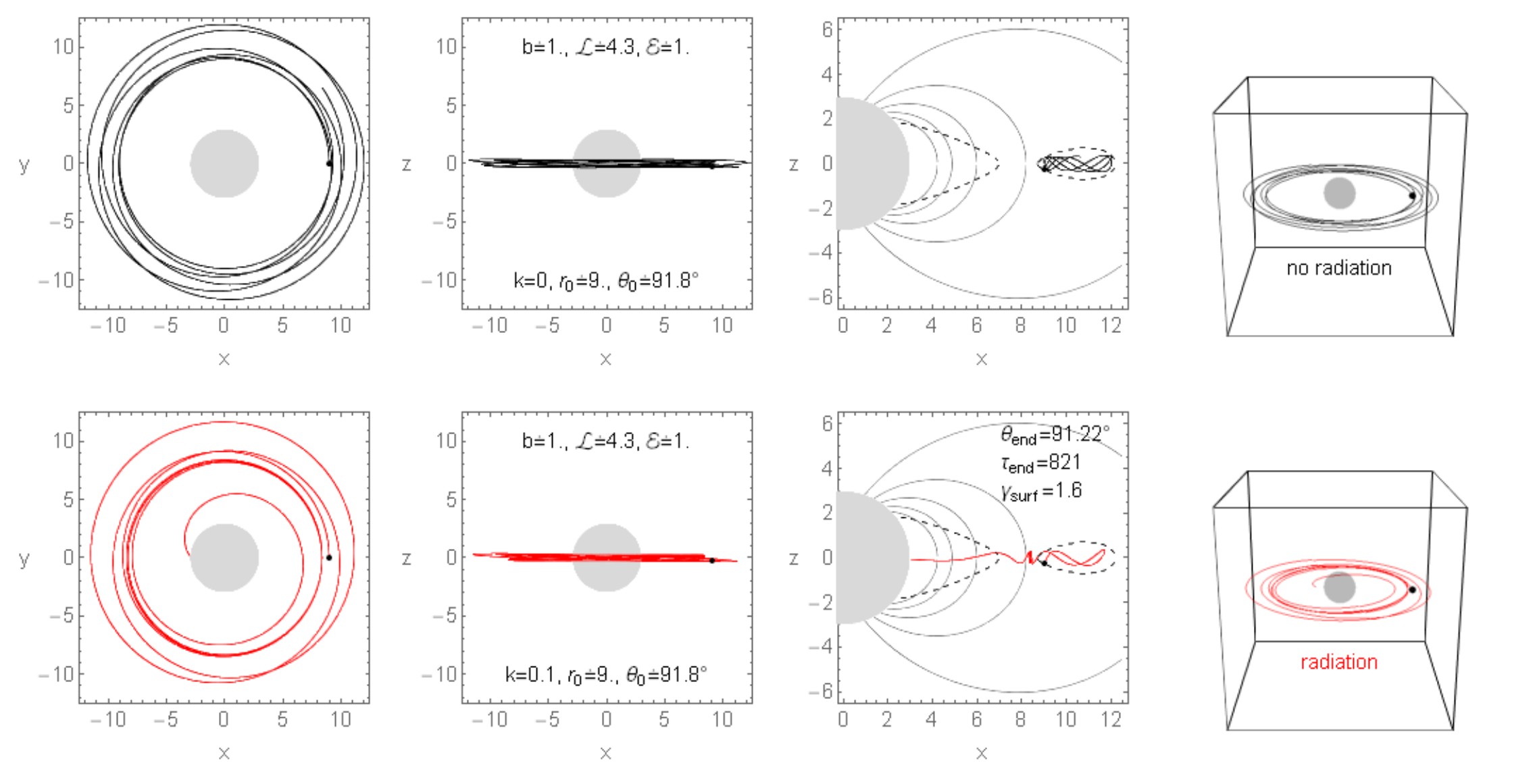}
	\caption{\label{f:f29b}The RR influence for the motion under attractive Lorentz force with effective potential barrier closed in both radial and latitudinal direction.}
\end{figure*}

In the first case, where the chaotic motion under the Lorentz force leads to a fall onto the NS surface, the inclusion of the RR force accelerates the fall and simplifies the particle's trajectory (see FIG.~\ref{f:f29}). In the second case, where the motion is restricted to a closed region by the Lorentz force, the inclusion of the RR force gradually reduces the size of the allowed region and diminishes the chaotic nature of the motion, causing eventually the particle fall onto the NS surface (see FIG.~\ref{f:f29b}).  
In both scenarios, this behavior is driven by successive losses of the particle's specific energy and specific angular momentum due to radiation processes and the corresponding RR force. The RR force acts as a dumping force, reducing the chaotic nature of charged particle motion and driving the dynamics toward a system attractor \cite{Kol-Mis-Tur:2023:EPJC:}. 

%

\subsection{Repulsive Lorentz force}

For particle motion above the NSs with surface fixed at $R=3$, there are three regimes of motion under the Lorentz force. 

First regime ($0 < b <-0.654$): in this range of magnetic parameter, the properties of particle motion are resemble those under the attractive magnetic force. 

Second regime ($-0.654< b < -0.751$): in this range, all equatorial circular orbits are stable against both radial and vertical perturbations. 

Third regime ($-0.751< b < -\infty$): in this range, equatorial circular orbits are unstable with respect to vertical perturbations at $r < r_{\tisco}$; the radius $r_{\tisco}$ marks the boundary where stable off-equatorial orbits begin to appear. 

Notice that with increasing surface radius of the NS, the first regime becomes more and more restricted and disappears for $R=6$, when the third region starts at $b\doteq-2.11$. For the NS surface at $R=4$, the critical values of the magnetic parameter are $b_{\risco}(r=4)=-0.48$ and $b_{\tisco}(r=4)=-1.12$.  

\subsubsection{Equatorial circular orbits: widening}

In the case of equatorial circular orbits, a phenomenon unique to the presence of the repulsive Lorentz force can be observed --  orbital widening. Under the combined influence of the Lorentz force and the RR force, the orbits slowly expand, in contrast to the usual shrinking observed with the attractive Lorentz force. This orbital widening in the field of Schwarzschild black holes, immersed in an asymptotically uniform magnetic field, has been explored in \cite{Tur-Kol-Stu:2018:AN:,Stu-etal:2020:Uni:,San-Car-Nat:2023:PRD:,San-Car-Nat:2024:PRD:,Stu-Kol-Tur-Gal:2024:Uni:}.

\begin{figure*}
    \centering
	\includegraphics[width=0.9\hsize]{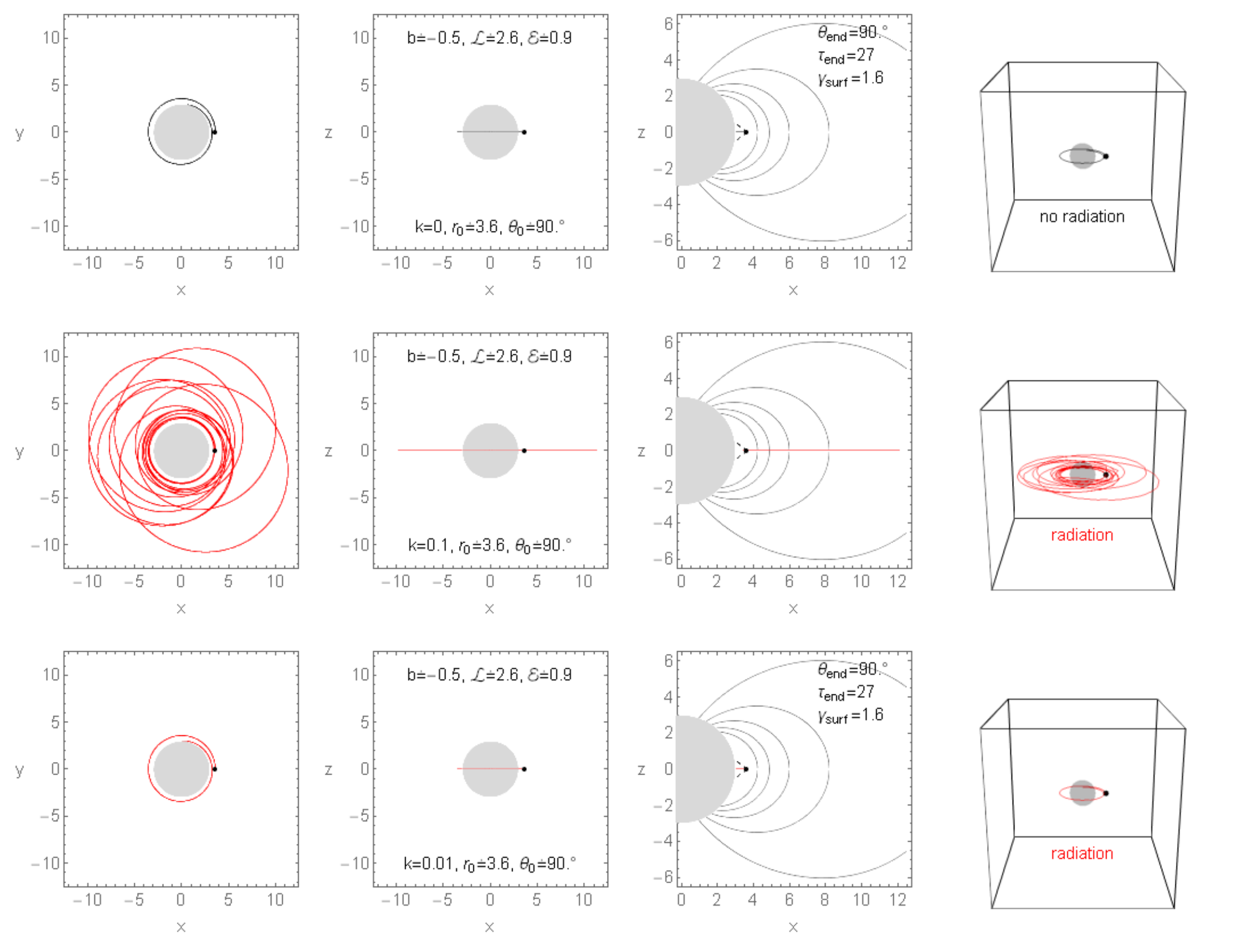}
	\caption{\label{f:f37a}The RR influence on the particle following an unstable equatorial circular orbit with $b$ parameter from the first region ($b=-0.5$) with repulsive Lorentz force, demonstrated for trajectories constructed with (red) and without (black) the RR force. The role of the reaction parameter $k$ is now crucial, leading to the widening for $k=0.1$, and to the fall for $k=0.01$.}
\end{figure*}

Widening due to magnetic repulsion appears even for the low values of the magnetic parameter, within the first regime ($-0.654 < b <0$). In this regime, the Lorentz force alone produces behavior of the equatorial circular orbits with similar magnetic attraction, as demonstrated in FIG.~\ref{f:f37a}. However, when unstable equatorial circular orbits exist close to the NS surface, the widening occurs only for sufficiently high RR parameter ($k=0.1$; see the second row of FIG.~\ref{f:f37a}). For small values of the reaction parameter ($k=0.01$), the particle falls onto the NS surface (see the third row of FIG.~\ref{f:f37a}). 

\begin{figure*}
    \centering
	\includegraphics[width=0.9\hsize]{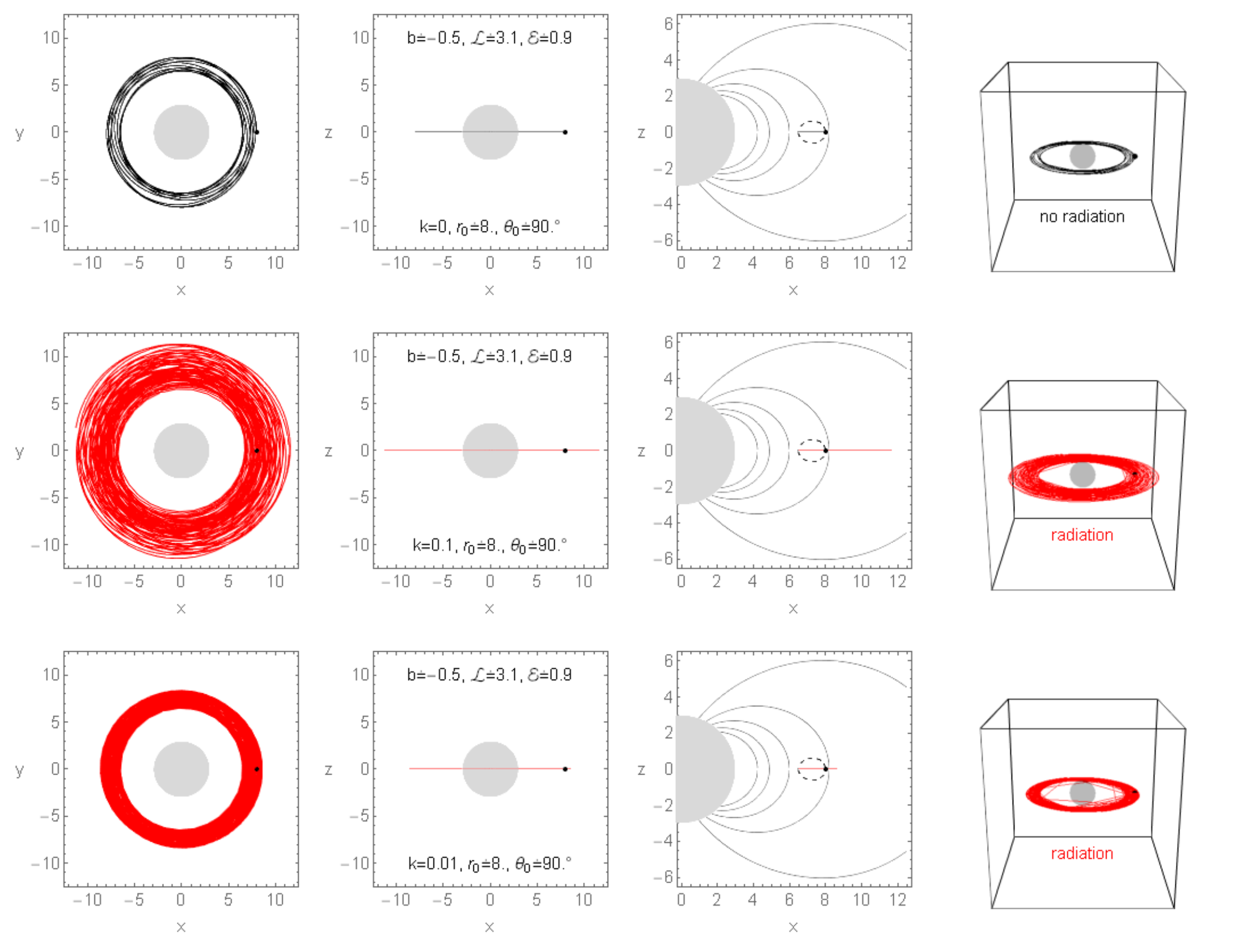}
	\caption{\label{f:f37b}The RR influence on the particle following a stable equatorial circular orbit with $b$ parameter from the first region ($b=-0.5$) with repulsive Lorentz force, demonstrated for trajectories constructed with (red) and without (black) the RR force. The role of the reaction parameter $k$ is now fixed to the widening for both $k=0.1$ and $k=0.01$, but efficiency of the widening decreases with decreasing $k$.}
\end{figure*}

When the orbital evolution due to the RR force begins at stable equatorial circular orbits, orbital widening is observed for all negative values of the magnetic parameter $b$, including the second and third regimes. This widening occurs regardless of the magnitude of the reactive parameter $k$, although its effectiveness decreases as $k$ decreases (see second and third row of FIG.~\ref{f:f37b}). 

\begin{figure*}
    \centering
	\includegraphics[width=0.9\hsize]{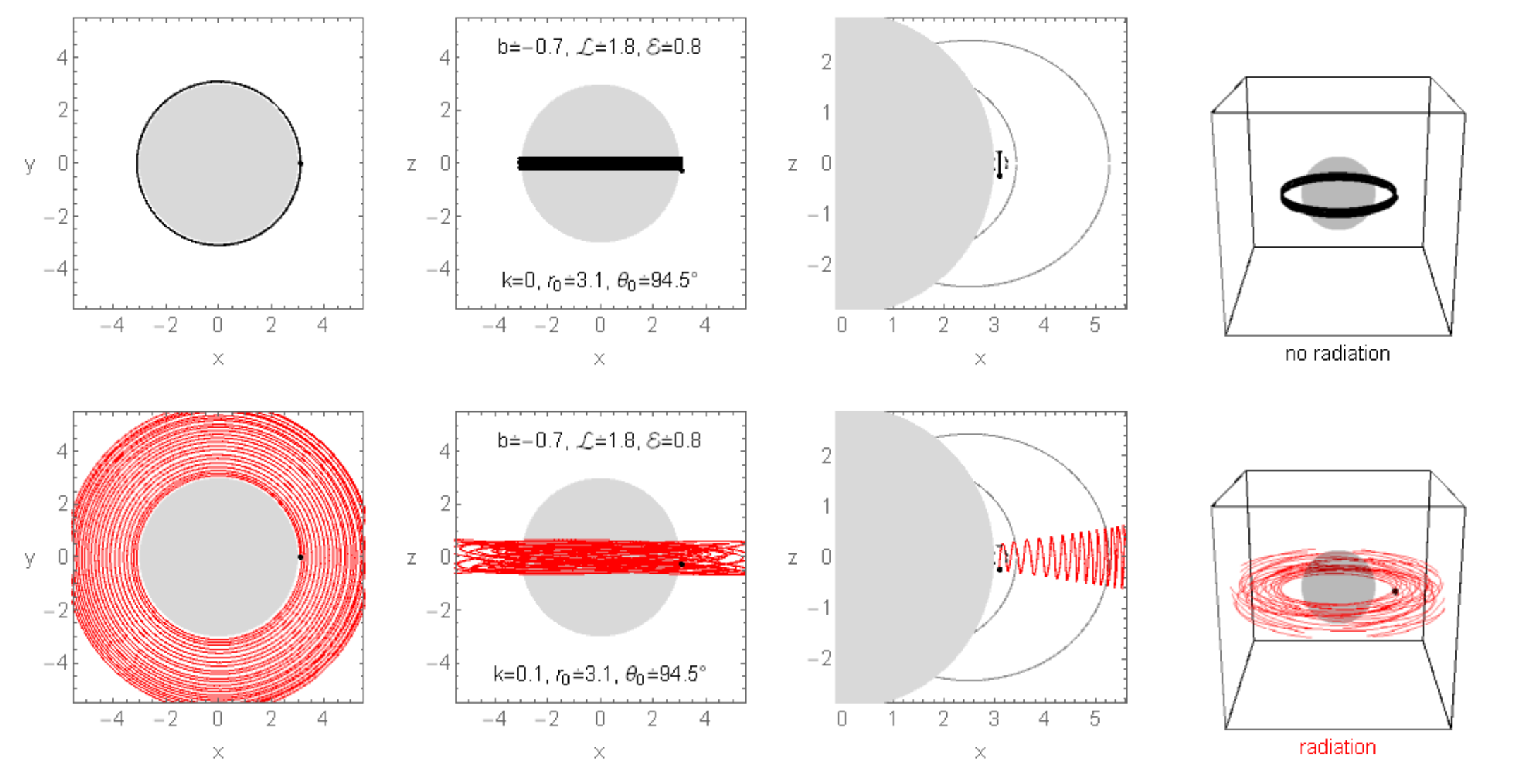}
	\caption{\label{f:f33}The RR influence on the particle oscillating around a stable equatorial circular orbit with $b$ parameter from the second region ($b=-0.7$) corresponding to the repulsive Lorentz force giving stable circular orbits at all $r>3$, demonstrated for trajectories constructed with (red) and without (black) the RR force. We observe widening with increasing magnitude of oscillatory motion around the equatorial plane.}
\end{figure*}

Orbital widening occurs similarly in the case of epicyclic motion for all values of the magnetic parameter $b<0$ (FIG.~\ref{f:f33}). In this scenario, the motion retains its oscillatory character around the equatorial circular orbits, with amplification of the oscillations in the vertical ($\theta$) direction. This vertical orbital widening is being reported for the first time and is likely caused by the inhomogeneity of the dipole magnetic field. Notably, the vertical orbital widening is not observed in the uniform magnetic field \cite{Tur-etal:2018:ApJ:}. 
%

\subsubsection{Off-equatorial circular orbits: widening vs fall}

Particles following off-equatorial circular orbits under the influence of the repulsive Lorentz force can exhibit two qualitatively different behaviors when the RR force is added. 

Orbital widening: In this scenario, particles are initially forced to shift toward the equatorial plane along a sequence of the off-equatorial. Once near the equatorial plane, the particles follow oscillatory motion around the equatorial plane that gradually expands to larger radii.

Direct fall: Alternatively, particles may approach the NS surface along the sequence of the off-equatorial, eventually ending by direct fall onto the surface. 

\begin{figure*}
    \centering
	\includegraphics[width=0.9\hsize]{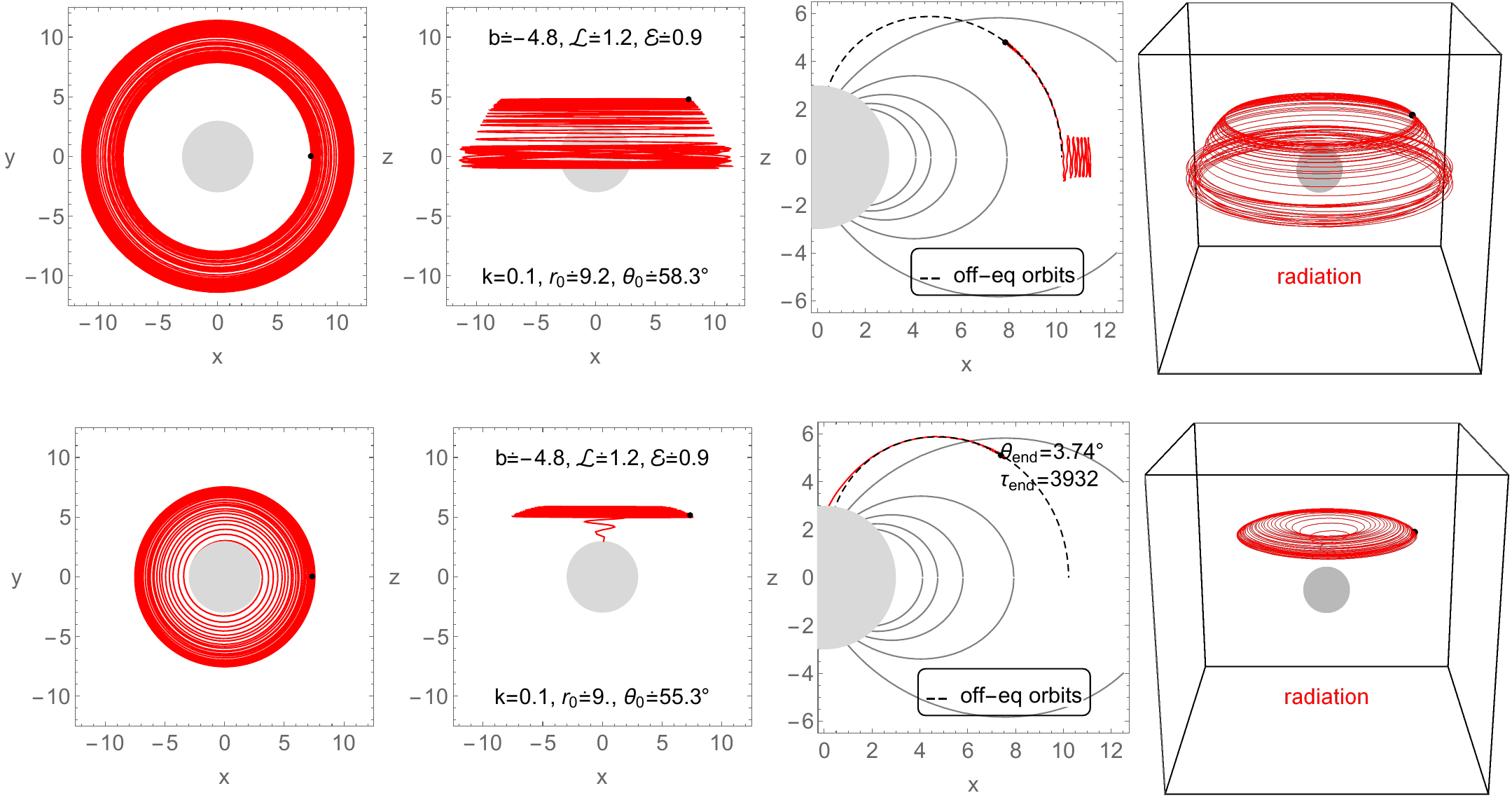}
	\caption{\label{f:f34}Widening vs. fall from off-equatorial circular orbits. Depending on the initial latitudinal angle $\theta_i$ of an off-equatorial circular orbit, the particle under a repulsive Lorentz force for the magnetic parameter $b<-0.751$ in the third region evolves under the RR force along the sequence of the off-equatorial circular orbits (black curve) constructed for the given value of $b$ to the NS surface for $\theta_i < \theta_{w/f}(b) < \pi/2$ and directly falls on to the surface in the final stages of the evolution, or evolves to the equatorial plane for $\theta_{w/f}(b) < \theta_i < \pi/2$ and follows oscillatory widening orbital motion around the equatorial plane.}
\end{figure*}

Under the RR force, a particle starting from an off-equatorial circular orbit slowly transitions along a sequence of off-equatorial circular orbits, either moving toward the equatorial plane or the NS surface. Small deviations occur in the vicinity of the equatorial plane, or near the NS surface (see FIG.~\ref{f:f34}). 

\begin{figure}
    \centering
	\includegraphics[width=0.66\hsize]{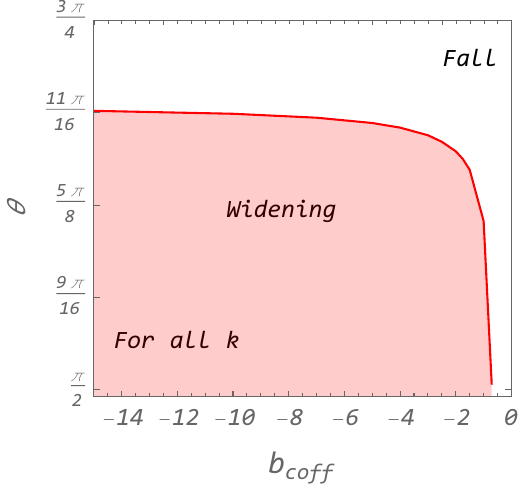}
    	\caption{\label{f:f36}Critical latitudinal angle $\theta_{w/f}(b)$ of the widening vs fall effect related to the off-equatorial circular orbits. The critical angle function is given by the red curve, the widening region is in red color, in white is the fall region. The critical angle depends only on the magnetic parameter $b$ being independent of the reaction parameter $k$. It is also relevant for the widening vs. fall effect related to the oscillatory and chaotic motion around the off-equatorial circular orbits.}
\end{figure}

These two behaviors under the influence of the RR force are separated by the critical latitude $\theta_{w/f}(b)$, which depends on the magnetic parameter $b$. For $\theta < \theta_{w/f}(b)$ ($\theta > \theta_{w/f}(b)$) the orbital widening (direct fall) occurs. The critical function $\theta_{w/f}(b)$ is numerically determined and illustrated in FIG.~\ref{f:f36}. This critical latitude depends on the magnetic parameter $b$, but it is independent of the RR parameter $k$. 

\begin{figure*}
    \centering
	\includegraphics[width=0.9\hsize]{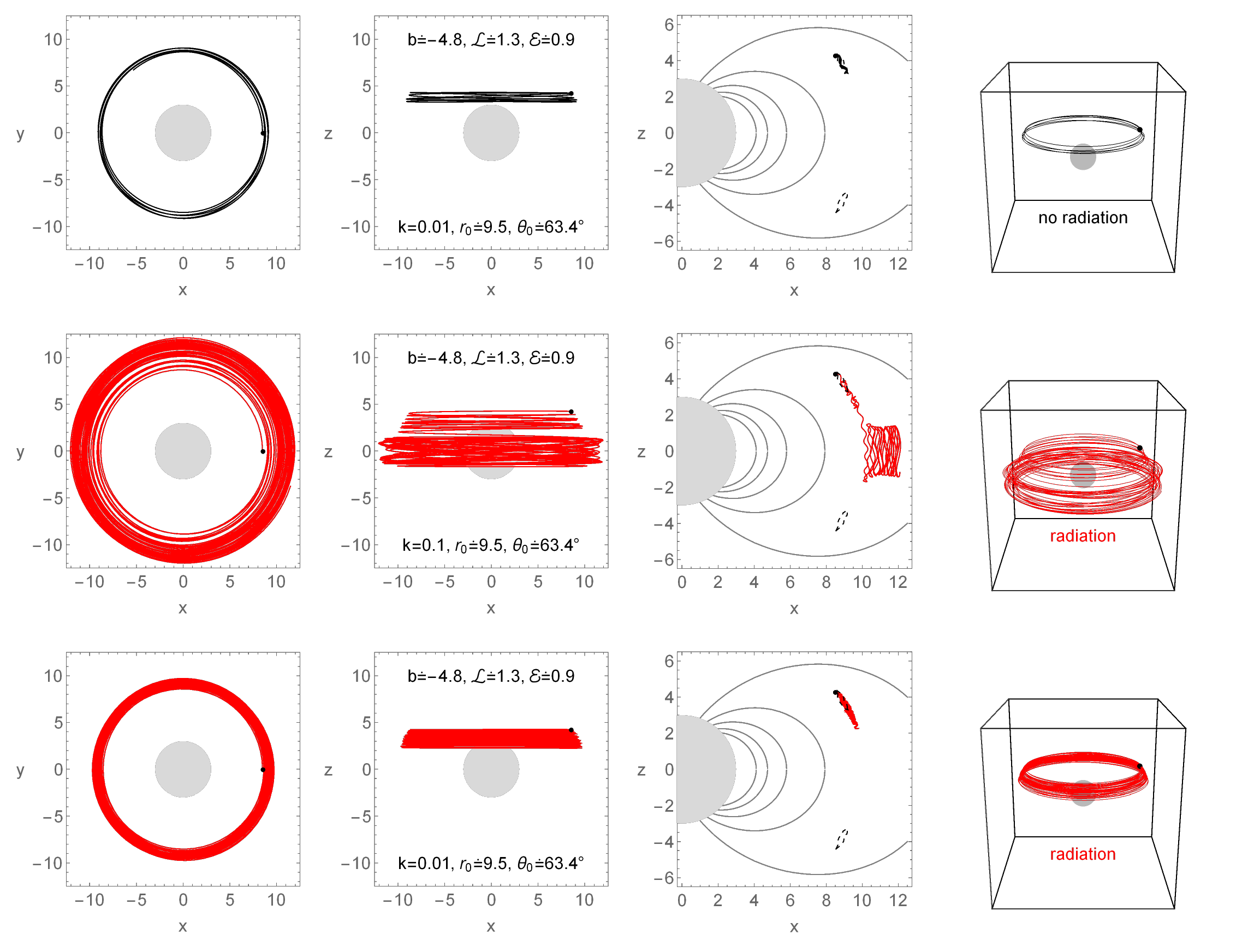}
	\caption{\label{f:f35c}Widening of the oscillatory motion around an off-equatorial circular orbit at $\theta_{w/f}(b) < \theta_i < \pi/2$. Motion under the repulsive Lorentz force is illustrated by black curves, with additional RR force with red curves. The role of the reaction parameter $k$ is also demonstrated.}
\end{figure*}

\begin{figure*}
    \centering
	\includegraphics[width=0.9\hsize]{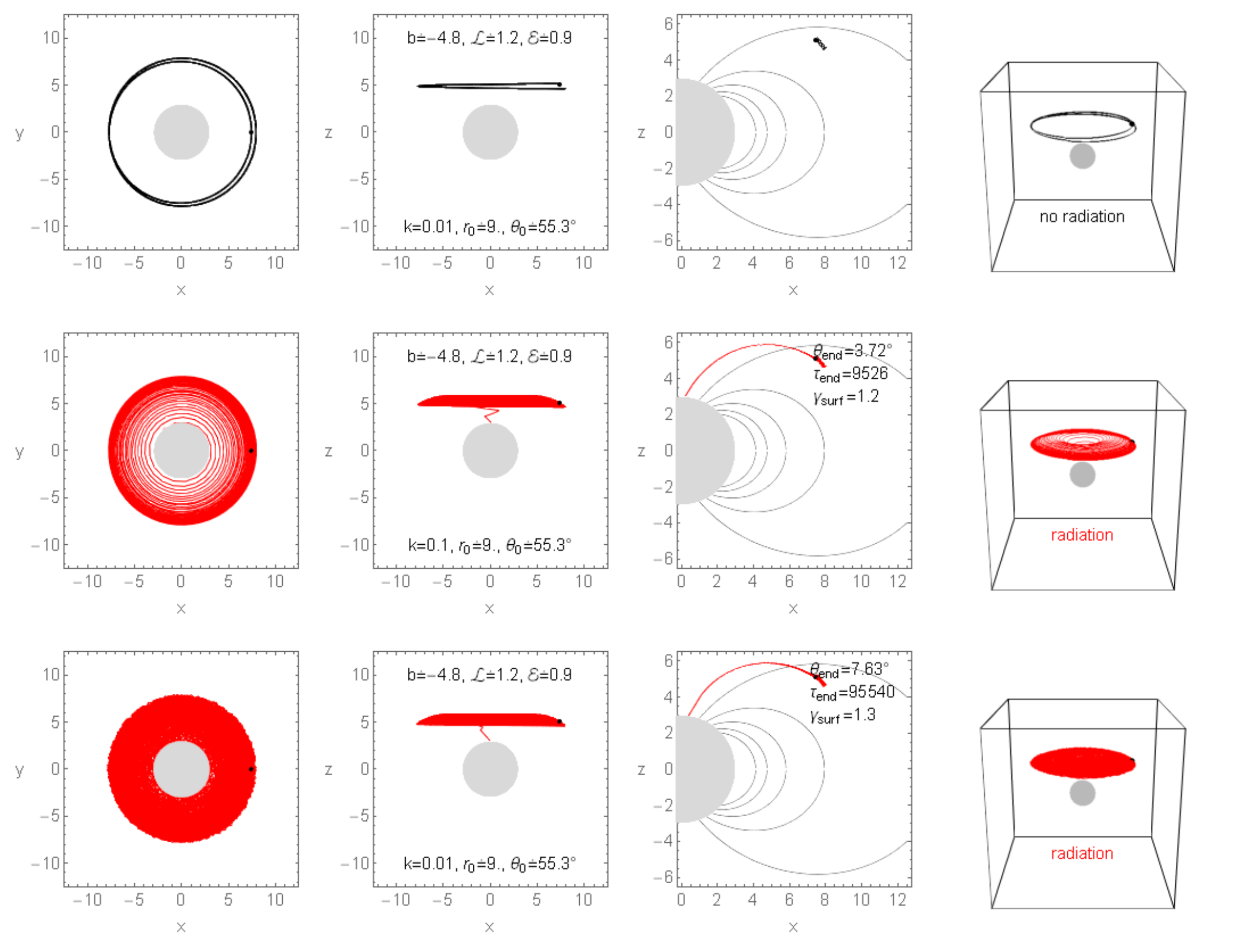}
	\caption{\label{f:f35d}Descend and fall onto the NS surface of the oscillatory motion around an off-equatorial circular orbit at $\theta_i < \theta_{w/f}(b) < \pi/2$. Motion under the repulsive Lorentz force is illustrated by black curves, with additional RR force with red curves. The role of the reaction parameter $k$ is also demonstrated.}
\end{figure*}

When the motion of a charged particle starts near an off-equatorial circular orbit, initially following limited epicyclic oscillatory motion, its evolution depends on the latitude $\theta_i$ of the initial off-equatorial circular orbit. If $\theta_i > \theta_{w/f}(b)$ ($\theta_i < \theta_{w/f}(b)$), the motion evolves in direction to the equatorial plane (to the NS surface), as demonstrated in FIGs~\ref{f:f35c} and \ref{f:f35d}. 
In the case of orbital widening, the slight oscillatory shifts along the sequence of the off-equatorial circular orbits toward the vicinity of the equatorial plane gradually transforms into chaotic oscillatory motion around the equatorial plane, accompanied by a slight outward shift from the NS (see FIG.~\ref{f:f35c}). 

In contrast, in the case of a fall, the initial epicyclic motion around the off-equatorial circular orbit shifts progressively along the sequence of the off-equatorial circular orbits, ultimately leading to impact with the NS surface (FIG.~\ref{f:f35d}). Notably, the time to fall onto the NS surface depends on the RR parameter $k$. As the RR parameter decreases from $k=0.1$ to $k=0.01$, the time to fall increases by an order of magnitude, while the latitude of the final impact point at the NS surface nearly doubles. The total evolution time increases as the RR parameter $k$ decreases. 
%

\subsubsection{Chaotic motion in the belts}

The large-scale chaotic motion in belts, initially governed solely by the Lorentz force, demonstrates similar effects under the additional influence of the RR force as in the previous cases of motion near circular orbits. 

\begin{figure*}
    \centering
	\includegraphics[width=0.9\hsize]{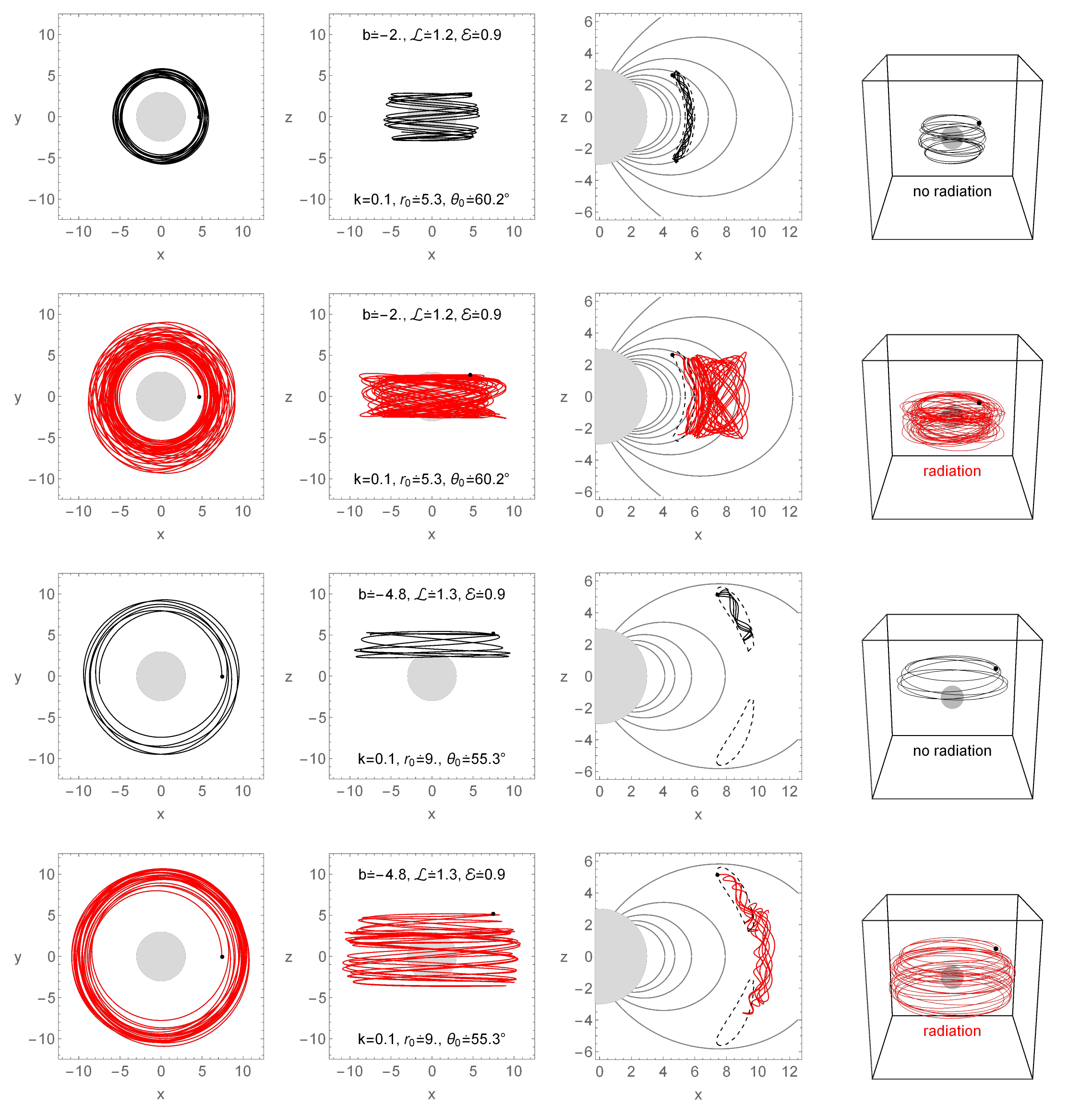}
	\caption{\label{f:f32a}The orbital widening related to the chaotic motion under repulsive Lorentz force allowing for existence of the off-equatorial circular orbit in the widening region. The orbits under both the Lorentz and RR forces are in red, the orbits under the Lorentz force only are in black. The widening effect occurs for both cases when the barrier of the motion under the Lorentz force is of island type (two bottom rows) or it is crossing the equatorial plane (two upper rows).}
\end{figure*}

When chaotic motion crosses the equatorial plane, the RR force gradually converts this motion into an orbital widening of chaotic nature, provided that the corresponding off-equatorial circular orbit, governing the barrier at the effective potential, lies in the region where the orbital widening occurs (FIG.~\ref{f:f32a}). 

In cases where the chaotic motion under the Lorentz force is confined by an island-type barrier that does not cross the equatorial plane, and the corresponding off-equatorial circular orbits of the orbital-widening type, the RR force causes the motion to eventually cross the equatorial plane. Afterward, the chaotic motion undergoes slow widening. 

\begin{figure*}
    \centering
	\includegraphics[width=0.85\hsize]{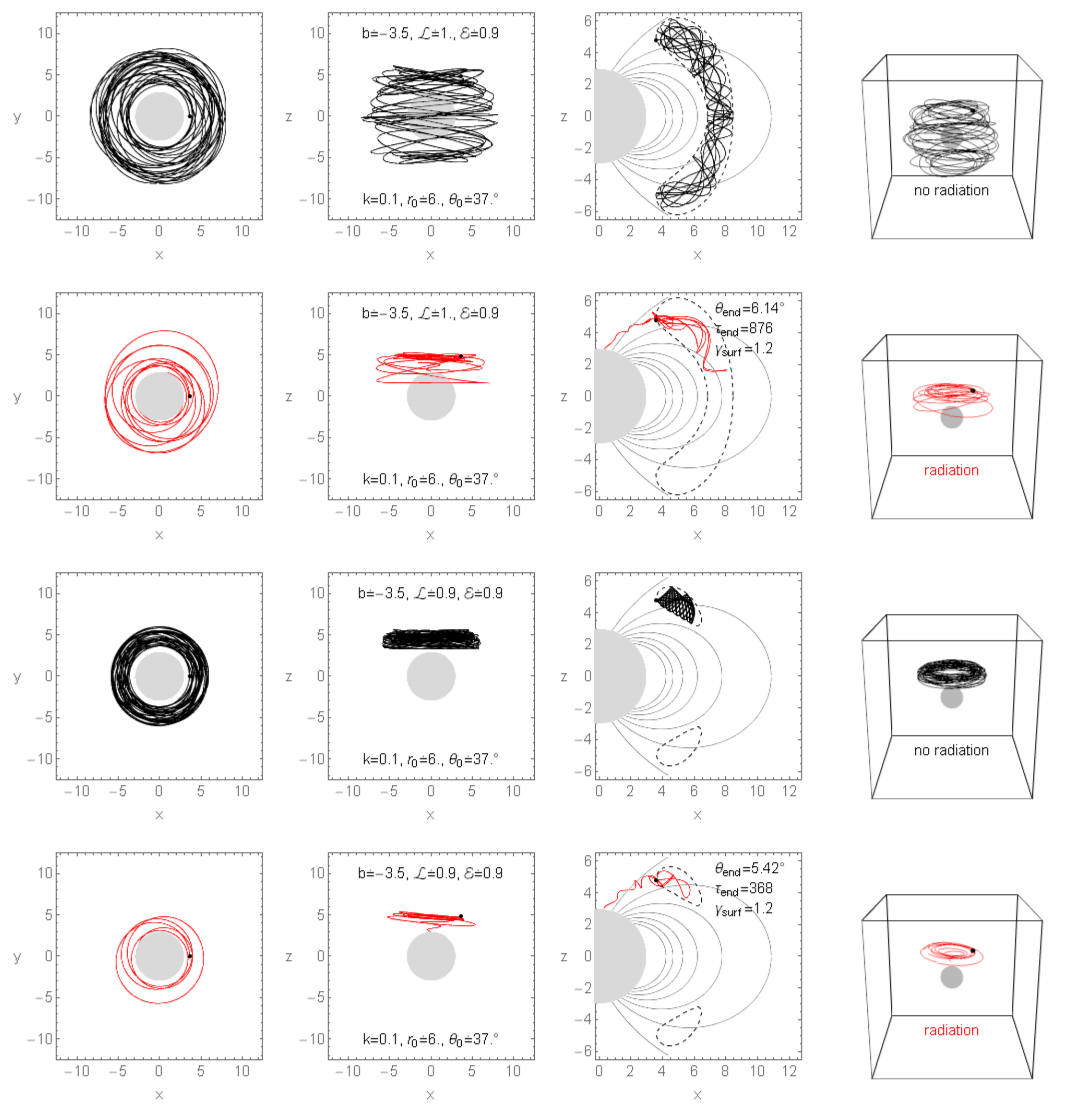}
    \includegraphics[width=0.85\hsize]{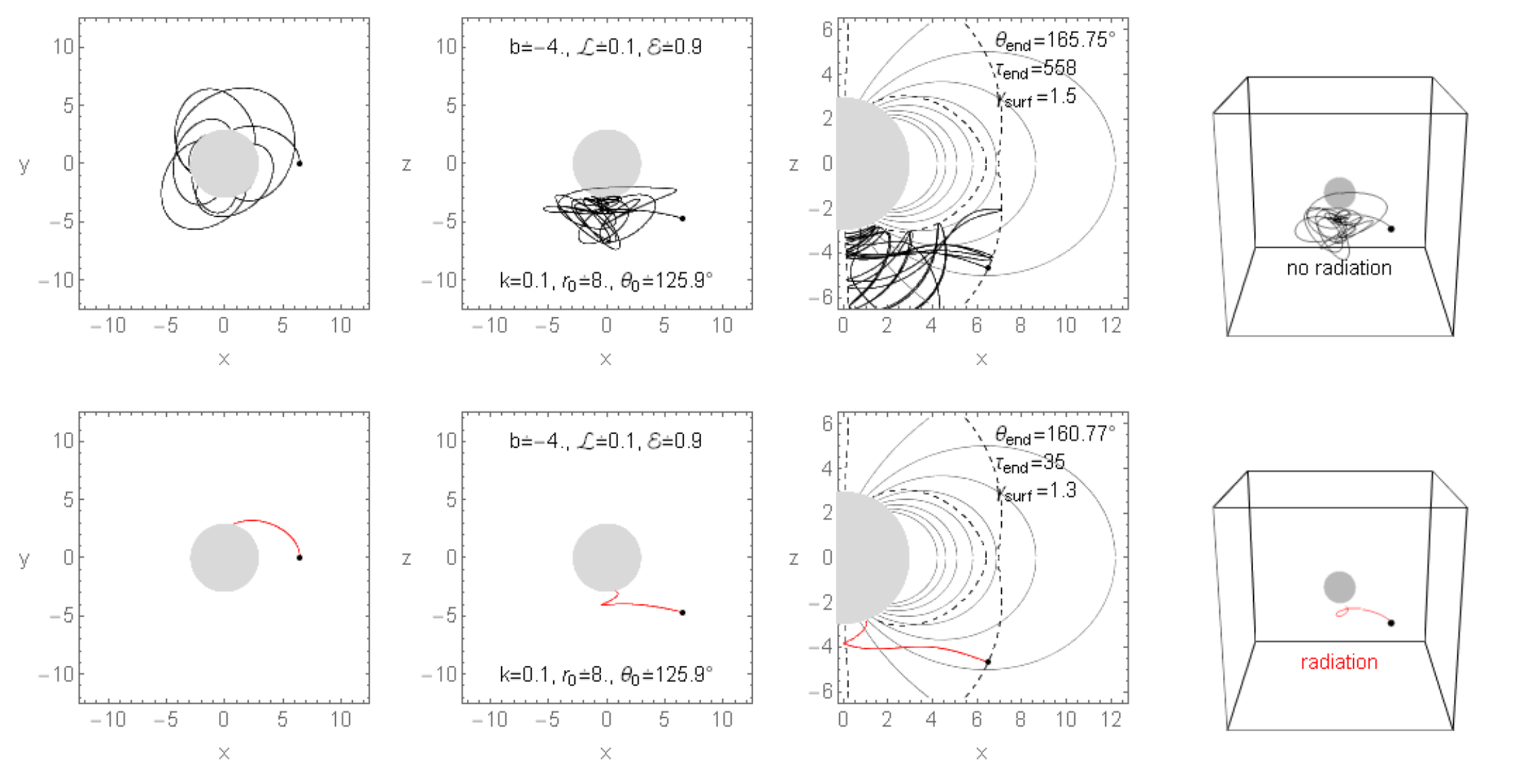}
	\caption{\label{f:f32b}The fall regime of the chaotic motion under the repulsive Lorentz force and additional RR force for the related off-equatorial circular orbit with initial angle in the fall regime. The fall is illustrated for both the cases when the effective barrier related to the Lorentz force is crossing the equatorial plane or is of the island type. For completeness, we also present the case where the barrier is open to the NS surface.}
\end{figure*}

On the other hand, when the barrier of the effective potential has a corresponding off-equatorial circular orbit at the "fall" region, the particle inevitably falls onto the NS surface due to the RR force effect. This outcome occurs whether the barrier forms an island off the equatorial plane or crosses the equatorial plane (FIG.~\ref{f:f32b}). For completeness, this figure also illustrates the case where the effective potential is open to the NS surface, allowing for complex, chaotic motion that ultimately results in a fall. In such cases, the RR force can significantly accelerate the fall of the particle. 

\begin{figure*}
    \centering
	\includegraphics[width=0.85\hsize]{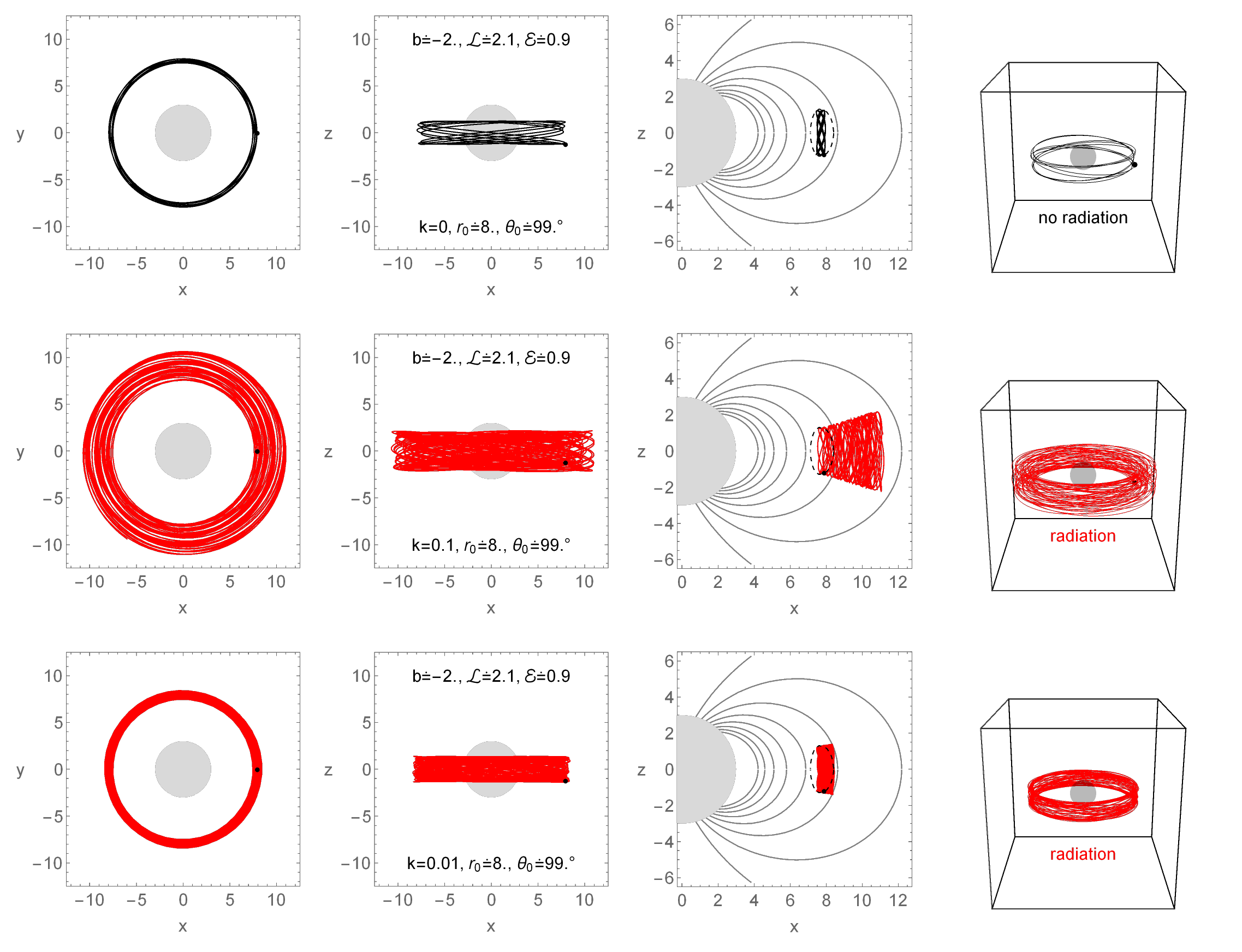}
	\caption{\label{f:f38}Orbital widening of the chaotic motion under the repulsive Lorentz force giving the equatorial barrier around a stable equatorial circular orbit.}
\end{figure*}

When the effective potential barrier is centered around an equatorial circular orbit, and no off-equatorial circular orbit exists for given parameters of the motion, orbital widening of the chaotic motion is naturally observed due to the influence of the RR force. This widening is accompanied by an increase in the amplitude of the oscillatory motion (FIG.~\ref{f:f38}). As expected, the efficiency of orbital widening decreases as the RR parameter $k$ decreases. Notably, the amplitude of oscillatory motion increases during the widening, a behavior characteristic of the dipole magnetic field. In contrast, this increase was not observed in the case of the orbital widening in the asymptotically uniform magnetic field around BHs \cite{Tur-Kol-Stu:2018:AN:}.

The magnitude of the dimensionless parameter $b$ used in this section ranges from $0.5$ to $5$. In this range, the attractive gravitational force and the repulsive or attractive Lorentz forces have comparable magnitudes, allowing us to describe the chaotic regime of the charged particle dynamics. The values of the RR parameter, $k=0.1$ or $k=0.01$, are deliberately chosen to be very high to clearly and effectively demonstrate the basic effects of the RR force on particle motion. For realistic values, refer to Table I. \footnote{For the possible role of the tail term in the RR force see \cite{Stu-Kol-Tur-Gal:2024:Uni:}.} Of course, such high values of the RR parameter $k$ are not realistic for describing the motion of protons, ions, or electrons in strong magnetic fields of NSs. 

When the Lorentz force is weak compared to the gravitational force ($|b|<<1$), the particle dynamics is primarily governed by the spacetime structure, leading to near-regular motion and a suppression of chaotic behavior. In this regime, the motion around stable circular orbits is typically epicyclic and harmonic, which can be used to explain the twin high-frequency quasiperiodic oscillations observed in binary systems containing a NS \cite{Vrb-Kol-Stu:2024:submitted:}. \footnote{An epicyclic harmonic motion is, of course, possible in a sufficiently small vicinity of stable circular orbits also in the chaotic regime of the motion.} However, when the Lorentz force dominates ($|b|>>1$), the particle dynamics is primarily controlled by the electromagnetic field whose structure is, of course, generally determined by the spacetime structure. In this case, motion along the Larmor circles is expected to dominate, and the chaotic character of the particle dynamics is once again suppressed. In such situation, the role of the RR force will be relatively weak (strong) for the case $|b|<<1$ ($|b|>>1$).  
%

\subsubsection{Evolution of the motion parameters due to the RR force}

To gain a deeper understanding of the role of the RR force in the motion of charged test particles, we need to study the time evolution of the key motion parameters, such as the specific energy and specific angular momentum under the combined influence of the gravitational, Lorentz, and RR forces. The time evolution of these characteristics revels the signatures of the RR force's influence on the motion. 

\begin{figure*}
    \centering
	\includegraphics[width=0.85\hsize]{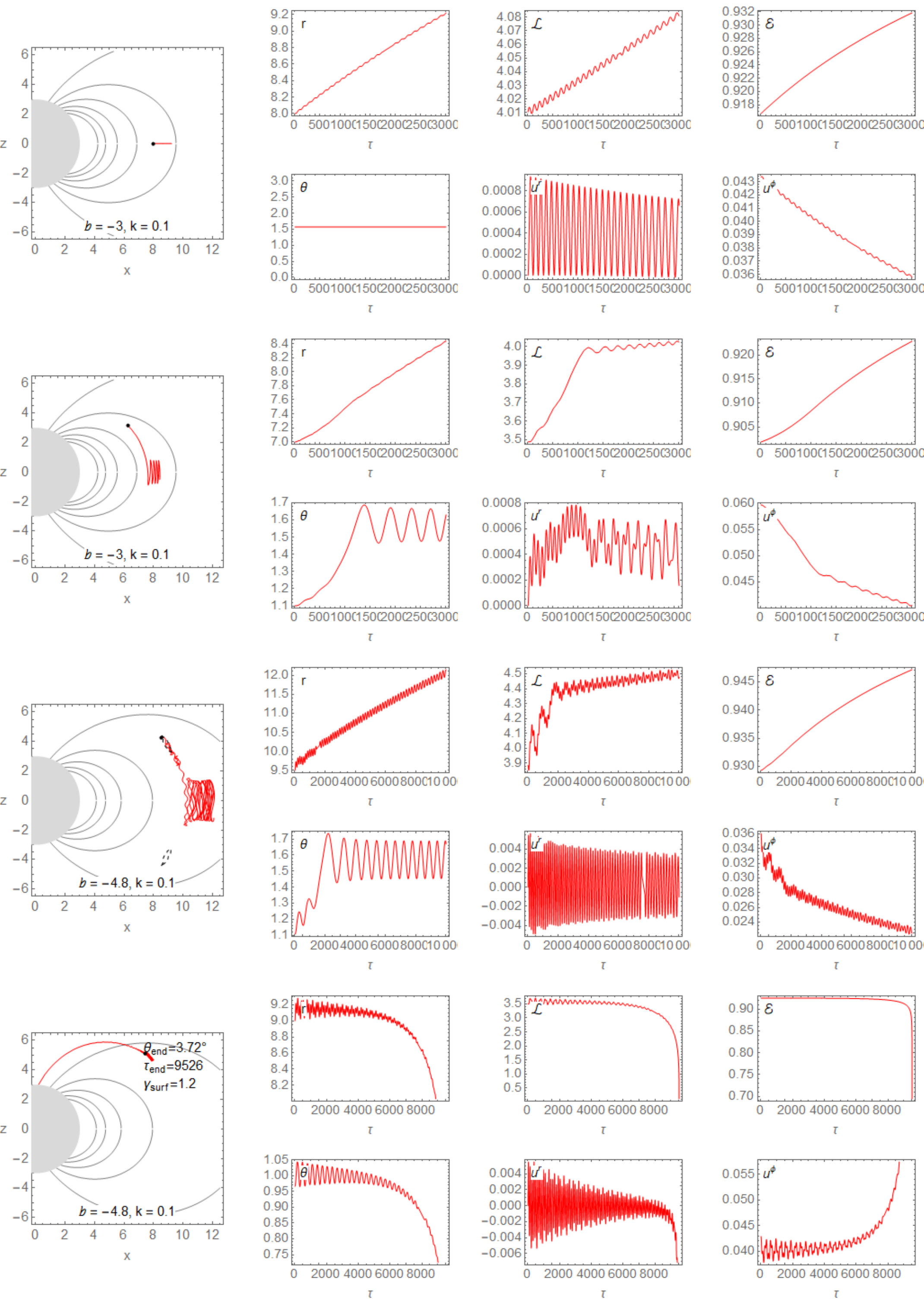}
    \caption{\label{f:f30}Demonstration of the effect of RR force on charged particles moving in a dipole field. The main panel on the left shows the particle's trajectory and the potential barrier for the particle without RR. The associated parameters as functions of time on the panels -- from the top left: the radial coordinate ($r$), the specific angular momentum ($\cl$), the specific energy ($\ce$), from the bottom left: latitudinal coordinate ($\theta$), the contravariant radial component ($u^r$) and the axial ($u^\phi$) components of the velocity. The top three rows present orbital widening, while the bottom row shows a fall onto the NS.}
\end{figure*}

We present the time evolution of these parameters for several selected cases from the classification discussed earlier, with a focus on the role of the RR force in the evolution of off-equatorial circular motion and the associated oscillatory or chaotic motion. The results are illustrated in FIG.~\ref{f:f30}.

It is immediately apparent that the orbital widening effect corresponds to an increase in both the specific energy and specific angular momentum of the particle, while the descent from off-equatorial circular orbits and subsequent fall onto the NS surface is related to a decrease in these two motion parameters. 

The increase of the particle specific energy and specific angular momentum during orbital widening is somewhat counter-intuitive and unexpected in the context of magnetic field background \cite{San-Car-Nat:2023:PRD:,San-Car-Nat:2024:PRD:}. Nevertheless, the De Witt-Brehme (the Landau-Lifshitz) equations which involve the second derivative of the $4-velocity$ (or covariant derivative of the Maxwell tensor) may account for this increase. In some sense, the "magnetic field can work", or an electric component of the field might be generated by the derivative of the Maxwell tensor, leading to an increase in the particle's specific energy. In fact, our preliminary calculations confirm that only the presence of the term $\frac{q\,k}{m} F^{\alpha}_{\,\,\,\beta ; \alpha} u^\beta u^\mu$ in the RR equations of motion is crucial for enabling the orbital widening effect.

Of course, the role of the tail term should also be investigated in detail, particularly in BH backgrounds. We plan to explore this in future studies, which will also address the approximations used in this paper and examine the astrophysical relevance of the orbital widening and its associated effects. 

\section{Radiation reaction influencing charged particles in the magnetic regime of motion}

Condition $|b|>>1$ is assumed for the magnetic regime of the charged particle motion. Numerical calculations show entering to the magnetic regime for $|b|>|b_{mag}|=10$, if motion near the NS surface is considered. However, for protons (ions) or electrons moving in the field of realistic magnetized NS, having $B_{surf} \geq 10^8$G, the magnetic parameter is by many orders exceeding the starting value of $|b_{mag}|=10$. To understand the behavior of protons and ions near the NS surface, we assume in this section very large magnitudes of the magnetic parameter $b$, related to the surface strength of the dipole magnetic field, and very low values of the RR parameter $k$. This arrangement allows to reflect nature of the proton motion along the symmetry axis near the NS surface, in comparison with the proton motion near the equatorial plane far away from the NS surface, where the effect of the magnetic field, represented by the parameter $b$, is suppressed as the dipole magnetic field intensity decreases faster in comparison with the decrease of the gravitational field intensity. 

In our treatment of the influence of the RR force on charged particles in the chaotic regime of motion, we used values of the parameters characterizing the motion that are by many orders smaller in comparison with parameters that correspond to elementary particles, if realistic, strongly magnetized, NSs are considered. We thus demonstrated the basic characteristics of the radiative phenomena in the chaotic regime of motion, but the obtained results are relevant for motion of charged dust, not of elementary particles. 

Results obtained in the present section are expected to give a view of the nature of proton (ion) motion near the surface of strongly magnetized NSs, presenting thus a more realistic illustration of the high-energy astrophysical processes near realistic NSs. 

Therefore, we consider motion of protons (ions) along the symmetry axis of the dipole magnetic field near the NS surface, for a particle moving in the magnetic regime under magnetic repulsion, starting from vicinity of the NS surface with energy allowing for the particle fall onto the NS surface. This case is compared with the particle motion at large distances from the NS surface with trajectories crossing the equatorial plane, where the chaotic regime of the motion is expected. We thus illustrate decreasing of the local influence of the magnetic parameter $b$ with increasing distance from the NS surface. 

First, we consider a scenario corresponding to a stable off-equatorial circular orbit in close vicinity of the NS surface near the symmetry axis of the magnetic dipole field, assuming the parameter $b$ corresponding to the Lorentz force acting on an proton or ion in the field of the test magnetized NS, and assuming very low values of the RR parameter $k$, approaching the realistic values for the test NS. Namely, we consider $b=-10^{5}$ and $k=10^{-8}$ and $k=10^{-10}$. 

\begin{figure*}
\begin{center}
    \includegraphics[width=0.9\hsize]{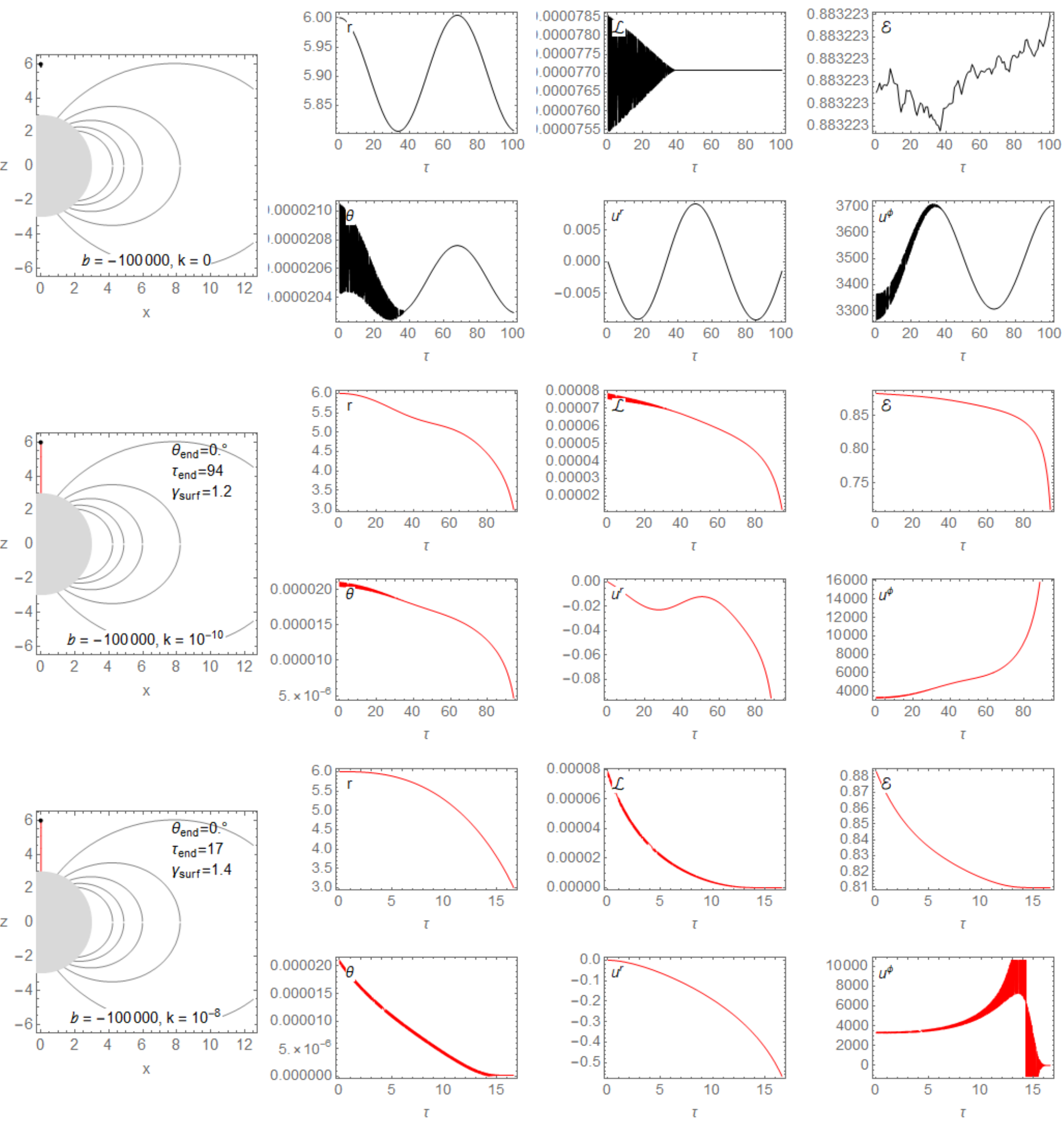}
\end{center}
	\caption{\label{f:f39}Trajectories representing magnetic regime under magnetic repulsion near the NS surface, constructed for high magnitude of the magnetic parameter $b$. The case without the RR force is on the first row (in black). The cases with the RR force (in red) are presented in the second and third rows, with different values of the RR parameter $k$, demonstrating thus its role.}
\end{figure*}

Results presented in FIG.~\ref{f:f39} demonstrate clearly the character of magnetic regime of motion: under the Lorentz force the particle slightly oscillates around the circular orbit, while small Larmor oscillations arise along the decreasing trajectory for the RR force included. The RR force causes an almost direct fall onto the NS surface, with an extremely small oscillatory character that decreases with decreasing RR parameter $k$, while the time of the fall increases with decreasing $k$; specific energy also decreases for the falling particle. These results are in agreement with our previous results demonstrating the classification of the RR force role in particle dynamics. 

\begin{figure*}
\begin{center}
    \includegraphics[width=0.85\hsize]{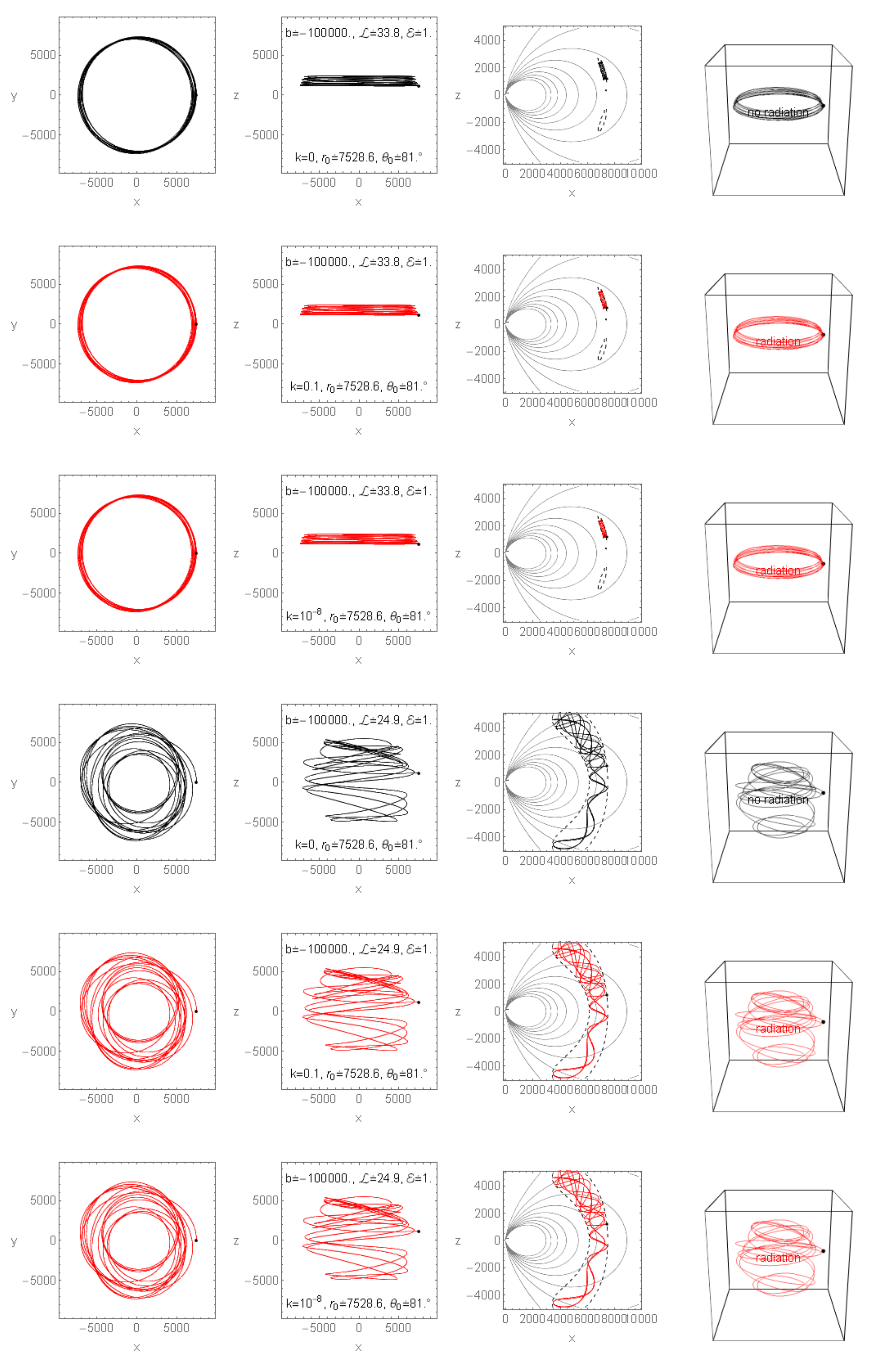}
	\caption{\label{f:f45}Trajectories representing magnetic regime under magnetic repulsion far away from the NS surface, constructed for high magnitude of the magnetic parameter $b$. The case without the RR force is on the first row (in black). The cases with the RR force (in red) are presented in the second and third rows, with different values of the RR parameter $k$. Two possibilities of the behavior of the potential barrier are considered: (i) barrier confined to region above the equatorial plane, (ii) barrier crossing the equatorial plane.}
\end{center}
\end{figure*}

\begin{figure*}
\begin{center}
    \includegraphics[width=0.85\hsize]{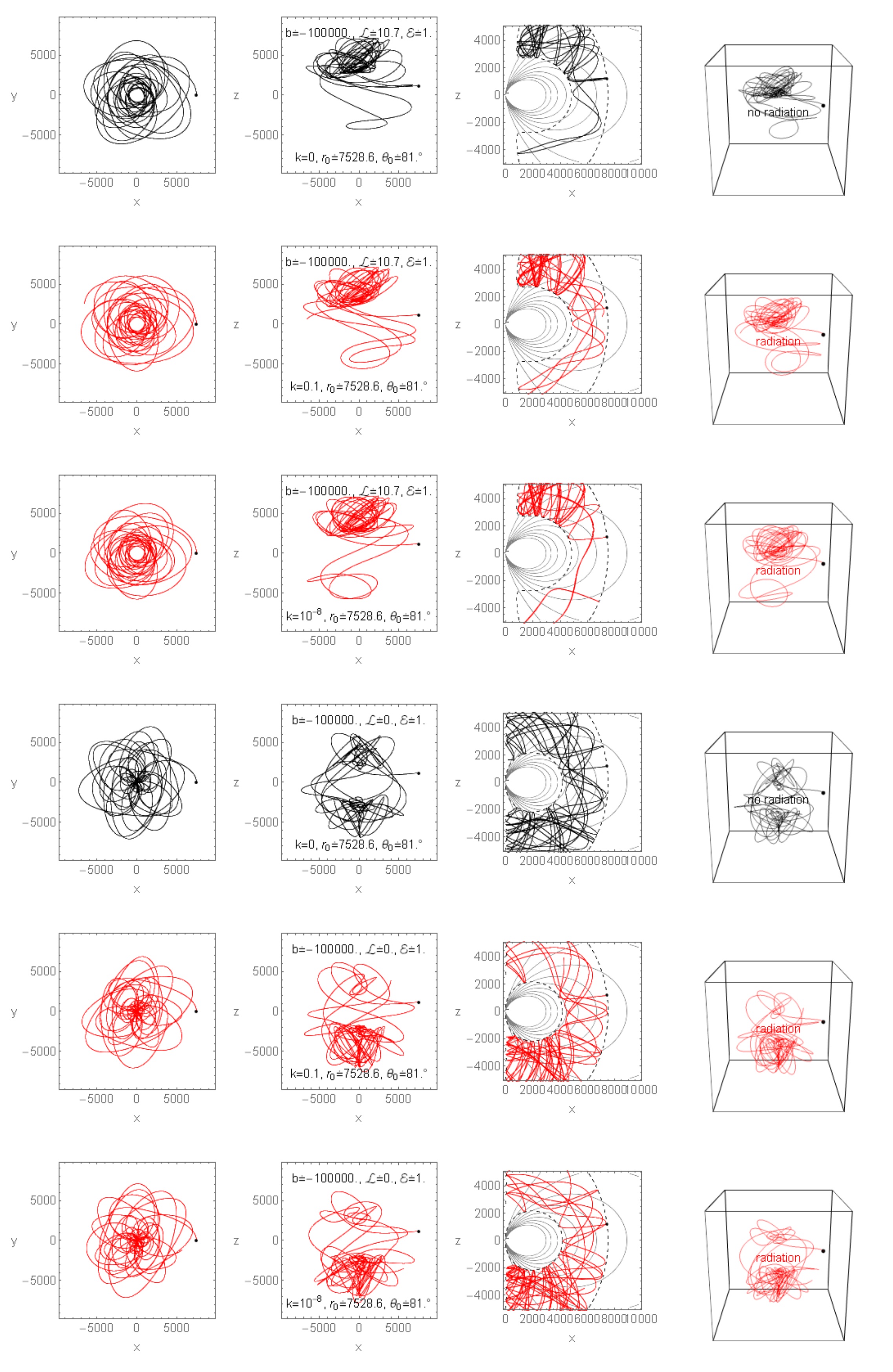}
	\caption{\label{f:f46}Trajectories representing magnetic regime under magnetic repulsion far away from the NS surface, constructed for high magnitude of the magnetic parameter $b$. The case without the RR force is on the first row (in black). The cases with the RR force (in red) are presented in the second and third rows, with different values of the RR parameter $k$. Two possibilities of the behavior of the potential barrier are considered: (i) barrier crossing the equatorial plane but separated from the NS surface, (ii) barrier crossing the equatorial plane and reaching the NS surface.}
\end{center}
\end{figure*}

Second, using in the case of magnetic repulsion the same parameters $b=-10^{5}$ as in the previous case, we consider the proton (ion) motion, starting near the marginally stable equatorial circular orbit of the given background, with various values of the particle specific angular momentum that governs the shape of the potential barrier determining the region allowed for the motion under the Lorentz force. We compare motion in a closed barrier located above the equatorial plane, with those in a narrow closed barrier crossing the equatorial plane in the first case (FIG.~\ref{f:f45}), and motion in an extended closed barrier crossing the equatorial plane with those in an extended barrier open to the NS surface in the second case (FIG.~\ref{f:f46}). The role of the RR parameter is tested for a large range of values, namely for $k=10^{-8}$ and $k=0.1$.  
  
FIGs \ref{f:f45} and \ref{f:f46} demonstrate clearly strongly chaotic character of the motion, i.e., we see the local decrease of the relevance of the magnetic parameter $b$ and shift from the magnetic regime near the surface to the chaotic regime in the large distance. Surprisingly, the influence of the magnitude of the RR parameter is rather suppressed. An interesting phenomenon is observed: even for equipotential surfaces (barrier) allowing for the fall onto the NS surface, it is very difficult to find the narrow throat allowing for this fall, even if the high values of the RR parameter are applied in calculations. Furthermore, we are not observing the orbital widening even in the case of narrow barriers. The reason is hidden in the low Lorentz factor of the motion in large distances from the NS surface, and the fact that the part of the LL equation responsible for the widening effect depends on the square of the Lorentz factor, implying thus strong suppression of its relevance. 


\begin{figure*}
\begin{center}
    \includegraphics[width=0.85\hsize]{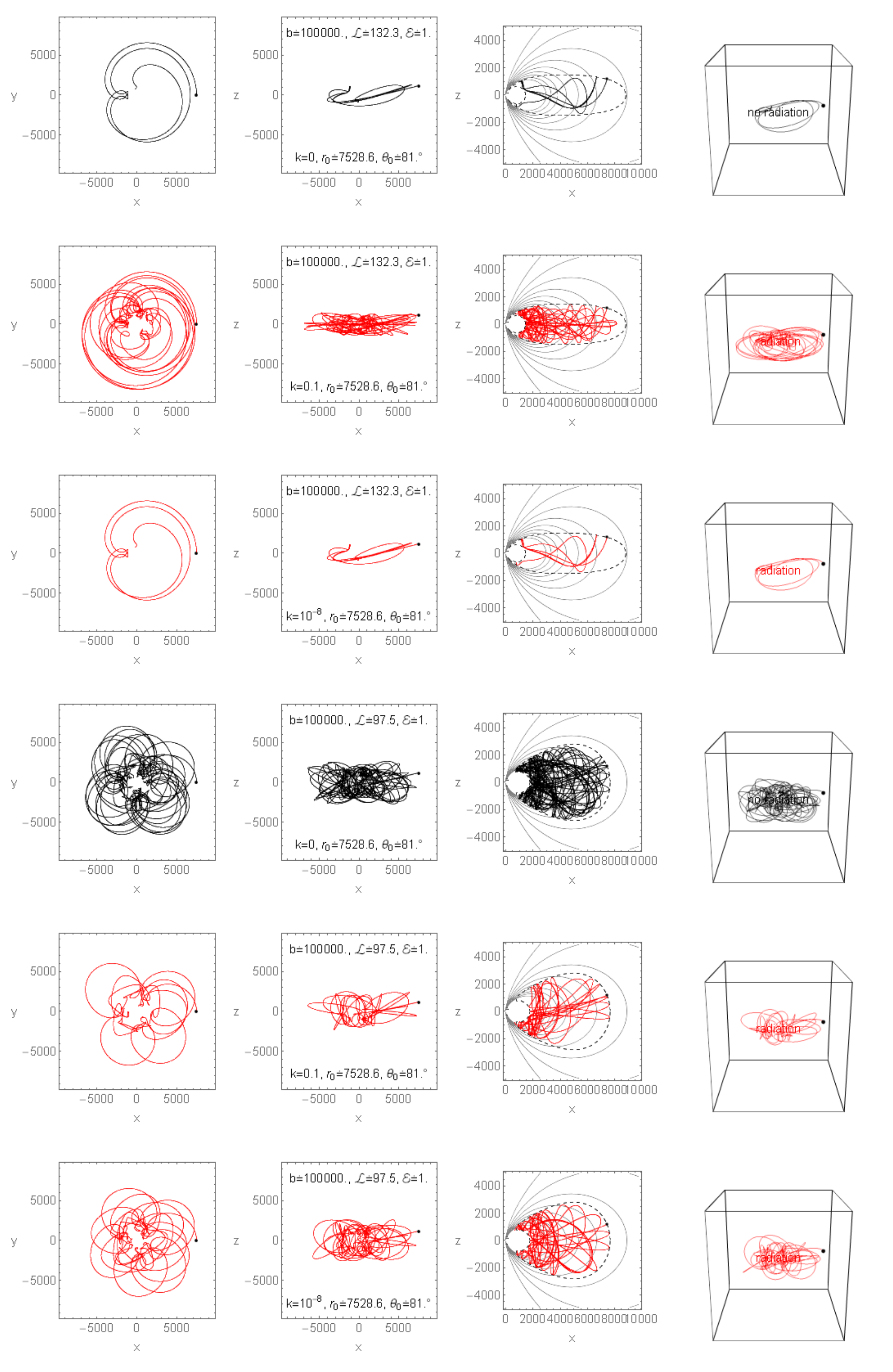}

	\caption{\label{f:f47}Trajectories representing magnetic regime under magnetic attraction, constructed for high magnitude of the magnetic parameter $b$ and two potential barriers extending from large distance to the NS surface along the equatorial plane. The case without the RR force is on the first row (in black). The cases with the RR force (in red) are presented in the second and third rows, with different values of the RR parameter $k$.}
\end{center}
\end{figure*}

\begin{figure*}
\begin{center}
    \includegraphics[width=0.85\hsize]{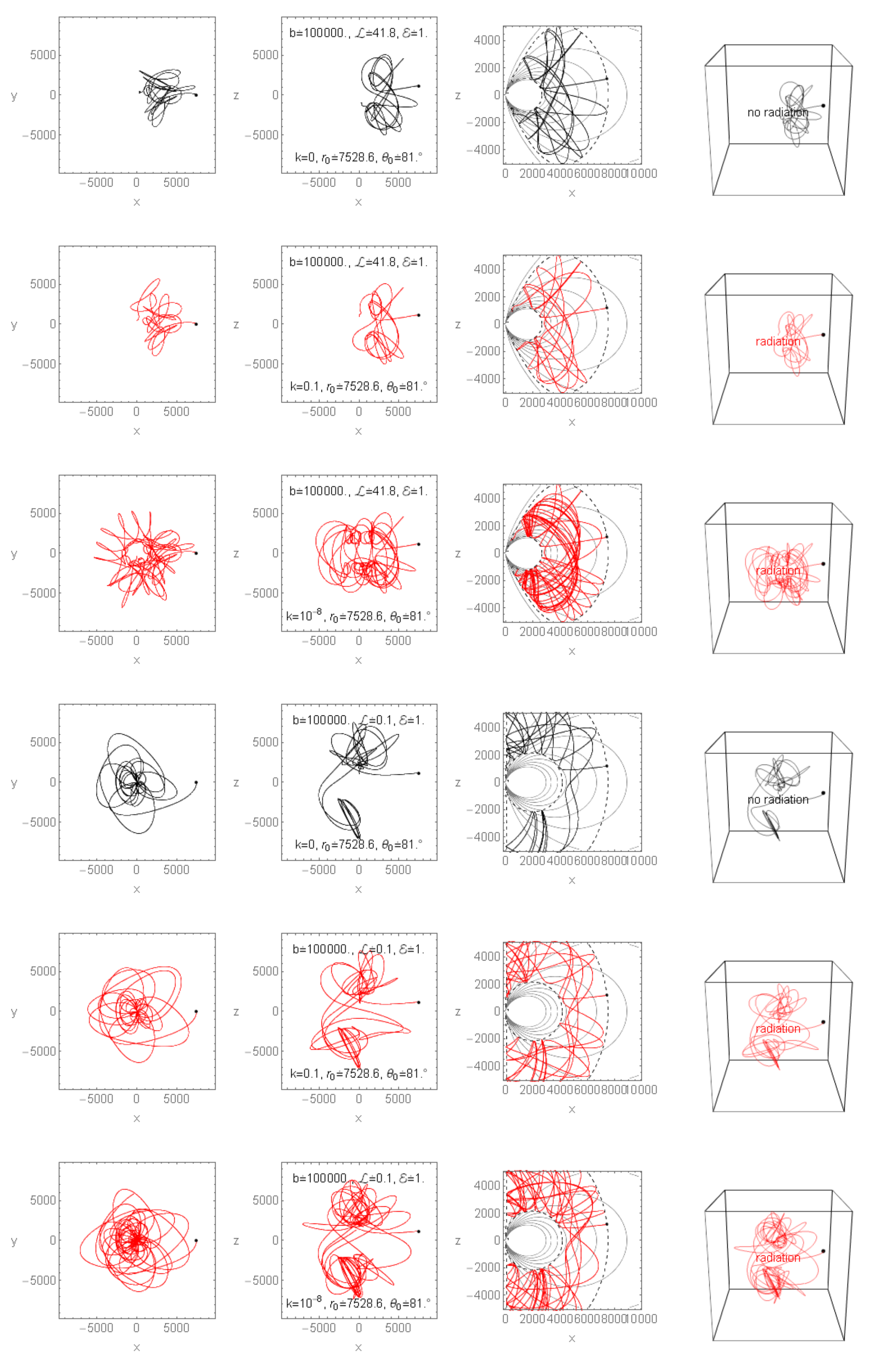}
	\caption{\label{f:f48}Trajectories representing magnetic regime under magnetic attraction, constructed for the high magnitude of the magnetic parameter $b$ and two potential barriers extending from large distance to the NS surface along the equatorial plane and extended across all possible values of latitudinal coordinate $\theta$. The case without the RR force is on the first row (in black). The cases with the RR force (in red) are presented in the second and third rows, with different values of the RR parameter $k$.}
\end{center}
\end{figure*}

Finally, we present an illustration of proton (ion) motion in the magnetic regime under magnetic attraction. Here we again assume a particle with parameter $b=10^{5}$ starting near the marginally stable equatorial circular orbit of the given background. The particle specific angular momentum, governing the shape of the potential barrier determining the region allowed for the motion under the Lorentz force, is chosen to give representative cases. We compare motion in a barrier located along the equatorial plane and open to the NS surface in the equatorial plane (FIG.~\ref{f:f47}), and motion in a barrier open to the NS surface near the symmetry axis (FIG.~\ref{f:f48}). The role of the RR parameter is again tested for a large range of values, namely for $k=10^{-8}$ and $k=0.1$.  
  
FIGs \ref{f:f47} and \ref{f:f48} demonstrate again strongly chaotic character of the motion. In the case of the motion in vicinity of the equatorial plane (FIG.~\ref{f:f47}) signatures of the magnetic regime appear in vicinity of the NS surface, as we observe a tendency to Larmor precession there. The influence of the magnitude of the RR parameter $k$ seems to be complex, as its increase can both decrease or increase the chaotic character of the motion. For motion approaching the symmetry axis of the magnetic field (FIG.~\ref{f:f48}), the chaotic character of the motion prevails in all considered cases. Increasing values of the parameter $k$ tends to slightly increase the chaotic character of the motion. 

Our results indicate that it could be very interesting and important to realize a detailed study of the radiation belts constituted from protons (ions) and electrons, orbiting the magnetized NSs, using e.g. out testing background. We plan to realize a detailed study in a future paper, where the situation observed in the case of van Allen radiation belts around the Earth has to be compared to the relativistic case of the testing NS. We predict that similar radiation belts, such as those recently observed around a brown dwarf \cite{2023Sci...381.1120C} should also be potentially observable in neutron star systems. We can expect some strong similarities between the NS background and the Earth's situation near the equatorial plane, where the belts of elementary particles are so distant that the gravity is very weak, the spacetime is nearly flat, and the magnetic field is significantly weakened. However, for particles moving in belts centered around the off-equatorial circular orbits approaching the NS surface and the symmetry axis of the dipole magnetic field, we expect relativistic phenomena significantly modifying the regime of dynamics in the nearly-flat spacetime.

\section{Conclusions}
The presence of a magnetic field around NS is of utmost importance in numerous astrophysical contexts, as its effects cannot be disregarded. Given the significant strength of these magnetic fields in the vicinity of observed NSs, ranging from $10^8$~G to $10^{15}$~G, they exert a substantial influence on the dynamics of charged test particles, particularly concerning their specific charge $\bar{q}/m$. In this article, we have extended previous studies of the dynamics of charged particles in the magnetosphere of non-rotating neutron stars, modelled using Schwarzschild spacetime and a dipole magnetic field that establishes the symmetry of the NS system. By employing the full framework of General Relativity, we expanded the investigation of the electromagnetic interaction, represented by the Lorentz force, on the equatorial and off-equatorial circular orbits, as well as the motion of charged particles within belts concentrated around the stable off-equatorial and equatorial orbits. However, the primary focus of our study was on the effect of radiative back-reaction on the motion of particles in both the equatorial and off-equatorial circular orbits and the associated belts. 

Our study focuses on the chaotic regime of the motion, characterized by $|b| \geq 1$, where the electromagnetic forces slightly exceed the gravitational forces. This provides a clear insight into the phenomena occurring in the close vicinity of the NS surface. In such cases, charged dust particles (or plasmoids) must be considered for realistic NS with $B > 10^{8}$ G. Nevertheless, the regime of magnetic dominance, where $b >> 1$, is also considered for the motion of protons or ions near the NS surface and along the symmetry axis of the magnetic dipole field. It is important to emphasize that, in the equatorial plane of realistic NSs, the effects that occur for charged dust near the NSs surface would occur at much larger distances from the surface for protons and ions, where the influence of the magnetic field is significantly reduced. The regime of gravitational dominance, where $b << 1$, was studied in detail in our previous work \cite{Vrb-Kol-Stu:2024:submitted:}.

We studied different types of bound or trapped trajectories for negative (positive) values of the magnetic parameter $b$, which correspond to the cases where gravitational force and Lorentz force are oppositely (identically) orientated. Several basic types of trajectories were presented. Particles falling onto the NS under the influence of a repulsive magnetic force do not fall radially onto the surface; their trajectories do not intersect the magnetic field lines. Instead, these particles are guided by the magnetic field towards the NS poles, where they can more easily penetrate the surface. This phenomenon is analogous to polar auroras observed on Earth or Jupiter.

We have investigated both equatorial and non-equatorial circular orbits. The non-equatorial corotating circular orbits occur only for negative values of the magnetic parameter $b$ and they smoothly transition into equatorial orbits, which loose their stability against vertical perturbations. The radial and latitudinal epicyclic frequencies, as well as the orbital (Keplerian) frequency associated with the epicyclic orbital motion around both the equatorial and off-equatorial circular orbits, were studied and applied to explain observational data from NS binaries in \cite{Vrb-Kol-Stu:2024:submitted:}. In this work, we have extended the study by examining the energy relations between the equatorial and off-equatorial circular orbits, which are crucial for understanding the motion in the belts surrounding the NSs. 

Using the Landau-Lifshitz equation to describe the influence of the radiative back-reaction on the motion of the charged particles, we were able to classify the effects of back-reaction based on the qualitative character of the motion, which is primarily governed by the Lorentz equation. 

Due to the role of the RR force, we distinguish between the cases of direct fall onto the NS surface and the cases of the orbital widening. 

A direct fall onto the NS surface can occur in the radial direction, close to the equatorial plane, and only in the case of an attractive Lorentz force. However, a direct fall along magnetic field lines, with impact near the symmetry axis of the magnetized NSs, can occur for both repulsive and attractive Lorentz forces. 

The widening of the circular orbits is associated with particles under magnetic repulsion. It is noteworthy that orbital widening is a general phenomenon related to motion under the repulsive Lorentz force, even in cases where the magnetic parameter $b$ is so small that the motion resembles that under an attractive Lorentz force, regardless of the magnitude of the RR parameter $k$. There is one interesting exception: in the case of motion along unstable circular orbits near the NS surface, a sufficiently high reaction coefficient $k$ enables orbital widening, while a weaker reaction coefficient leads to the particle falling onto the NS. 

For off-equatorial motion originating in some "island" region of the effective potential, the distribution of the "direct fall" and "widening" regimes is determined by a critical angle, which is related to the location of the off-equatorial circular orbit determined by the magnetic parameter $b$ and initial angular momentum of the considered particle. As intuitively expected, the widening effect occurs for positions sufficiently close to the equatorial plane. We have shown that the orbital widening occurs not only for the case of an asymptotically uniform magnetic field around a black hole \cite{Tur-etal:2020:ApJ:,Tur-Kol-Stu:2018:AN:,Stu-etal:2020:Uni:}, but also in the dipole magnetic field around NSs. The widening effect exhibits broader characteristic features of the widening in the dipole field compared to the asymptotically uniform field. For instance, in the dipole magnetic background we observe an increasing chaotic nature of the widening effect with increasing RR force. This contrasts with the behavior in the asymptotically uniform magnetic fields, where the RR force reduces the chaotic nature of the motion during orbital widening \cite{Tur-etal:2020:ApJ:,Tur-Kol-Stu:2018:AN:}. 

The results presented in the article pertain to the widening of the circular orbits caused by the combined effect of magnetic repulsion and radiative back-reaction, which warrant further detailed studies. Special attention should be given to the boundary between the widening and fall effects of the RR force. In the case of magnetic attraction, the focus should remain on direct fall onto the NS surface due to the RR force. Future studies will address both the chaotic nature of the off-equatorial motion and the potential to fit the observational data of HF QPOs through detailed analyses of epicyclic oscillations of dust or plasmoids in the dipole magnetic field around NSs. Another intriguing area for future studies involves the behavior of radiation belts constituted by elementary particles orbiting magnetized NSs, with possible similarities to the radiation belts observed around Earth, which could offer insights in the regime of weak gravitational and magnetic fields at large distances from the NSs.

A fundamental open problem is the physical origin of the counter-intuitive orbital widening effect, particularly in relation to the energy balance and the role of the tail term \cite{San-Car-Nat:2023:PRD:,Stu-Kol-Tur-Gal:2024:Uni:,San-Car-Nat:2024:PRD:}. The origin of the orbital widening is clearly linked to the second derivative of the $4-velocity$ in the De Witt-Brehme basic equation and the crucial role played by the term containing the Maxwell tensor derivative in the LL equation. Another important open question concerns the role of the tail term, which requires detailed future studies, along with the broader relevance of the radiative back-reaction in various astrophysical contexts. Other interesting challenges are linked to the inclusion of collective and quantum phenomena for processes in strong dipole magnetic fields.

\begin{acknowledgments}
The authors acknowledge the institutional support of the Institute of Physics, Silesian University in Opava. M.K. acknowledges the project (GA\v{C}R) No.~\mbox{23-07043S} of the Czech Science Foundation.
\end{acknowledgments}


%
\bibliography{references}

\end{document}